\documentclass[12pt]{article}
\usepackage{amssymb,amsmath,amscd}
\usepackage{epsfig}
\usepackage{epstopdf}
\usepackage{latexsym}
\usepackage{graphicx}
\usepackage{booktabs}
\usepackage{bbm}
\usepackage[margin=15pt,small]{caption}
\usepackage{subcaption}
\usepackage{enumitem}
\usepackage{float}
\usepackage[normalem]{ulem} 
\usepackage[numbers,compress]{natbib}
\usepackage{tikz-cd}
\usepackage{tikz}
\usetikzlibrary{matrix,arrows,backgrounds}

\usepackage[T1]{fontenc}

\usetikzlibrary{positioning}

\usepackage{xcolor}

\usepackage{datetime}

\usepackage[
      colorlinks=false,
      linkcolor=darkblue,  
      urlcolor=blue,    
      filecolor=blue,     
      citecolor=red,
linktocpage=true,
      pdfstartview=FitV,
      bookmarksopen=true    
      ]{hyperref}

\setlist{nolistsep}

\numberwithin{equation}{section}

\ifpdf
\fi

\let\oldbibliography\thebibliography
\renewcommand{\thebibliography}[1]{\oldbibliography{#1}
\setlength{\itemsep}{0pt}} 

\newcommand*{\boxedcolor}{red}
\makeatletter
\renewcommand{\boxed}[1]{\textcolor{\boxedcolor}{%
  \fbox{\normalcolor\m@th$\displaystyle#1$}}}
\makeatother

\definecolor{cardinal}{rgb}{0.6,0,0}
\definecolor{darkgreen}{rgb}{0,0.5,0}
\definecolor{golden}{rgb}{0.92, 0.7, 0}
\definecolor{midnight}{rgb}{0, 0, 0.5}
\definecolor{darkblue}{rgb}{0.2, 0, 0.8}

\def\Re{{\rm Re}} \def\Im{{\rm Im}}

\def\coeff#1#2{\relax{\textstyle {#1 \over #2}}\displaystyle}

\def\ZZ{\mathbb{Z}}

\def\cN{{\cal N}}

\def\eql{~=~}
\def\eeql{~\equiv~}
\def\cals#1{\mathcal{#1}}

\def\eql{=}

\def\SO{{\rm SO}}

\def\SU{{\rm SU}}

\def\fg{\frak{g}}

\def\fn{\frak{n}}

\def\fq{\frak{q}}

\def\fF{\frak F}
\def\fG{\frak G}
\def\fW{\frak W}

\def\bfs#1{{\boldsymbol{#1}}}

\def\cals#1{\mathcal{#1}}

\def\bDelta{\Delta}
\def\i{{\rm i}}
\def\r{\Delta}
\def\cfb{{\Phi}}
\def\fGG{{\frak G}}

\begin{document}  

\begin{titlepage}
 
\medskip
\begin{center} 
{\Large \bf   Mass-deformed ABJM and Black Holes in  AdS$_{\bfs 4}$}

\bigskip
\bigskip
\bigskip
\bigskip

{\bf Nikolay Bobev,${}^{(1)}$ Vincent S. Min,${}^{(1)}$ and  Krzysztof Pilch${}^{(2)}$  \\ }
\bigskip
${}^{(1)}$
Instituut voor Theoretische Fysica, KU Leuven,\\ 
Celestijnenlaan 200D, B-3001 Leuven, Belgium
\vskip 5mm
${}^{(2)}$ Department of Physics and Astronomy \\
University of Southern California \\
Los Angeles, CA 90089, USA  \\
\bigskip
\tt{nikolay.bobev@kuleuven.be,~vincent.min@kuleuven.be,~pilch@usc.edu}  \\
\end{center}

\bigskip
\bigskip
\bigskip
\bigskip

\begin{abstract}

\noindent  
\end{abstract}

\noindent We find a class of new supersymmetric dyonic black holes in four-dimensional maximal gauged supergravity which are asymptotic to the ${\rm SU(3)}\times {\rm U(1)}$ invariant AdS$_4$ Warner vacuum. These black holes can be embedded in eleven-dimensional supergravity where they describe the backreaction of M2-branes wrapped on a Riemann surface. The holographic dual description of these supergravity backgrounds is given by a partial topological twist on a Riemann surface of a three-dimensional $\cN=2$ SCFT that is obtained by a mass-deformation of the ABJM theory. We compute explicitly the topologically twisted index of this SCFT and show that it accounts for the entropy of the black holes.

\vfill

\end{titlepage}

\newpage

\setcounter{tocdepth}{2}
\tableofcontents

\newpage

\section{Introduction}

Holography has evolved into an indispensable tool to study the dynamics of strongly coupled quantum field theories. In addition, this duality can be used  to learn new lessons about the structure of black holes. For a long time, an important outstanding question in black hole physics has been to account microscopically for the entropy of asymptotically AdS black holes in more than three dimensions.\footnote{Here we are focusing on black holes and do not discuss higher-dimensional black branes.} While this problem still remains open for black holes in five or more dimensions, recently there has been a rapid  progress in understanding the microstate counting for supersymmetric black holes in AdS$_4$ \cite{Benini:2015eyy,Benini:2016rke,Hosseini:2016tor,Hosseini:2016ume,ABCMZ}. These  developments were triggered by employing the tools of supersymmetric localization (see \cite{Pestun:2016zxk} for a recent review) to define and compute a suitable partition function, called ``topologically twisted index'' \cite{Benini:2015noa,Benini:2016hjo,Closset:2016arn}, which can be used to count the microstates of these black holes.

The basic idea of the recent work is to engineer a black hole in M-theory\footnote{See also \cite{ABCMZ,Guarino:2017eag,Guarino:2017pkw,Hosseini:2017fjo,Benini:2017oxt} for an extension of these results to asymptotically AdS$_4$ black holes in massive IIA string theory.} which is asymptotic to an AdS$_4\times M_7$ solution, where  $M_7$ is a Sasaki-Einstein manifold. The horizon of such  four-dimensional black holes is a compact Riemann surface, $\Sigma_{\fg}$. This gravitational background in turn is holographically dual to a three-dimensional $\mathcal{N}=2$ SCFT of the ABJM type \cite{Aharony:2008ug,Bagger:2007jr,Bagger:2007vi} placed on $\mathbb{R}\times \Sigma_{\fg}$ with a partial topological twist. For such twisted three-dimensional SCFTs the supersymmetric partition function was studied in \cite{Benini:2015noa,Benini:2016hjo,Closset:2016arn} and it reduces to a matrix model due to supersymmetric localization. In the planar limit of a large number, $N$, of coincident M2-branes, one can solve this matrix model and obtain the free energy of the twisted SCFT to leading order in $N$.\footnote{See \cite{Liu:2017vll,Jeon:2017aif,Liu:2017vbl} for recent attempts to account for subleading corrections in $N$.} This in turn reproduces the entropy of the black hole. This procedure is best studied for black holes in  eleven-dimensional supergravity compactified on $S^7$ \cite{Benini:2015eyy,Benini:2016rke}, i.e.\ for the ABJM theory at $k=1,2$, but it can also be generalized to other manifolds $M_7$ \cite{Hosseini:2016tor,Hosseini:2016ume,ABCMZ}. The black holes in AdS$_4$ can also be viewed as holographic duals of RG flows across dimensions in the spirit of Maldacena-Nu\~nez \cite{Maldacena:2000mw,Gauntlett:2001qs,BC}.

In this work we will follow a slightly different approach. Our starting point is the well-known observation that the ABJM theory admits a mass deformation that preserves $\mathcal{N}=2$ supersymmetry and leads to an interacting SCFT in the IR \cite{Benna:2008zy} (see also \cite{Klebanov:2008vq}).  We refer to this SCFT as mABJM. Although this theory is strongly coupled, some information about its physics can be obtained using symmetries and supersymmetric localization. For example, the partition function of the theory on $S^3$ was computed in \cite{Jafferis:2011zi} (see, in particular, Section 5). In addition, mABJM has a holographic dual which was constructed in four-dimensional gauged-supergravity by Warner (W) in \cite{Warner:1983vz,Warner:1983du} (see also \cite{Ahn:2000aq,Ahn:2000mf}) and uplifted to eleven dimensions in~\cite{Corrado:2001nv}. The situation here is akin to the well-known $\mathcal{N}=1$ Leigh-Strassler fixed point arising from a supersymmetric mass-deformation of four-dimensional $\mathcal{N}=4$ SYM \cite{Leigh:1995ep}. The gravity dual of this four-dimensional $\mathcal{N}=1$ SCFT was studied in \cite{Freedman:1999gp,Pilch:2000ej}

There are  two main objectives that we have in mind. On one hand, we are interested in studying the topologically twisted index of \cite{Benini:2015noa,Benini:2016hjo,Closset:2016arn} for the mABJM $\mathcal{N}=2$ SCFT. On the other hand we want to construct new four-dimensional supersymmetric black holes that are asymptotic to the AdS$_4$ Warner vacuum \cite{Warner:1983vz,Warner:1983du} (or alternatively the CPW solution of eleven-dimensional supergravity \cite{Corrado:2001nv}) and have a near-horizon AdS$_2$ region. The large $N$ limit of the topologically twisted index should then reproduce the Bekenstein-Hawking entropy of these black holes. It is worth emphasizing that the CPW AdS$_4$ solution in eleven-dimensional supergravity is not of the usual Freund-Rubin type and thus the class of black holes that we study is different from the ones explored recently in the literature \cite{Benini:2015eyy,Benini:2016rke,Hosseini:2016tor,Hosseini:2016ume,ABCMZ}.

The calculation of the topologically twisted index in the planar limit proceeds similarly as in \cite{Benini:2015eyy,Benini:2016rke,Hosseini:2016tor,Hosseini:2016ume,ABCMZ}. However, there are several subtle points related to the electric charge parameters of the index, which we emphasize and clarify along the way. 

The construction of the new black hole solutions is  more involved. We start with the maximal $\SO(8)$ gauged supergravity in four-dimensions \cite{deWit:1982bul}, which  is a consistent truncation to the lowest-lying KK modes of the eleven-dimensional supergravity on $S^7$ \cite{deWit:1986oxb,Nicolai:2011cy}. The three-dimensional mABJM SCFT of interest is dual to the $\mathcal{N}=2$ AdS$_4$ vacuum discovered by Warner \cite{Warner:1983vz,Warner:1983du}. It has the usual ${\rm U(1)}_R$ R-symmetry along with an ${\rm SU(3)}_F$ flavor symmetry which is manifested on the supergravity side by the presence of a massless ${\rm SU(3)}\times {\rm U(1)}$ gauge field in the AdS$_4$ Warner vacuum. 
The supersymmetric black hole solutions of interest are similar to the ones found in \cite{Romans:1991nq,Caldarelli:1998hg,Cacciatori:2009iz}. In particular, they have non-vanishing  gauge fields lying in the Cartan subalgebra of ${\rm SU(3)}\times {\rm U(1)}$. This allows us to simplify the construction by focusing 
on an ${\rm U(1)}^3$-invariant consistent truncation of the maximal supergravity. In addition to the metric and three Abelian gauge fields, the bosonic sector of that truncation contains also eight real scalars. By analyzing the BPS equations and the equations of motion, we construct a plethora of magnetic and dyonic supersymmetric black holes in this truncated theory.

The  ${\rm U(1)}^3$-invariant truncation can be embedded into a larger ${\rm U(1)}^2$-invariant truncation of the four-dimensional $\mathcal{N}=8$ supergravity. The advantage of doing  that is that the resulting theory is a fully-fledged four-dimensional $\mathcal{N}=2$ gauged supergravity coupled to three vector multiplets and one hypermultiplet. The ten real scalars in this truncation parametrize the coset
\begin{equation}\label{MVMHintro}
\cals M_{V}\times \cals M_{H} \eql \left[{\rm SU(1,1)\over U(1)}\right]^3\times {\rm SU(2,1)\over SU(2)\times U(1)}\,.
\end{equation}
Recasting our black hole solutions in the $\cals N=2$  language offers some additional insights and allows us to use  the existing results on black holes in four-dimensional $\mathcal{N}=2$ gauged supergravity, see \cite{DallAgata:2010ejj,Hristov:2010ri,Gnecchi:2013mta,Halmagyi:2013sla,Halmagyi:2013qoa,Halmagyi:2014qza} for a non-exhaustive list of references.

We note that our current set-up is very similar to the one  in \cite{Bobev:2014jva}, where a partial topological twist of the Leigh-Strassler theory  \cite{Leigh:1995ep} placed on $\mathbb{R}^2\times \Sigma_{\fg}$ led to a holographic RG flow from the four-dimensional $\mathcal{N}=1$ SCFT to a two-dimensional $(0,2)$ SCFT. The holographic dual to this setup is a family of black string solutions with an AdS$_3$ near-horizon region which are asymptotic to the AdS$_5$ fixed point of the five-dimensional $\mathcal{N}=8$ gauged supergravity found in \cite{Khavaev:1998fb}.

\subsection{Synopsis}

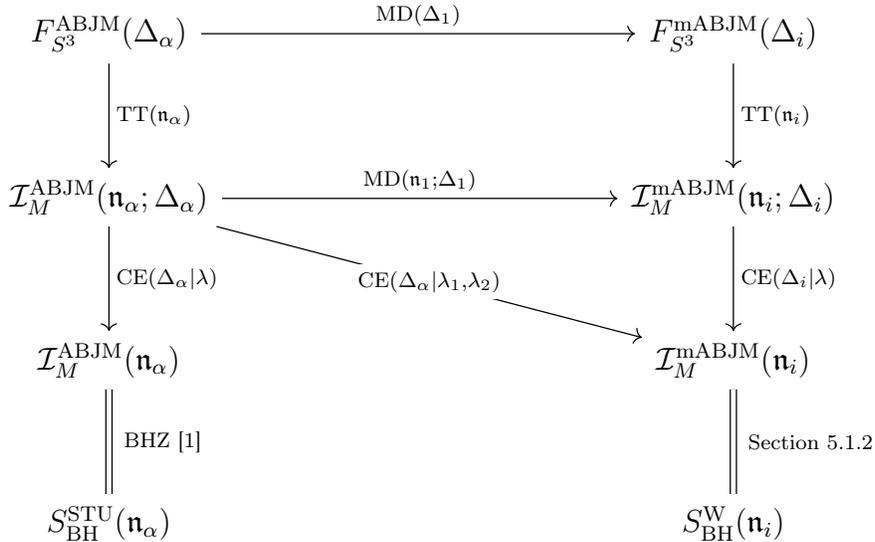
\begin{figure}[t]
\captionsetup{width=.85\linewidth}
\begin{center}
\begin{tikzcd}[row sep=40 pt, column sep = 75 pt]
&\hyperref[FS3]{F_{S^3}^\text{ABJM}(\Delta_\alpha)} 
\arrow[rr,"{\text{MD}(\Delta_1)}"] 
\arrow[d,"{\text{TT}(\fn_\alpha)}"] 
&&  \hyperref[FS3CPWgen]{F_{S^3}^\text{mABJM}(\Delta_i)}
\arrow[d,"\text{TT}(\fn_i)"] 
\\
&\hyperref[IABJM]{\cals I_M^\text{ABJM}(\fn_\alpha;\Delta_\alpha)}
\arrow[rr, "{\text{MD}(\fn_1;\Delta_1)}"]
\arrow[drr,"{  {\text{CE}(\Delta_\alpha|\lambda_1,\lambda_2)}}" description ]
\arrow[d,"{\text{CE}(\Delta_\alpha|\lambda)}" ]
&& \hyperref[IgenCPW]{\cals I_M^\text{mABJM}(\fn_i;\Delta_i)}
\arrow[d,"\text{CE}(\Delta_i|\lambda)" ]
\\
& \hyperref[{topindmagABJM}]{\cals I_M^\text{ABJM}(\fn_\alpha)} 
\arrow[d, equal,"\text{~BHZ~\cite{Benini:2015eyy}}"]
&&
 \hyperref[twistedCPW]{\cals I_M^\text{mABJM}(\fn_i)} 
\arrow[d, equal, "{\hyperref[subsec:genBHmag]{\text{~Section~\ref{subsec:genBHmag}}}}" ]& 
\\
& S_\text{BH}^\text{STU}(\fn_\alpha)  && \hyperref[{SBHgenmag}]{S_\text{BH}^\text{W}(\fn_i)}& 
\end{tikzcd}
\end{center}
\caption{ \label{Figone} Magnetically charged AdS$_4$ black holes: Formal relations between the free energy, $F_{S^3}$, the  twisted topological index, $\cals I_M$, and the black hole entropy, $S_{\rm BH}$. The operations along the arrows are:  
MD - mass deformation, TT - topological twist, CE - constrained extremization.}
\end{figure}

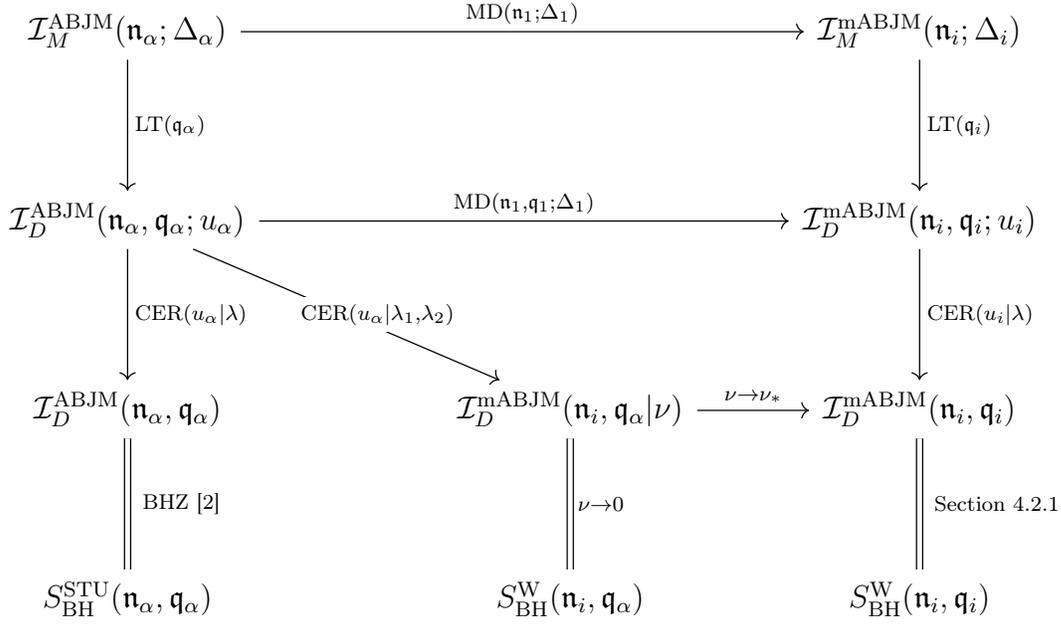
\begin{figure}[t]
\captionsetup{width=.85\linewidth}
\begin{center}
\begin{tikzcd}[row sep=50 pt, column sep = 35 pt]
\hyperref[IABJM]{\cals I_M^\text{ABJM}(\fn_\alpha;\Delta_\alpha)}
\arrow[d,"{\text{LT}(\fq_\alpha)}"]
\arrow[rrr, "{\text{MD}(\fn_1;\Delta_1)}"]
&&&  \hyperref[IgenCPW]{\cals I_M^\text{mABJM}(\fn_i;\Delta_i)}
\arrow[d,"{\text{LT}(\fq_i)}"]
\\
\hyperref[IdABJM]{\cals I_D^\text{ABJM}(\fn_\alpha,\fq_\alpha;u_\alpha)}
\arrow[rrr, "{\text{MD}(\fn_1,\fq_1;\Delta_1)}"]
\arrow[drr,"{\qquad \text{CER}(u_\alpha|\lambda_1,\lambda_2)}" description]
\arrow[d,"\text{CER}(u_\alpha|\lambda)"]
 &&& \hyperref[mabjmdind]{\cals I_D^\text{mABJM}(\fn_i,\fq_i;u_i)}
\arrow[d,"\text{CER}(u_i|\lambda)"]
\\
\hyperref[extdinabjm]{\cals I_D^\text{ABJM}(\fn_\alpha,\fq_\alpha)}
\arrow[d, equal,"\text{~BHZ~\cite{Benini:2016rke}}"]&&
 \cals I_D^\text{mABJM}(\fn_i,\fq_\alpha|\nu) 
 \arrow[r, "\nu\to\nu_*"] 
 \arrow[d, equal,"\nu\to 0"]
 &\hyperref[mABJMttind]{\cals I_D^\text{mABJM}(\fn_i,\fq_i)} 
\arrow[d, equal,"{\hyperref[sec:mabjmentropy]{\text{~Section~\ref{sec:mabjmentropy}}}}"]
& 
\\
 \hyperref[sec:abjmstu]{S_\text{BH}^\text{STU}(\fn_\alpha,\fq_\alpha)} && \hyperref[sec:eleccharg]{S_\text{BH}^\text{W}(\fn_i,\fq_\alpha)} 
  & \hyperref[sec:mabjmentropy]{S_\text{BH}^\text{W}(\fn_i,\fq_i)}
\end{tikzcd}
\end{center}
\caption{\label{Figtwo} Dyonic AdS$_4$ black holes. Formal relations between the topological twisted index, $\cals I_M$, the dyonic twisted index, $\cals I_D$, and the black hole entropy, $S_{\rm BH}$. The operations along arrows are: MD - mass defomration, LT - Legendre transform, CER - constrained extremization with reality conditions.}
\end{figure}

Since the paper is rather long and technical, let us highlight some of the main results first.
We begin in Section~\ref{sec:CFT} with a discussion of the field theory side of the duality, specifically the mABJM theory that is obtained from the ABJM theory by a mass deformation (MD), formally  captured by a constraint on the R-charge,\footnote{The ABJM theory has a global symmetry group of rank 4 and thus the magnetic fluxes, R-charges and fugacities are labelled by $\alpha=1,2,3,4$. The mass deformation in mABJM reduces the rank of the symmetry group to 3 and the parameters are labelled by $i=2,3,4$.} $\Delta_1$. 
The main object of interest  is the topologically twisted index, which is a partition function on $\mathbb{R}\times \Sigma_{\fg}$ that depends on electric charges and magnetic fluxes  as well as complex fugacities for the continuous global symmetries of the theory. The calculation of this observable in mABJM proceeds in several steps and has been schematically summarized in Figures~\ref{Figone} and \ref{Figtwo}. It  parallels a similar calculation in  ABJM \cite{Benini:2015eyy,Benini:2016rke}, see  the right and left columns in the figures, respectively. 

First we compute the ``magnetic index,'' $\mathcal{I}_M(\fn;\Delta)$, which depends on the magnetic topological twist parameters $\fn_{\alpha/i}$ and the real fugacities $\Delta_{\alpha/i}$.  We use the observation in \cite{Hosseini:2016tor} that the index, $\cals I _M(\fn;\Delta)$,  is directly related by a topological twist (TT) to the supersymmetric partition function, $F_{S^3}$, of the CFT on $S^3$,  where  the fugacities, $\Delta_{\alpha/i}$, are identified with  the R-charges on $S^3$.
 The  topologically twisted index, $\cals I_M(\fn)$,  which is a function of the magnetic fluxes, $\fn_{\alpha/i}$, only,
is then obtained from $\mathcal{I}_M(\fn;\Delta)$ by  extremization (CE) with respect to the fugacities, $\Delta_{\alpha/i}$,  subject to an algebraic constraint (with the corresponding real Lagrange multiplier,~$\lambda$) that is imposed by supersymmetry~\cite{Benini:2015eyy}. We show that the end result for the magnetic index, $\cals I^\text{mABJM}_M(\fn_i)$,  in the mABJM theory is the same irrespective of whether one first applies the mass deformation to the ABJM twisted index, $\cals I^\text{ABJM}_M(\fn_\alpha;\Delta_\alpha)$, to obtain the corresponding twisted index, $\cals I_M^\text{mABJM}(\fn_i;\Delta_i)$, which is then extremized with respect to its fugacities, or, equivalently, one extremizes  $\cals I^\text{ABJM}_M(\fn_\alpha;\Delta_\alpha)$ while imposing simultaneously two constraints on the fugacities: the one for the  mass deformation and the one for the topological twist. The resulting extremized index, $\mathcal{I}_M^{\rm mABJM}(\fn_i)$, is shown in Section~\ref{subsec:genBHmag} to match the entropy of the new family of magnetic black holes that we construct in Section~\ref{sec:solspace}.

In general, the topologically twisted index is dyonic, it depends on both electric charges, $\fq_{\alpha/i}$, and magnetic fluxes, $\fn_{\alpha/i}$,  as well as complex fugacities, $u_{\alpha/i}$. To include these extra parameters we follow the approach in \cite{Benini:2016rke} which is summarized in Figure~\ref{Figtwo}.  We start with $\mathcal{I}_{M}(\fn;\Delta)$,   analytically continue it from real fugacities, $\Delta_{\alpha/i}$, to complex fugacities,  $u_{\alpha/i}$, and introduce the   electric charges, $\fq_{\alpha/i}$, by a Legendre transformation (LT). This yields  the dyonic index, $\mathcal{I}_D(\fn,\fq;u)$, which is   then extremized (CER)  with respect to constrained fugacities, $u_{\alpha/i}$. 

This calculation is more subtle than for the purely magnetic index in Figure~\ref{Figone}. The reason is that there should be a linear relation between the electric charges   to ensure supersymmetry, however,  it is not entirely clear how to find it. It was proposed in \cite{Benini:2016rke} that to  reproduce correctly  the entropy of a macroscopic black hole, the imaginary part of   the dyonic index,  $\mathcal{I}_{D}(\fn,\fq)$, should vanish. This provides exactly one additional constraint that serves as the expected relation between the electric charges. 

It is straightforward to implement this procedure in the ABJM theory, see \cite{Benini:2016rke} and Section~\ref{sec:dabjm}, as  summarized by the left column in Figure~\ref{Figtwo}. In mABJM there is a further subtlety 
at which step of the calculation one should  impose the massive deformation that eliminates one of the global ${\rm U}(1)$ symmetries. One possibility, see Section~\ref{sec:mabjmdind}, is to  perform the mass deformation first. This leaves three ${\rm U}(1)$ global symmetries the corresponding electric charges, $\fq_i$, magnetic fluxes, $\fn_i$, and  fugacities, $u_i$. Then the extremization with respect to those fugacities together with the reality constraint yields an unambiguous result for the dyonic twisted index, $\cals I_D^\text{mABJM}(\fn_i,\fq_i)$, that matches the entropy of the new dyonic black holes constructed in Section~\ref{subsec:dyonicBPS}. 
The other possibility, suggested by the corresponding calculation of the magnetic index, is to extremize the ABJM dyonic index, $\cals I_D^\text{ABJM}(\fn_\alpha,\fq_\alpha;u_\alpha)$, while imposing two constraints on the fugacities using two  complex Lagrange multipliers, $\lambda_1$ and $\lambda_2$, see the diagonal arrow in  Figure~\ref{Figtwo}. Indeed, in Section~\ref{sec:manjmditc} we find the extremized dyonic index and the magnetic fluxes are the same as above.
However, unlike before, this  extremization does not yield a unique result for the electric charges because of a shift symmetry that involves the imaginary parts of the Lagrange multipliers. In Sections~\ref{sec:manjmditc} and \ref{sec:eleccharg} we show that this symmetry can be fixed consistently in two ways: (i) one can set the electric charge $\fq_1$ to zero, thus reducing the calculation to the one in the mABJM theory above, and (ii) set the imaginary parts of both Lagrange multipliers to zero, which gives a consistent match with the dyonic black holes in the dual supergravity with four vector fields. 

Throughout the paper we work with a consistent truncation of the maximal $\SO(8)$ gauged supergravity, which is discussed in Section~\ref{sec:Sugra} and Appendices~\ref{appendixA} and \ref{appendixB}. We use the truncation to construct supersymmetric  AdS$_2\times \Sigma_{\fg}$ solutions which should be thought of as the near-horizon geometry of a class of dyonic black holes asymptotic to the AdS$_4$ Warner vacuum.  In Section~\ref{sec:DyonicComp}, we show explicitly that the Bekenstein-Hawking entropy of these black holes is the same as the topologically twisted index.  In Section~\ref{sec:magneticBH}, we study the magnetically charged black holes in more detail. 
We conclude in Section \ref{sec:Conclusions} with a short discussion and some open questions for future work. In Appendix \ref{appconv} we summarize our notation and conventions. In Appendix~\ref{appendixA} we  also show how to formulate our truncation  in the canonical language of four-dimensional $\mathcal{N}=2$ gauged supergravity. In Appendix~\ref{appendixB} we present some details on the derivation of the near-horizon BPS equations used in Section~\ref{sec:Sugra}.  Finally, in Appendix~\ref{appattractor} we show how these BPS equations can be written in a form similar to the ``attractor mechanism'' equations discussed in \cite{DallAgata:2010ejj,Hristov:2010ri}.

\section{Field theory}
\label{sec:CFT}

\subsection{ABJM and a mass deformation}
\label{subsec:ABJMmass}

Here we offer a short summary on the ABJM SCFT \cite{Aharony:2008ug} and a particular supersymmetric mass deformation studied in \cite{Benna:2008zy} (see also \cite{Klebanov:2008vq,Jafferis:2011zi}). The ABJM theory is a double Chern-Simons theory with gauge group ${\rm U(N)}\times {\rm U(N)}$ and equal and opposite levels for the two gauge groups $(k,-k)$. This theory describes the low-energy dynamics of $N$ coincident M2-branes probing a $\mathbb{C}^4/\mathbb{Z}_k$ singularity in M-theory. The dual holographic description at large $N$ is in terms of an AdS$_4\times S^7/\mathbb{Z}_k$ solution in M-theory. For $k>2$ the theory has only $\mathcal{N}=6$ supersymmetry which gets enhanced to $\mathcal{N}=8$ for $k=1,2$. In the following we will focus on $k=1$ where there is no orbifold singularity and the gravitational solution is well-described by eleven-dimensional supergravity.\footnote{We expect that most of our results should hold for more general values of $k$.} For $k=1$ the theory has an ${\rm SO(8)}$ R-symmetry which, however, is not manifest at the level of the ABJM Lagrangian. 

The ABJM theory can be succinctly described using the  $\mathcal{N}=2$ superspace formalism. In addition to the two vector multiplets, there are four chiral multiplets denoted by $A_a$ and $B_c$,  $a,c=1,2$, transforming  in the $(\overline{\textbf{N}},\textbf{N})$ and $(\textbf{N},\overline{\textbf{N}})$ representation of ${\rm U(N)}_k\times {\rm U(N)}_{-k}$, respectively, with the superpotential 
\begin{equation}\label{ABJMsuperp}
W \sim \text{Tr}\left(\epsilon^{ab}\epsilon^{cd}A_aB_cA_bB_d\right)\,.
\end{equation}
The R-charges of these chiral superfields,\footnote{As usual, the R-charge of an $\cals N=2$ chiral supermultiplet is defined as the R-charge of its lowest component which is a complex scalar.} 
$R[A_a] \equiv \Delta_{A_a}$ and $R[B_c] \equiv \Delta_{B_c}$,  must satisfy the constraint
\begin{equation}
\label{DeltaconstrABJM}
\Delta_{A_1}+\Delta_{A_2}+\Delta_{B_1}+\Delta_{B_2}=2\,,
\end{equation}
so that the total R-charge of the superpotential \eqref{ABJMsuperp} is equal to 2.

In this formulation only an ${\rm U(1)}_R\times {\rm SU(2)}\times {\rm SU(2)}\times {\rm U(1)}_b$ subgroup of the global symmetry is manifest. It is enhanced to ${\rm SU(4)}_R\times {\rm U(1)}_b$ when the Lagrangian is written in components, see for example \cite{Jafferis:2011zi}. The ${\rm U(1)}_b$ global symmetry has a topological nature characteristic of three-dimensional QFTs and is generated by the current $*_3\text{Tr}(F+\tilde{F})$, where $F$ and $\tilde{F}$ are the field strengths of the two ${\rm U(N)}$ gauge fields and $*_3$ is the Hodge star in three dimensions. Due to this topological current there are gauge invariant monopole operators, $T^{(q)}$, in the theory, which turn on $q$ units of flux for the topological current through an $S^2$ surrounding the insertion point. When $k=1$ the operator $T^{(1)}$ transforms in the $(\textbf{N},\overline{\textbf{N}})$ and the operator $T^{(-1)}$ transforms in the $(\overline{\textbf{N}},\textbf{N})$ representation of the gauge group. This is ultimately responsible for the enhancement of the supersymmetry to $\mathcal{N}=8$ and of the R-symmetry to ${\rm SO(8)}_R$. In the dual holographic description for $k=1$, given by the AdS$_4\times S^7$ solution of M-theory, the ${\rm SO(8)}_R$ is realized as the isometry group of $S^7$. The metric of $S^7$ can be written as a circle fibration over $\mathbb{CP}^3$, then the ${\rm SU(4)}_R$ is the isometry group of $\mathbb{CP}^3$ and ${\rm U(1)}_b$ is realized as the isometry of the fibre.

The $S^3$ free energy of the ABJM SCFT can be computed using supersymmetric localization and is given by the following function of the R-charges:\footnote{See \cite{Freedman:2013ryh} where a discussion on $F_{S^3}$ for the ABJM theory as a function of $\Delta_a$ can be found. Note also that we define the free energy as $F_{S^3} = -\log Z_{S^3}$, where $Z_{S^3}$ is the supersymmetric partition function of the theory on $S^3$.}
\begin{equation}
\label{FS3}
F_{S^3}=\frac{4\sqrt{2}\pi}{3}N^{3/2}\sqrt{\Delta_{A_1}\Delta_{A_2}\Delta_{B_1}\Delta_{B_2}} \;.
\end{equation}
Using $F$-maximization \cite{Jafferis:2010un,Closset:2012vg} while satisfying the second relation in \eqref{DeltaconstrABJM}, one finds the values of the R-charges at the superconformal point,
\begin{equation}\label{DeltaABJM}
\Delta_{A_1}=\Delta_{A_2}=\Delta_{B_1}=\Delta_{B_2}=\frac{1}{2}\;,
\end{equation}
so that the free energy on $S^3$ for ABJM reads
\begin{equation}
\label{FS3ABJM}
F_{S^3}^{\text{ABJM}} =\frac{\sqrt{2}\pi}{3}N^{3/2}\;.
\end{equation}
Note that the values of $\Delta_{A_{1,2}}$ and $\Delta_{B_{1,2}}$ in \eqref{DeltaABJM} can also be obtained as a condition for  enhanced supersymmetry of the SCFT. When the ABJM theory is placed on $S^3$, for values of $\Delta_{A_{1,2}}$ and $\Delta_{B_{1,2}}$ that obey \eqref{DeltaconstrABJM}, but not \eqref{DeltaABJM}, the theory preserves $\mathcal{N}=2$ supersymmetry but is not conformal, see \cite{Freedman:2013ryh} for a  discussion. 

The ABJM superpotential \eqref{ABJMsuperp} can be deformed by a mass term that preserves $\mathcal{N}=2$ supersymmetry
\begin{equation}\label{CPWsuperp}
\Delta W \sim \text{Tr}(T^{(1)}A_1)^2\;.
\end{equation}
Adding this deformation triggers an RG flow from the ABJM theory in the UV to an interacting $\mathcal{N}=2$ SCFT in the IR.  This was studied in \cite{Benna:2008zy} (see also \cite{Klebanov:2008vq,Jafferis:2011zi}) from a field theory perspective. The holographic description of this mABJM SCFT is given by the Warner vacuum of four-dimensional maximal ${\rm SO(8)}$ gauged supergravity \cite{Warner:1983vz,Warner:1983du} (see also \cite{Nicolai:1985hs}) which was uplifted to eleven-dimensional supergravity in \cite{Corrado:2001nv}. There have been several consistency checks of this proposed duality including a match between the spectrum of protected operators \cite{Klebanov:2008vq} (see also \cite{Nicolai:1985hs} for earlier work) as well as the free-energy to leading order in $N$ \cite{Jafferis:2011zi}. It is worth pointing out that, while the ABJM theory is parity invariant, the $\mathcal{N}=2$ SCFT obtained by the mass deformation in \eqref{CPWsuperp} breaks parity. In the supergravity description this breaking of parity is manifested by the fact that one of the four-dimensional $\cN=8$ supergravity pseudoscalars has a non-vanishing value at the Warner vacuum.

The mABJM SCFT has the following symmetries: The four chiral superfields of the ABJM theory, ordered as $(A_1,A_2,B_1,B_2)$, transform in the ${\bf 4}_{1}$ of ${\rm SU(4)}_R\times {\rm U(1)}_b$ (see, e.g., \cite{Freedman:2013ryh} for a   summary). The superpotential \eqref{CPWsuperp} breaks   ${\rm SU(4)}_R$ to ${\rm SU(3)}_F$ and only a linear combination of the ${\rm U(1)}_c$ commutant of ${\rm SU(3)}_F$ inside of ${\rm SU(4)}$ and the ${\rm U(1)}_b$ is preserved. 
We will call that linear combination ${\rm U(1)}_{R}^{\text{W}}$ since it is the  superconformal R-symmetry of the mABJM conformal fixed point, which in turn is dual to the Warner vacuum in supergravity. The ${\rm SU(3)}_F$ symmetry does not act on the supercharges and thus deserves the name flavor symmetry. The linear combination of ${\rm U(1)}_b$ and ${\rm U(1)}_c$ orthogonal to ${\rm U(1)}_{R}^{\text{W}}$ is broken by the quadratic superpotential \eqref{CPWsuperp} and corresponds to the massive ${\rm U(1)}_m$ vector field in the supergravity discussion below.

The superpotential deformation   \eqref{CPWsuperp} modifies the R-charge assignments in the theory.\footnote{One can always choose a gauge in which the monopole operators have vanishing R-charge, see for example \cite{Jafferis:2011zi} for a discussion on this.} In particular, the value of the R-charge for $A_1$ is set to unity. Combining this with \eqref{DeltaconstrABJM} one finds
\begin{equation}\label{DeltaconstrCPW}
\Delta_{A_1}=1 \, , \qquad \Delta_{A_2}+\Delta_{B_1}+\Delta_{B_2}=1\;.
\end{equation}
The $S^3$ free energy for general values of the three R-charges can be computed by localization \cite{Jafferis:2011zi}. The final result can be obtained by formally setting $\Delta_{A_1}=1$ in \eqref{FS3} and reads
\begin{equation}
\label{FS3CPWgen}
F_{S^3} =\frac{4\sqrt{2}\pi}{3}N^{3/2}\sqrt{\Delta_{A_2}\Delta_{B_1}\Delta_{B_2}} \;.
\end{equation}
Applying $F$-extremization to \eqref{FS3CPWgen} and enforcing the second constraint in \eqref{DeltaconstrCPW}, we find that at the mABJM fixed point 
\begin{equation}\label{DeltaCPW}
\Delta_{A_2}=\Delta_{B_1}=\Delta_{B_2}=\frac{1}{3}\; .
\end{equation}
This is compatible with the ${\rm SU(3)}_F$ flavor symmetry of the model and leads to the following $S^3$ free energy of the mABJM SCFT:
\begin{equation}
\label{FS3CPW}
F_{S^3}^{\text{mABJM}} =\frac{4\sqrt{2}\pi}{9\sqrt{3}}N^{3/2}\;.
\end{equation}
Thus one finds that the UV (ABJM) and IR (mABJM) SCFTs have the following relation between their $S^3$ partition functions
\begin{equation}
F_{S^3}^{\text{mABJM}} =\frac{4}{3\sqrt{3}} F_{S^3}^{\text{ABJM}}\; .
\end{equation}
As expected this is compatible with the $F$-theorem, namely $F_{S^3}^{\text{ABJM}}>F_{S^3}^{\text{mABJM}}$ \cite{Jafferis:2011zi,Casini:2012ei}.

\subsection{The topologically twisted index}
\label{ssec:twistedindex}

A three-dimensional $\mathcal{N}=2$ SCFT can be placed on the manifold $\mathbb{R} \times \Sigma_{\mathfrak{g}}$, where $\Sigma_{\mathfrak{g}}$ is a closed Riemann surface of genus $\mathfrak{g}$,\footnote{See, Appendix~\ref{appconv} for our conventions.} while preserving at least two supercharges by employing the topological twist of Witten \cite{Witten:1988ze}. The procedure amounts to turning on a background gauge field for the ${\rm U(1)}$ R-symmetry of the SCFT with a finely tuned magnitude so as to cancel the curvature of the Riemann surface. In addition, one is free to turn on any appropriately quantized flux for the background gauge fields that couple to the continuous flavor symmetry currents in the CFT. This procedure can be applied to both the ABJM and the mABJM theories discussed above.

An interesting supersymmetric observable  {which} captures non-trivial information about a topologically twisted three-dimensional $\mathcal{N}=2$ SCFT on $\mathbb{R} \times \Sigma_{\mathfrak{g}}$ is the topologically twisted index
\begin{equation}\label{}
\cals I(\fn_i;\Delta_i)\equiv \log Z_{\mathbb{R}\times \Sigma_{\fg}} (\Delta_i,\fn_i),
\end{equation}
defined in \cite{Benini:2015noa,Benini:2016hjo,Closset:2016arn}.  This is a supersymmetric partition function that depends on the theory at hand, the genus, $\mathfrak{g}$, of the Riemann surface, the magnetic fluxes, $\mathfrak{n}_i$, for the background magnetic fields as well as the fugacities, $\Delta_i$, for those global symmetries. We use the same notation for the fugacities and the R-charges in \eqref{DeltaconstrABJM} since they obey formally the same constraint \cite{Hosseini:2016tor}. The general form of this partition function is quite complicated, but it simplifies dramatically in an appropriate large $N$ limit which will be the focus of our discussion.

The large $N$ limit of the twisted index was first studied in \cite{Benini:2015eyy} for $\mathfrak{g}=0$. Here we use mostly the results of \cite{Hosseini:2016tor}, combined with the ones in Section 6 of \cite{Benini:2016hjo}, which are applicable for the large $N$ limit of the so called non-chiral quiver gauge theories that include both the ABJM and mABJM theories. The resulting  formula for the twisted index can be expressed in terms of the partition function on  $S^3$  and the background magnetic fluxes $\fn_i$,\footnote{We use the shorthand notation $\Delta_{A_1,A_2} \to \Delta_{1,2}$ and $\Delta_{B_1,B_2} \to \Delta_{3,4}$.}
\begin{equation}\label{index}
\cals I(\fn_i;\Delta_i) =  (\fg-1) \bigg ( F_{S^3} (\Delta_i) + \sum_{i=1}^{r_{G}} \left [ \left( \frac{\fn_i}{\fg-1} - \Delta_i  \right) \frac{1}{2} \frac{\partial F_{S^3} (\Delta_i)}{\partial\Delta_i} \right ] \bigg ) \,,
\end{equation}
where $r_G$ is the rank of the continuous global symmetry group. For the ABJM theory, $r_G=4$ since the global symmetry is ${\rm SO(8)}$. For the mABJM theory, the global symmetry is ${\rm SU(3)}_F\times {\rm U(1)}_R^{\text{W}}$ and thus $r_G=3$. As argued in \cite{Benini:2015eyy}, the topologically twisted index can be found by extremizing \eqref{index} with respect to $\Delta_i$, subject to the constraint \eqref{DeltaconstrABJM} for ABJM and \eqref{DeltaconstrCPW} for mABJM. This means that to use the formula in \eqref{index} one should first fix the background fields (the genus, $\fg$, and the magnetic fluxes, $\fn_i$), then solve the constrained extremization problem to find the extremal values $\bDelta _i(\fn)$ that are finally plugged back into \eqref{index} to obtain the topologically twisted index as a function of the background fields. 

\subsubsection{The ABJM twisted index}

Let us illustrate this procedure in some detail for the ABJM theory. We begin by turning on background magnetic fields along the four  Cartan generators, $T_\alpha$, of the ${\rm SO(8)}$ global symmetry, 
\begin{equation}\label{so8mag}
F\eql F^{(\alpha)}T_\alpha\,,\qquad F^{(\alpha)} = \fn_\alpha {\rm vol}_{\Sigma_{\fg}}\;, \qquad \alpha=1,\ldots,4\,.
\end{equation}
To preserve supersymmetry we have to impose the following relation between the magnetic fluxes: 
\begin{equation}\label{mfabjm}
\fn_1+\fn_2+\fn_3+\fn_4\eql 2(\fg-1)\,,
\end{equation}
which implements the topological twist.
In addition, we must ensure the  proper flux quantization for the magnetic fields piercing the Riemann surfaces. In our conventions this amounts to  $\fn_\alpha \in \mathbb{Z}$.

The general formula for the topologically twisted index in \eqref{index}, after using the explicit expression for the free energy on $S^3$ in \eqref{FS3},   is 
\begin{equation}\label{IABJM}
\cals I_M(\fn_\alpha;\Delta_\alpha)\eql {\sqrt 2\pi\over 3}\,N^{3/2}\sqrt{\Delta_1\Delta_2\Delta_3\Delta_4}\,\Big({\fn_1\over\Delta_1} +{\fn_2\over \Delta_2}+{\fn_3\over\Delta_3}+{\fn_4\over\Delta_4}\Big)\,.
\end{equation}
One should  extremize it as a function of $\Delta_\alpha$ subject to  the constraint in \eqref{DeltaconstrABJM}. To this end we introduce the Lagrange multiplier, $\lambda$, and extremize
\begin{equation}\label{}
\cals I(\fn_\alpha;\Delta_\alpha|\lambda)\eql \cals I_M(\fn_\alpha;\Delta_\alpha)+\pi\lambda  (\Delta_1+\Delta_2+\Delta_3+\Delta_4-2) \,.
\end{equation}
This yields the system of equations
\begin{equation}\label{extreqmag}
{\partial\cals I_M\over\partial\Delta_\alpha}+\pi \lambda\eql 0\,,\qquad \alpha\eql 1,\ldots,4\,,
\end{equation}
which can be solved for the magnetic charges
\begin{equation}\label{magabjmnns}
\begin{split}
\fn_\alpha &= -\frac{3\lambda}{2\sqrt{2} N^{3/2}} \frac{\sum_\beta  \sigma_{\alpha\beta} \Delta_\alpha \Delta_\beta}{\sqrt{\Delta_1\Delta_2\Delta_3\Delta_4}} \,,
\end{split}
\end{equation}
where 
\begin{equation}\label{defsigab}
\sigma_{\alpha\beta}\eql \begin{cases}-1 & \text{for~~ $\alpha=\beta$} \,,\\ \phantom{-} 1 &  \text{for~~ $\alpha\not =\beta$}\,. \end{cases}
\end{equation}
Plugging this back into \eqref{IABJM} and using \eqref{DeltaconstrABJM}, we find
\begin{equation}
\mathcal{I}(\fn_\alpha;\Delta_\alpha|\lambda) = -2\pi \lambda \,.
\end{equation}

Imposing the topological twist condition \eqref{mfabjm} on the magnetic fluxes \eqref{magabjmnns}, we solve for the Lagrange multiplier,
\begin{equation}
\begin{split}
\lambda = -2(\fg-1)\frac{2\sqrt{2}N^{3/2}}{3} \frac{\sqrt{\Delta_1\Delta_2\Delta_3\Delta_4}}{\sum_{\alpha,\beta} \sigma_{\alpha\beta}\Delta_\alpha \Delta_\beta}\,.
\end{split}
\end{equation}
Then \eqref{magabjmnns} can be rewritten as 
\begin{equation}\label{nofDelta}
\begin{split}
\fn_\alpha &= 2(\fg-1) \frac{\sum_\beta\sigma_{\alpha\beta} \Delta_\alpha \Delta_\beta}{\sum_{\gamma,\delta} \sigma_{\gamma\delta} \Delta_\gamma \Delta_\delta} \,,
\end{split}
\end{equation}
and the topologically twisted index as a function of extremal fugacities, $\Delta_\alpha=\Delta_\alpha^\text{ext}$, is
\begin{equation}\label{topindmagABJM}
\mathcal{I}(\fn_\alpha(\Delta);\Delta_\alpha)\eql \frac{8 \sqrt{2}\pi}{3}(\fg-1) N^{3/2} \frac{\sqrt{\Delta_1\Delta_2\Delta_3\Delta_4}}{\sum_{\alpha,\beta}  \sigma_{\alpha\beta}\Delta_\alpha \Delta_\beta}\,.
\end{equation}

To find the final expression for the topologically twisted index as a function of the magnetic fluxes, $\fn_\alpha$, one has to solve the algebraic equations in \eqref{nofDelta} for $ \Delta _\alpha(\fn)$ and plug the result in \eqref{topindmagABJM} to obtain, $\cals I_M(\fn_\alpha)$. Clearly, given the nonlinearity of \eqref{nofDelta}, this is in general a complicated algebraic problem that one would rather avoid.  So, instead we will work with the implicit formulae above for the magnetic fluxes 
and the twisted topological index as functions of the extremal fugacities.

There is a special topological twist, the so-called universal twist \cite{ABCMZ,BC}, for which one can perform the algebraic calculations above explicitly.  This twist is characterized by having background magnetic fields that extend only along the unique $\mathcal{N}=2$ superconformal R-symmetry of the ABJM theory. In our conventions this amounts to setting\footnote{Note that due to the quantization condition $\fn_\alpha\in\ZZ$, the universal twist is well defined only on Riemann surfaces for which $\fg$ is odd.} $\fn_1=\fn_2=\fn_3=\fn_4=(\fg-1)/2$. Solving \eqref{nofDelta} for these values of the background fluxes one finds $\bDelta _{\alpha}=1/2$ which in turn leads to the following simple expression for the topologically twisted index
\begin{equation}
\mathcal{I}_{\text{univ}}^{\text{ABJM}} = (\fg-1) \frac{\sqrt{2}\pi}{3} N^{3/2} = (\fg-1) F_{S^3}^{\text{ABJM}}\;,
\end{equation}
where we have used \eqref{FS3ABJM}. Note that only for $\fg>1$ one finds a positive topologically twisted index at leading order in $N$.

\subsubsection{The mABJM twisted index}
\label{ssec:mABJMmagI}

Now we apply the same procedure to the mABJM theory obtained by  a mass deformation of the  ABJM   superpotential  in \eqref{CPWsuperp}. As we discussed in Section~\ref{subsec:ABJMmass}, this breaks the global $\SO(8)$ symmetry to ${\rm SU(3)}_F\times {\rm U(1)}_{R}^{\text{W}}$. In terms of the $\SO(8)$ Cartan generators, $T_{\alpha}$, $\alpha=1,\ldots,4$, the Cartan subalgebra of the  new global symmetry group is spanned by %
\begin{equation}\label{newCart}
T^{(1)}\eql {1\over 2}(T_2-T_3)\,,\qquad T^{(2)}{1\over 2\sqrt 3}(T_2+T_3-2 T_4)\,, \qquad T^{(R)}\eql {1\over 3}(3T_1+T_2+T_3+T_4)\,,
\end{equation}
where the first two are Cartan generators of $\SU(3)_F$ and the third one is the generator of the new R-symmetry, ${\rm U(1)}_{R}^{\text{W}}$. If we start with a general $\SO(8)$ magnetic field \eqref{so8mag}, the symmetry breaking along the RG-flow restricts it   to the Cartan subalgebra of the new global symmetry, which is enforced by the condition 
\begin{equation}\label{massconstr}
\fn^{(m)}\equiv \fn_1-\fn_2-\fn_3-\fn_4\eql 0\,,
\end{equation}
while the topological twist along the new R-symmetry generators gives
\begin{equation}\label{toptwist}
\fn^{(R)}\equiv \frac{1}{2}(3 \fn_1 + \fn_2 + \fn_3 + \fn_4)\eql 2(\fg-1)\,.
\end{equation}
It is illuminating to rewrite the two constraints as 
\begin{equation}\label{magconst}
\fn_1=\fg-1\,,\qquad \fn_2+\fn_3+\fn_4\eql \fg-1\,,
\end{equation}
which is analogous to the condition \eqref{DeltaconstrCPW} on the R-charges. Using \eqref{FS3CPWgen} and \eqref{index} we then find the following expression for the topologically twisted index
\begin{equation}\label{IgenCPW}
\cals I(\fn_i;\Delta_i) \eql 
{\sqrt 2\pi\over 3}\,N^{3/2}\sqrt{\Delta_2\Delta_3\Delta_4}\,\Big(\fg-1 +{\fn_2\over \Delta_2}+{\fn_3\over\Delta_3}+{\fn_4\over\Delta_4}\Big)\,,
\end{equation}
which must be extremized as a function of $\Delta_{2,3,4}$ satisfying  the constraint \eqref{DeltaconstrCPW}. Introducing a Lagrange multiplier and extremizing as above, we obtain the following relations between the extremal values $\bDelta _i=\Delta_i^\text{ext}$ and the magnetic fluxes: 
\begin{equation}\label{n234CPW}
\begin{split}
\fn_2 & \eql (\fg-1)\bDelta _2\bigg[{\bDelta _3+\bDelta _4\over \bDelta _2\bDelta _3+\bDelta _3\bDelta _4+\bDelta _4\bDelta _2}-1\bigg]\,,\\[6 pt]
\fn_3 & \eql (\fg-1)\bDelta _3\bigg[{\bDelta _2+\bDelta _4\over \bDelta _2\bDelta _3+\bDelta _3\bDelta _4+\bDelta _4\bDelta _2}-1\bigg]\,,\\[6 pt]
\fn_4 & \eql (\fg-1)\bDelta _4\bigg[{\bDelta _2+\bDelta _3\over\bDelta _2\bDelta _3+\bDelta _3\bDelta _4+\bDelta _4\bDelta _2}-1\bigg]\,.\\
\end{split}
\end{equation}
Finding the twisted index as a function of $\fn_{2,3,4}$ again amounts to solving the algebraic equations in \eqref{n234CPW} for $\bDelta _i$ and plugging the result in \eqref{IgenCPW} which is difficult. Hence, we proceed as previously and express the final result in terms of the extremal values $\bDelta _i=\Delta_i^\text{ext}$,
\begin{equation}\label{twistedCPW}
\cals I(\fn_i(\bDelta );\bDelta_i )\eql {2\sqrt{2}\pi\over 3}(\fg-1)N^{3/2}\sqrt{\bDelta _2\bDelta _3\bDelta _4}\,\bigg[{1\over \bDelta _2\bDelta _3+\bDelta _3\bDelta _4+\bDelta _4\bDelta _2}-1\bigg]\,.
\end{equation}

We have obtained   \eqref{IgenCPW} for mABJM  from \eqref{index} for ABJM by imposing the constraints \eqref{DeltaconstrCPW} and \eqref{magconst} on $\Delta_1$ and $\fn_1$, respectively. However, 
implementing those constraints does not commute with the extremization of the topologically twisted index. Indeed, 
 \eqref{twistedCPW} differs  from the result one would have obtained by evaluating the topologically twisted index for ABJM in \eqref{topindmagABJM} with $\fn_1$  and $\Delta_1$ set to their mABJM values. If one wants to start with the ABJM index \eqref{IABJM}, the correct procedure is to extremize it with both  constraints in  \eqref{DeltaconstrCPW}. 

\subsection{Explicit examples}
\label{subsec:magnexpl}

It is instructive to discuss  two examples in which we can solve the algebraic equations in \eqref{n234CPW} and obtain the twisted index in a compact form as an explicit function of the magnetic fluxes. 

Our first example is the  universal twist which amounts to turning on the magnetic flux only along the R-symmetry generator  in \eqref{newCart}. This leads to the following values for $\fn_{2,3,4}$\footnote{Due to the quantization condition $\fn_{i} \in \mathbb{Z}$, we find that the universal twist for the mABJM theory is well defined only on Riemann surfaces for which $\fg$ is a multiple of 4.} 
\begin{equation}
\fn_{2}=\fn_{3}=\fn_{4}= {1 \over 3}\,(\fg-1)\,.
\end{equation}
Plugging this in \eqref{n234CPW} one finds the solution $\bDelta _2=\bDelta _3=\bDelta _4=1/3$. As expected on general grounds, see \cite{ABCMZ,BC}, the topologically twisted index is then
\begin{equation}\label{IunivCPW}
\mathcal{I}_{\text{univ}}^{\text{mABJM}} = (\fg-1) \frac{4\sqrt{2}\pi}{9\sqrt{3}}N^{3/2} = (\fg-1) F_{S^3}^{\text{mABJM}}\;,
\end{equation}
where for the second equality we used \eqref{FS3CPW}.

The second  example is more involved. We impose the following relation between the magnetic fluxes:
\begin{equation}
\fn_2= \fn_3 \equiv (\fg-1)\,\fn \, .
\end{equation}
The remaining magnetic fluxes are then fixed by \eqref{magconst}. Solving the equations in \eqref{n234CPW} with these restrictions leads to the following four branches of solutions for $\bDelta _{2,3}$:
\begin{equation}\label{branchesCPW}
\begin{split}
\text{Branch 1$_{\pm}$:} \quad & \bDelta _2 = \frac{1 - \fn \pm \sqrt{(1+\fn)(1-3\fn)} }{2} \, , \quad \bDelta _3 = \frac{1 - \fn \mp \sqrt{(1+\fn)(1-3\fn)} }{2} \, , \\ 
\text{Branch 2$_{\pm}$:} \quad & \bDelta _2 = \bDelta _3 = \frac{1 - \fn \pm \sqrt{(1+\fn)(\fn-1/3)}}{2} \, .
\end{split}
\end{equation}
Note that $\bDelta _4$ is fixed uniquely by the linear relation in \eqref{DeltaconstrCPW} once a choice of a branch of solutions in \eqref{branchesCPW} has been made. The corresponding twisted index reads: 
\begin{equation*}\label{twistSU2}
\begin{split}
\text{Branch 1$_{\pm}$:} \quad & \mathcal{I}(\fn) = \frac{3\sqrt{3\fn} (1-\fn)}{2} \; (\fg-1) F_{S^3}^{\text{mABJM}} \, , \\[6 pt] 
\text{Branch 2$_{\pm}$:} \quad & \mathcal{I}(\fn) = \frac{3}{2\sqrt{2}} \frac{(1 - 2\fn)\left(1+3 \fn^2 \pm(1-3 \fn) \sqrt{(1+\fn) \left(\fn-\frac{1}{3}\right)}\right)}{\sqrt{(1 - 2\fn)\left(1-3 \fn^2 \mp(1-3 \fn) \sqrt{(1+\fn) \left(\fn-\frac{1}{3}\right)}\right)}} (\fg-1) F_{S^3}^{\text{mABJM}} \, ,
\end{split}
\end{equation*}
where once again we have expressed the result in terms of the mABJM free energy on  $S^3$, see \eqref{FS3CPW}. Interestingly, we find that the two branches of solutions $1_{\pm}$ have the same twisted index. 

The extremized values $\bDelta _{i}$ play the role of R-charges in the one-dimensional quantum mechanical system arising at low energies after the twisted compactification on $\Sigma_{\fg}$. We thus have to impose that $\bDelta _{i}$ are real. In addition the twisted index is expected to reproduce the entropy of the black hole that describes this twisted compactification holographically. For that reason we also have to find $\mathcal{I}>0$. Imposing these two constraints in the expression \eqref{twistSU2} restricts the value of the magnetic flux for Branch $1_\pm$ to the range $0<\fn<1/3$ and for Branch $2_\pm$ to $1/3<\fn<1/2$. At the special value $\fn=1/3$, the R-charges reduce to $\bDelta _i=1/3$ and one should recover the universal twist. Indeed, when evaluated at $\fn=1/3$, the twisted index \eqref{twistSU2} reduces to the universal relation \eqref{IunivCPW} for all branches. In addition we find that the Riemann surface has to be hyperbolic, i.e. $\fg>1$.

\subsection{Dyonic generalization}
\label{ssec:dyonicindex}

So far we have limited ourselves to turning on a background metric and magnetic fluxes on the Riemann surface $\Sigma_{\fg}$. There are, however, more background parameters that can be turned on while preserving the supersymmetry of the topologically twisted index \cite{Benini:2015noa,Benini:2016hjo,Benini:2016rke}. In the context of holography,  these additional parameters correspond to electric charges that can be in general non-vanishing in the dual supersymmetric AdS$_4$ black holes, see \cite{Benini:2016rke} and references therein.

A generalization of the topologically twisted  index to include electric charges has been proposed  in \cite{Benini:2016rke}. The new ``dyonic'' index is defined as a Legendre transform  of the  ``magnetic'' index discussed in the previous sections and is explicitly given by
\begin{equation}\label{Idyonic}
\mathcal{I}_D(\fn_i,\fq_i;u_i) \equiv \cals I_M(\fn_i;u_i) - \i\, \pi \sum_{i=1}^{r_G}u_i \mathfrak{q}_i\,,
\end{equation}
where $\fn_i$ and $\fq_i$ are the background magnetic fluxes and electric charges, respectively, while $u_i$  are 
 complex  fugacities  replacing the real fugacities $\Delta_i$. The magnetic index, $ \cals I_M(\fn_i;u_i)$,  as a function of the complex $u_i$ is defined by an analytic continuation.   As usual, the magnetic fluxes, $\fn_i$, satisfy the   topological twist condition that preserves supersymmetry, such as \eqref{mfabjm} or \eqref{magconst}. However,   it is a priori not known how to impose the corresponding supersymmetry constraint on the electric charges $\fq_i$. 
 
It was argued in \cite{Benini:2016rke} that in order to obtain the leading saddle point approximation to the dyonic topologically twisted index in the limit of large $N$, one must first fix the values of the electric and magnetic charges $(\mathfrak{q_i},\mathfrak{n_i})$ and then extremize $\mathcal{I}_D(\fn_i,\fq_i;u_i)$ with respect to the complex variables $u_i$, subject to the same constraints as the corresponding $\Delta_i$.   The entropy of the dual dyonic black hole, $S_{\rm BH}(\mathfrak{q},\mathfrak{n})$, should then be identified with the real part of the dyonic index at this extremum. Furthermore, it was conjectured in \cite{Benini:2016rke} that when the index scales with $N$ such that there is a classical dual AdS$_4$ black hole with a regular horizon, i.e.\ $N^{3/2}$ for the {ABJM} and mABJM SCFTs, the supersymmetry constraint on the electric charges is equivalent to $\mathcal{I}_D(\fn_i,\fq_i;u_i)$ being real after the extremization.

In the following subsections we will illustrate this procedure in detail for the two theories of interest and obtain explicit formulae for the twisted index that can be compared directly with the entropy of the dual black holes.

\subsubsection{The ABJM dyonic twisted index}
\label{sec:dabjm}

We start by specializing \eqref{Idyonic} to the ABJM theory. Using \eqref{IABJM}, 
\begin{equation}\label{IdABJM}
\cals I_D(\frak n_{\alpha},\frak q_{\alpha};u_{\alpha})\eql  {\sqrt 2\pi\over 3}\,N^{3/2}\sqrt{u_1u_2u_3u_4}\,\Big({\fn_1\over u_1} +{\fn_2\over u_2}+{\fn_3\over u_3}+{\fn_4\over u_4}\Big)-\i\,\pi\sum_{\alpha=1}^4 u_{\alpha}\frak q_{\alpha}\,,
\end{equation}
where the magnetic fluxes satisfy the supersymmetric twist condition \eqref{mfabjm} while the complex fugacities are constrained by, cf.\ \eqref {DeltaconstrABJM}, 
\begin{equation}\label{uabjmconstr}
u_1+u_2+u_3+u_4\eql 2\,.
\end{equation}
The extremization equations now read
\begin{equation}\label{exdyabjm}
{\partial \cals I_M\over\partial u_{\alpha}}-\i\,\pi\,\fq_{\alpha}+\pi \lambda \eql 0\,,\qquad {\alpha}\eql1,\ldots,4\,,
\end{equation}
where $\lambda=\mu+\i\,\nu$ is a complex Lagrange multiplier. Solving \eqref{exdyabjm} for the electric charges, $\fq_{\alpha}$, and substituting the result back into \eqref{IdABJM}, one finds that the extremized  index is simply given by 
\begin{equation}\label{LagIndabjm}
\cals I_D(\frak n_{\alpha},\frak q_{\alpha};u_{\alpha})\Big|_{u_{\alpha}\eql u_{\alpha}^\text{ext}}\eql -2\pi\lambda(\frak n_{\alpha},\frak q_{\alpha};u_{\alpha}^\text{ext})\,.
\end{equation}
Hence, by imposing the reality condition on the extremal index, we conclude that $\lambda$ must be real.

Next, we go back to \eqref{exdyabjm} and decompose the equations into their real and imaginary parts. This yields eight real equations that are linear in the magnetic fluxes, $\fn_{\alpha}$, and the electric charges, $\fq_{\alpha}$, but highly nonlinear with respect to the complex fugacities, $u_{\alpha}$. Hence, just as before, determining the extremal fugacities, $u_{\alpha}^\text{ext}$, as functions of $\fn_{\alpha}$ and $\fq_{\alpha}$ is a daunting task. Instead, we solve the linear system \eqref{exdyabjm} for the magnetic fluxes and electric charges. 

To present the result in a compact form, it is convenient to set
\begin{equation}\label{upolar}
u_{\alpha}\eql \r_{\alpha}\, e^{\i\,\theta_{\alpha}}\,,\qquad  \theta_\alpha\in (-\pi,\pi)\,,
\end{equation}
and
\begin{equation}\label{analcont}
\sqrt{u_1u_2u_3u_4}\eql  \sqrt{\Delta_1\Delta_2\Delta_3\Delta_4}\, \,e^{{\i\over 2}(\theta_1+\theta_2+\theta_3+\theta_4)}\,,
\end{equation}
which fixes the analytic continuation we are working with.\footnote{The restriction used in \cite{Benini:2016rke} to avoid the square-root sign ambiguity in  \eqref{LagIndabjm} is $0<\Re\, u_\alpha<2$. 
}  Let us define  the following linear combinations of the phases:
\begin{equation}\label{}
\begin{split}
\theta_{\alpha\beta}\eql  \theta_\alpha-\theta_\beta\,,\qquad \theta^*_{\alpha\beta}\eql {1\over 2}\epsilon_{\alpha\beta\gamma\delta}\theta_{\gamma\delta}\,,
\end{split}
\end{equation}
and 
\begin{equation}\label{}
\vartheta_\alpha\eql {1\over 2}(4\theta_\alpha-\theta_1-\theta_2-\theta_3-\theta_4)\,.
\end{equation}

The solution to the linear system \eqref{exdyabjm} can be simplified using the constraint \eqref{uabjmconstr}. This yields 
\begin{equation}\label{thenabjm}
\fn_\alpha\eql -{3\sqrt 2\over N^{3/2}}{\mu\over C(\vartheta)}{\r_\alpha\over \sqrt{\r_1\r_2\r_3\r_4}}\sum_{\beta=1}^4 \sigma_{\alpha\beta}\r_\beta\cos\theta^*_{\alpha\beta}\,,\qquad \alpha=1,\ldots,4\,,
\end{equation}
and  
\begin{equation}\label{theqabjm}
\fq_\alpha\eql -{\mu\over C(\vartheta)}\,\Big[S(\vartheta)+{2\over \r_\alpha}\sum_{\beta=1}^4\sigma_{\alpha\beta}\r_\beta\sin\coeff 1 2(\vartheta_\alpha+\vartheta_\beta)\Big]\,,\qquad \alpha=1,\ldots,4\,,
\end{equation}
where
\begin{equation}\label{}
\begin{split}
C(\vartheta) & \eql \cos \vartheta_1+\cos\vartheta_2+\cos\vartheta_3+\cos\vartheta_4\,,\\[6 pt]
S(\vartheta) & \eql \sin \vartheta_1+\sin\vartheta_2+\sin\vartheta_3+\sin\vartheta_4\,,
\end{split}
\end{equation}
and $\sigma_{\alpha\beta}$ is defined in \eqref{defsigab}. 
We can now use the  topological  twist condition \eqref{mfabjm} to determine the Lagrange multiplier $\lambda=\mu$ and the twisted index as a function of extremal fugacities,
\begin{equation}\label{extdinabjm}
\cals I_D^\text{ABJM}(\fn_\alpha(u),\fq_\alpha(u);u_\alpha)\eql {2\sqrt 2\pi\over 3}N^{3/2}(\frak g-1)\,C(\vartheta)\,{\sqrt{\r_1\r_2\r_3\r_4}\over \sum_{\alpha,\beta}\sigma_{\alpha\beta}\r_\alpha\r_\beta\cos\theta^*_{\alpha\beta}}\,.
\end{equation}
Finally, one can use \eqref{LagIndabjm} to eliminate $\mu$ from \eqref{thenabjm} and \eqref{theqabjm}, to obtain a complete solution to the extremization problem.

It should be clear that the reality of the index provided the ``missing equation'' needed to determine the Lagrange multiplier and hence the electric charges. Somewhere within the solution \eqref{theqabjm} there is a hidden supersymmetric twist condition one should impose ab initio on the electric charges. Identifying this condition more clearly within the field theory remains a puzzle. We will return to this issue  in Section~\ref{sec:abjmstu} when we discuss the  corresponding supergravity calculation. 

It is straightforward to check that in the pure magnetic limit, $\theta_\alpha\to0$, the electric charges \eqref{theqabjm} vanish, while the magnetic fluxes \eqref{thenabjm} and the extremized index \eqref{extdinabjm} reduce to \eqref{nofDelta} and \eqref{topindmagABJM}, respectively.

\subsubsection{The mABJM dyonic twisted index}
\label{sec:mabjmdind}

The extremization of the twisted dyonic index for mABJM  proceeds similarly as for the ABJM index above. We start with
\begin{equation}\label{mabjmdind}
\cals I_D(\fn_i,\frak q_i;u_i) \eql {\sqrt 2\pi\over 3}\,N^{3/2}\sqrt{u_2u_3u_4}\,\Big(\fg-1+\sum_{i=2}^4 {\fn_i\over u_i}\Big)-\i\,\pi\sum_{i=2}^4 u_i\frak q_i\,,
\end{equation}
that follows from \eqref{IdABJM} and \eqref{magconst}, the constraint
\begin{equation}\label{uconsmabjm}
u_2+u_3+u_4\eql 1\,,
\end{equation}
and the corresponding Lagrange multiplier $\lambda=\mu+\i\,\nu$. The extremization equations have the same form as in \eqref{exdyabjm}. Using them to simplify  \eqref{mabjmdind} yields the following relation between the extremized index and the Lagrange multiplier,
\begin{equation}\label{extindm}
\cals I_D(\frak n_i,\frak q_i;u_i)\eql
{2\over 3}\sqrt 2 \pi N^{3/2}(1-\fg)\sqrt{u_2u_3u_4}-2\pi\lambda+\i\,\pi (u_2\fq_2+u_3\fq_3+u_4\fq_4)\,,
\end{equation}
where ${u_i\eql u_i^\text{ext}}$. The more complicated form of this equation in comparison with \eqref{LagIndabjm} is due to the fact that unlike the ABJM index \eqref{LagIndabjm}, the mABJM index \eqref{mabjmdind} is not a homogenous function of $u_i$.
Still, the reality of the extremal index \eqref{extindm} provides an additional equation  that leads to a unique solution for the magnetic fluxes, $\fn_i$, the electric charges, $\fq_i$, and the index as functions of the fugacities.

It is convenient to use the polar parametrization \eqref{upolar} and the following linear combinations of the phases:\footnote{In this section, the indices $i,j,\ldots$ run over the set $2,3,4$. In particular, $\epsilon_{234}=1$, etc.}
\begin{equation}\label{thetastar}
 \theta_{ij}\eql\theta_i-\theta_j\,,\qquad \theta^*_i\eql {1\over 2}\epsilon_{ijk}\theta_{jk}\,,\qquad \theta^*_{ij}\eql \epsilon_{ijk}\theta_k\,,
\end{equation}
and
\begin{equation}\label{varthetam}
\tau_i\eql 3\theta_i-\theta_2-\theta_3-\theta_4\,.
\end{equation}
Let\footnote{$C(\tau)=C(\vartheta)|_{\theta_1=0}$ and $S(\tau)=S(\vartheta)|_{\theta_1=0}$.}
\begin{equation}\label{}
\begin{split}
C(\tau) & \equiv \sum_{i=2}^4\cos\coeff 1 2(\tau_i+\theta_i)+\cos\coeff 1 2(\theta_2+\theta_3+\theta_4)\,,\\
S(\tau) & \equiv \sum_{i=2}^4\sin\coeff 1 2(\tau_i+\theta_i)-\sin\coeff 1 2(\theta_2+\theta_3+\theta_4)\,.
\end{split}
\end{equation}
Then
\begin{equation}\label{nncpw}
\begin{split}
\fn_i \eql & (1-\fg)\r_i\cos\theta^*_i+ \mu {3\sqrt 2\over N^{3/2}}{1\over C(\tau)}{\r_i\over \sqrt{\r_2\r_3\r_4}}\Big[
2\r_i-\sum_{j=2}^4\r_j(\cos\theta^*_{ij}+\cos\theta^*_i\cos\theta^*_j)\Big]\,,
\end{split}
\end{equation}
and  
\begin{equation}\label{qqcpw}
\begin{split}
\fq_i\eql {\sqrt 2\over 3} N^{3/2} (1-\fg) & {\sqrt{\r_2\r_3\r_4}\over \r_i}  \sin\coeff 1 2(\tau_i-\theta_i) \\[6 pt]
 &  +{\mu\, }{\r_i(\sin\theta_i+\sin\tau_i)-\sum_j \r_j\sin(\tau_i-\theta_{ij}) 
\over\r_i(\cos\theta_i+\cos\theta^*_i)}+\nu\,.
\end{split}
\end{equation}

Substituting \eqref{nncpw} into the topological twist condition \eqref{magconst}, we can evaluate the real part of the Lagrange multiplier, $\mu$, and from the reality of the extremized index \eqref{extindm}, the imaginary part $\nu$. A tedious algebra yields the following result for the extremized index: 
\begin{equation}\label{mABJMttind}
\begin{split}
\cals I^\text{mABJM}_D (\fn_i(u),& \fq_i(u);u_i)\eql  -{\sqrt 2\pi\over 3} (\fg-1)N^{3/2} C(\tau) \sqrt{\r_2\r_3\r_4}\\[6 pt]
& \times
{1+ \sum_{i} \r_i\cos\theta^*_i-\sum_{i<j}\r_i\r_j[\cos\theta^*_{ij}+\cos(\theta_i-\theta_j)]\over
\sum_{i} \r_i^2\sin^2\theta^*_i- \sum_{i<j} \r_i\r_j[2\cos\theta^*_{ij}+\cos(\theta^*_i-\theta^*_j)+\cos(\theta_i-\theta_j)]
}\,,
\end{split}
\end{equation}
which has been further simplified using the constraint \eqref {uconsmabjm}.

\subsubsection{The mass deformed twisted dyonic ABJM index}
\label{sec:manjmditc}

At the end of Section~\ref{ssec:mABJMmagI}, we have observed that the extremized  mABJM twisted index \eqref{twistedCPW} could be obtained by starting with the ABJM index \eqref{IABJM} and extremizing it under two constraints \eqref{DeltaconstrCPW}. This is  equivalent to using the ABJM constraint \eqref{DeltaconstrABJM} together with the first constraint  in \eqref{DeltaconstrCPW}, where the latter  formally imposes the mass deformation from ABJM  to mABJM. In this section we discuss this extremization in more detail for the dyonic index, which is also more subtle.

We start with 
\begin{equation}\label{IdABJM2c}
\begin{split}
\cals I_D(\frak n_\alpha,\frak q_\alpha;u_\alpha|\lambda_r) & \eql  {\sqrt 2\pi\over 3}\,N^{3/2}\sqrt{u_1u_2u_3u_4}\,\Big({\fn_1\over u_1} +{\fn_2\over u_2}+{\fn_3\over u_3}+{\fn_4\over u_4}\Big)-\i\,\pi\sum_{\alpha=1}^4 u_\alpha\frak q_\alpha\\
&\qquad + \pi \lambda_1 (u_1-1)+\pi \lambda_2(u_2+u_3+u_4-1) \,,
\end{split}
\end{equation}
where the first line is the ABJM index \eqref{IdABJM} and the second line are the constraints with the corresponding Lagrange multipliers, $\lambda_r=\mu_r+\i\,\nu_r$, $r=1,2$. In addition we impose two conditions \eqref{magconst} on the magnetic fluxes. 

The same calculation as previously shows that the extremized index is 
\begin{equation}\label{exIhg}
\cals I_D(\frak n_\alpha,\frak q_\alpha;u_\alpha|\lambda_r)\eql -\pi (\lambda_1+\lambda_2)\,,
\end{equation}
and hence the reality condition sets,
\begin{equation}\label{nu12s}
\nu_1+\nu_2\eql 0 \,.
\end{equation}

The subtlety, which does not arise in any of the previous examples, is that the extremization does not  lead to a  unique  solution for the electric charges. This comes about from the flat direction in \eqref{IdABJM2c}. If we shift the electric charges by $\delta\frak q_\alpha$ and the imaginary parts of the Lagrange multipliers by $\delta\nu_1$ and $\delta\nu_2$, respectively, then \eqref{IdABJM2c} and \eqref{nu12s} remain invariant provided
\begin{equation}\label{shiftsymm}
\delta\frak q_1\eql -\delta\frak q_2\eql -\delta\frak q_3\eql -\delta\frak q_4\eql\delta\nu_1\eql -\delta\nu_2\,.
\end{equation}

Differentiating \eqref{nu12s} with respect to $u_\alpha$ and solving the imaginary parts of the resulting equations for the electric charges we find
\begin{equation}
\begin{split}\label{mdefqabjm}
\fq_1  = \nu_1 & - {2\over 3}\mu _1{1\over C(\tau)}\sum_{j=2}^4 \Big[\sin\theta_j \cos \coeff 1 2\left(\tau_j-\theta_j\right)-\sin\coeff{1}{2}\left(\tau_j-\theta_j\right) \cos \theta_j^*\Big]\\
& -2 \mu _2 {1\over C(\tau)}\sum_{i=2}^4 \Delta _i \sin \coeff{1}{2} (\tau_i-\theta_i)\,,\\[6 pt]
\fq_i = \nu_2 & -2\mu_1{1\over C(\tau)}{1\over \r_i}\sin\coeff 1 2 (\tau_i-\theta_i)\\
&
-\mu_2{1\over C(\tau)}\,\Big[S(\tau) -4\sin\coeff 1 2(\tau_i+\theta_i)+{2\over \r_i}\sum_{j=2}^4 \r_j\sin \coeff12(\theta_i+\tau_j+\theta_{ij})\Big]\,,
\end{split}
\end{equation}
where the various angles are the same as  in Section~\ref{sec:mabjmdind}, see  \eqref{thetastar} and \eqref{varthetam}.
Substituting \eqref{mdefqabjm} into the real part of the extremization equations we solve for the magnetic fluxes, 
\begin{equation}\label{thevns}
\begin{split}
\fn_1 &= \frac{3 \sqrt{2}}{N^{3/2}}{1\over C(\theta)} {1\over\sqrt{\Delta _2 \Delta _3 \Delta _4}} \Big(\mu_1-\mu _2 \sum_{i=2}^{4}\Delta _i \cos \theta_i^*\Big)
 \,,\\[6 pt]
\fn_i &= \frac{3 \sqrt{2}}{N^{3/2}}\frac{1}{C(\theta)}\frac{\Delta _i}{\sqrt{\Delta_2 \Delta _3 \Delta _4}}\Big[-\mu _1 \cos \theta _{i}^*+\mu _2 \big(2\Delta_i-\sum_{j=2}^{4}\Delta _j \cos \theta _{ij}^*\big)\Big] \,.
\end{split}
\end{equation}
Those depend only on the real parts of the Lagrange multipliers, $\mu_1$ and $\mu_2$, which in turn are determined using   \eqref{magconst},  
\begin{equation}
\begin{split}
\mu_1 &= {\sqrt 2\over 6 }(\fg-1) N^{3/2} {C(\theta)\over \cals D(\Delta,\theta)} \sqrt{\Delta_2\Delta_3\Delta_4}\Big(\sum_{i=2}^4\Delta_i\left(\Delta_i+ \cos \theta^*_i \right)-2\sum_{i<j}\Delta_i\Delta_j\cos\theta_{ij}^*\Big)\,,\\
\mu_2 &= {\sqrt 2\over 6 }(\fg-1) N^{3/2} {C(\theta)\over \cals D(\Delta,\theta)} \sqrt{\Delta_2\Delta_3\Delta_4} \Big(1+\sum_{i=2}^4 \cos \theta^*_i \Delta_i\Big)\,,
\end{split}
\end{equation}
where
\begin{equation}\label{}
\cals D(\Delta,\theta)\eql \sum_{j=2}^4 \sin^2\theta_j^* \Delta_j^2-\sum_{i<j}\Delta_i \Delta_j \left[2\cos\theta_{ij}^*+\cos\left(\theta^*_i-\theta_j^*\right)+\cos\theta_{ij}\right]\,.
\end{equation}

In the formulae above, we have implemented explicitly the constraint $u_1^\text{ext}=1$. 
One can check that as functions of the extremal fugacities, $u_i^\text{ext}$, $i=2,3,4$, subject to the constraint \eqref{uconsmabjm}, the magnetic fluxes, $\fn_2,\fn_3,\fn_4$, in \eqref{thevns} reproduce exactly the magnetic fluxes \eqref{nncpw} in mABJM in Section~\ref{sec:mabjmdind}. Similarly, the extremized twisted index \eqref{exIhg}, that depends only on the real parts of the Lagrange multipliers, is the same as the   dyonic twisted index \eqref{mABJMttind}. In fact, the present calculation yields \eqref{mABJMttind} without using the constraint \eqref{uconsmabjm} to simplify intermediate expressions, which is much simpler.

The electric charges \eqref{mdefqabjm} remain undetermined due the shift symmetry \eqref{shiftsymm}. One way to fix it, is to compare the four electric charges, $\fq_\alpha$, in \eqref{mdefqabjm} with the electric charges, $\tilde \fq_i$, in  
\eqref{qqcpw}, where we have introduced the  ``tilde'' to avoid any confusion. By direct calculation, one can check that $\fq_i=\tilde \fq_i$ for that value of $\nu_1=-\nu_2$ for which $\fq_1\eql 0$, precisely the result one would expect.

To summarize, we have shown that the extremization of the ABJM index with two constraints reproduces exactly the mABJM extremized dyonic twisted index and the corresponding magnetic fluxes and electric charges provided, in addition to \eqref{magconst}, we also impose the condition 
\begin{equation}\label{fq1cond}
\fq_1\eql 0\,,
\end{equation}
on the electric charges. However, one also has the option to fix the shift symmetry differently, which then results in four, typically non-vanishing,  electric charges. As we will see in Section~\ref{sselecchar}, this freedom will be crucial for matching our field theory results with supergravity calculations.

\section{Supergravity}
\label{sec:Sugra}

We expect that the holographic dual description of the twisted compactification of the mABJM SCFT on $\Sigma_{\fg}$ discussed in the previous section is provided by asymptotically AdS$_4$ supersymmetric black holes. In this section we study  these black hole solutions  within the maximal ${\rm SO(8)}$ gauged supergravity theory of de Wit and Nicolai \cite{deWit:1982bul}, which is a consistent truncation of eleven-dimensional supergravity compactified on $S^7$ \cite{deWit:1986oxb,Nicolai:2011cy}. In particular, this means that  our solutions can be uplifted to  M-theory. 

\subsection{The truncation}
\label{sec:trunc}

The four-dimensional $\cals N=8$ supergravity  has many bosonic fields, but to construct the black hole solutions of interest it  is sufficient  to work within a subsector of the theory that is invariant under  the symmetry of  the dual topologically twisted mABJM theory. 

The topological twist breaks the ${\rm SU(3)}_F$ symmetry of the mABJM SCFT,  and the corresponding AdS$_4$ Warner vacuum,  to the Cartan subgroup ${\rm U}(1)^2$. It is   natural to impose this ${\rm U(1)}^2$-invariance on the  $\cals N=8$ supergravity, which then yields  a consistent truncation  to a four-dimensional $\mathcal{N}=2$ supergravity coupled to three vector multiplets and one hypermultiplet. The bosonic fields of the resulting theory are the metric, the graviphoton gauge field along with three ${\rm U(1)}$ gauge fields in the vector multiplets. The  ten real scalars in   the  truncation  parametrize the manifold   \eqref{MVMHintro} and combine into three complex scalars, $z_i$, $i=1,2,3$, in the   vector multiplets and two complex scalars, $\zeta_{1}$ and $\zeta_2$, in the hypermultiplet. The details of the truncation and the geometric data of the resulting $\cals N=2$ supergravity are presented in Appendix~\ref{appendixA}.\footnote{A four-dimensional $\mathcal{N}=2$ supergravity with the same mater content was recently used in \cite{Hosseini:2017fjo,Benini:2017oxt,Kim:2018cpz} to construct asymptotically AdS$_4$ black holes which admit uplifts to massive type IIA supergravity. The supergravity theory we study here differs from the one in \cite{Hosseini:2017fjo,Benini:2017oxt,Kim:2018cpz} by the type of gauging performed on the vector multiplets.} 

The recasting of our truncation into the canonical formalism of $\mathcal{N}=2$ gauged supergravity lets us draw on some standard identities  (see, e.g., \cite{Andrianopoli:1996vr})  and may prove useful for a general analysis of black hole solutions using the results of \cite{Cacciatori:2009iz,DallAgata:2010ejj,Hristov:2010ri,Halmagyi:2013qoa}. However, given the simplicity of the truncation, we also opt for a more direct approach whenever possible. 

In particular, we observe that the topological twists in Section \ref{ssec:twistedindex} have additional invariance, namely  ${\rm U(1)}_R^{\text{W}}$. Imposing this symmetry on our supergravity theory at the level of the bosonic fields amounts to truncating half of the hypermultiplet by setting $\zeta_1=0$. The remaining complex scalar, $\zeta_2$, in the hypermultiplet will be denoted by  $z\equiv \zeta_2$.

The four Abelian gauge fields, $A_{\mu}^{\alpha}$,  $\alpha=0,1,2,3$,  are related to the standard Cartan gauge fields in $\SO(8)$ by 
\begin{equation}\label{Avfieldsmain}
\begin{split}
A^{12} & \eql {1\over 2}(A^0+A^1-A^2-A^3)\,,\qquad 
A^{34}  \eql {1\over 2}(A^0-A^1+A^2-A^3)\,,\\
A^{56} & \eql {1\over 2}(A^0-A^1-A^2+A^3)\,,\qquad 
A^{78}  \eql {1\over 2}(A^0+A^1+A^2+A^3)\,.
\end{split}
\end{equation}

The bosonic Lagrangian for the truncated fields comprises of the usual Einstein-Hilbert term, kinetic terms for the scalars, a Maxwell term and a scalar potential\footnote{The four dimensional metric, $g_{\mu\nu}$, has signature $(-+++)$ and  $e=\sqrt{-\det g_{\mu\nu}}$.}
\begin{equation}
e^{-1}\cals L= {1\over 2}\,R+\cals L_\text{kin}+\cals L_{\text{Max}}-g^2\cals P \,.
\end{equation}
The details of the derivation can be found in Appendix~\ref{appendixA} and here we present only the final result.

The scalar kinetic term is given by
\begin{equation}\label{Lkindef}
\begin{split}
\cals L_\text{kin} & \eql -\sum_{i=1}^3 {\partial_\mu z_i\partial^\mu\bar z_i\over (1-|z_i|^2)^2} -{\big[\partial_\mu z-{\rm i}gA^{(m)}_\mu\,z\big]\big[\partial^\mu \bar z+
{\rm i}gA^{(m)}_\mu\,\bar z\big]\over (1-|z|^2)^2}\,,
\end{split}
\end{equation}
where
\begin{equation}\label{massA}
A^{(m)}_\mu\equiv A^0_\mu-A^1_\mu-A^2_\mu-A^3_\mu\,.
\end{equation}
The scalars parametrize the coset manifold
\begin{equation}\label{}
\cals M\eql \left[\rm {\SU(1,1)\over U(1)}\right]^3\times {\SU(1,1)\over {\rm U}(1)}\,.
\end{equation}
From \eqref{Lkindef} we read-off the diagonal metrics, 
\begin{equation}\label{metrics}
g_{z_i\bar z_j}\eql {\delta_{ij}\over (1-|z_i|^2)^2}\,,\qquad g_{z\bar z}\eql {1\over (1-|z|^2)^2}\,,
\end{equation}
that come  from the K\"ahler potentials,
\begin{equation}\label{}
K_V\eql -\log\big[ (1-|z_1|^2)(1-|z_2|^2)(1-|z_3|^2)\big]\,,\qquad K_H\eql -\log(1-|z|^2)\,,
\end{equation}
respectively.

The contribution from the complex scalar fields to the Lagrangian, $\cals L_{\text{Max}}$,  for the gauge fields is quite complicated. To write it in a compact form it is convenient to use the standard scalar tensors from the $\cals N=2$ formalism \cite{Andrianopoli:1996vr} as summarized in Appendix~\ref{appendixA}. To this end we introduce the holomorphic sections, $X^\alpha$, 
\begin{equation}\label{theXsmain}
\begin{split}
X^0 & \equiv {1\over 2\sqrt 2}\,(1-z_1)(1-z_2)(1-z_3)\,,\qquad 
X^1   \equiv {1\over 2\sqrt 2}\,(1-z_1)(1+z_2)(1+z_3)\,,\\
X^2 & \equiv {1\over 2\sqrt 2}\,(1+z_1)(1-z_2)(1+z_3)\,,\qquad  
X^3   \equiv {1\over 2\sqrt 2}\,(1+z_1)(1+z_2)(1-z_3)\,,\\
\end{split}
\end{equation}
 and the prepotential
\begin{equation}\label{prepmain}
F\equiv -2\,\i\, \sqrt{X^0X^1X^2X^3}\,.
\end{equation}
The Maxwell Lagrangian is then\footnote{The dual field strength is $ \widetilde F^{\alpha}_{\mu\nu}={1\over 2}\eta_{\mu\nu}{}^{\lambda\sigma}F^\alpha_{\lambda\sigma}$, where $\eta_{0123}=e$.} 
\begin{equation}\label{Lmaxdef}
\cals L_{\text{Max}} = {1\over 4}\Big(\cals I_{\alpha\beta}F_{\mu\nu}^\alpha F^{\beta\,\mu\nu}-\cals R_{\alpha\beta}F_{\mu\nu}^\alpha \widetilde F^{\beta\,\mu\nu}\Big)\,,
\end{equation}
where  $\cals R_{\alpha\beta}$  and  $\cals I_{\alpha\beta}$ are  the real and imaginary parts of $\cals N_{\alpha\beta}=\cals R_{\alpha\beta}+\i\,\cals I_{\alpha\beta}$, 
\begin{equation}\label{Nalphabetadef}
\cals N_{\alpha\beta}\equiv \overline F_{\alpha\beta}+2 \,\i\,{(\Im F_{\alpha\gamma})(\Im F_{\beta\delta})X^\gamma X^\delta\over (\Im F_{\gamma\delta})X^\gamma X^\delta}\,,\qquad  F_{\alpha\beta} \equiv {\partial^2 F\over\partial X^\alpha\partial X^\beta}\,.\end{equation}
 
The potential for the scalars is 
\begin{equation}\label{potential}
\begin{split}
\cals P & \eql  {2\over (1-|z|^2)^2}\bigg(3-\sum_{i=1}^3 {2\over 1-|z_i|^2}\bigg)\\
& \quad +{2\,|z|^2\over  (1-|z|^2)^2}\bigg(\prod_{i=1}^3 {1\over 1-|z_i|^2}\bigg)\,\Big[4+4|z_1|^2|z_2|^2|z_3|^2 -(z_1+\bar z_1)(z_2+\bar z_2)(z_3+\bar z_3)
\\
&\hspace{100 pt} -(1+|z_1|^2)(z_2-\bar z_2)(z_3-\bar z_3)-(1+|z_2|^2)(z_1-\bar z_1)(z_3-\bar z_3)\\[6 pt]
&\hspace{100 pt} -(1+|z_3|^2)(z_1-\bar z_1)(z_2-\bar z_2)\,\Big]\;.
\end{split}
\end{equation}
Let us define  the $\cals N=1$ ``holomorphic'' superpotential,
\begin{equation}\label{holV}
\mathcal{V} = \frac{|z|^2}{1-|z|^2}(1-z_1)(1-z_2)(1-z_3)+\frac{2}{1-|z|^2}(z_1z_2z_3-1)\;.
\end{equation}
Then
\begin{equation}
\mathcal{P} = \frac{1}{2}e^{K_V}\big[\,g^{z_i\bar{z}_j}\,\nabla_{z_i}\mathcal{V}\nabla_{\bar{z}_j}\overline{\mathcal{V}}+4g^{z\bar z}\,\partial_{z}\mathcal{V}\partial_{\bar{z}}\overline{\mathcal{V}}-3\mathcal{V}\overline{\mathcal{V}}\,\big]\;,
\end{equation}
where
\begin{equation}\label{defnabla}
\nabla_{z_i}\mathcal{V} = \partial_{z_i}\mathcal{V}+(\partial_{z_i}K_V)\mathcal{V}\;,
\end{equation}
is a covariant derivative.

There are two supersymmetric AdS$_4$ solutions in this truncation corresponding to the  critical points of the potential \eqref{potential} and the superpotential \eqref{holV}.\footnote{In the sense that $\nabla_{z_i}\cals V=\partial_z\cals V=0$.}
 The first one is the ${\rm SO(8)}$-invariant vacuum at
\begin{equation}\label{AdS4SO8}
z_i=0\;,  \qquad z=0\;, \qquad \mathcal{P}_* = -6\;,
\end{equation}
where $\cals P_*$ is the value of the potential at the critical point, which uplifts to the AdS$_4\times S^7$ solutions of the eleven dimensional supergravity. This solution is dual to the conformal vacuum of the ABJM theory.

The second supersymmetric AdS$_4$ solution was found by Warner \cite{Warner:1983vz} and is dual to the mABJM theory. In our parametrization of the potential, it is at %
\begin{equation}\label{cpwz}
z\eql \pm {{\rm i}\over \sqrt 3}\,,\qquad z_1\eql z_2\eql z_3\eql \sqrt 3-2\;,\qquad \mathcal{P}_* = -\frac{9\sqrt{3}}{2}\,.
\end{equation}
It has an $\mathcal{N}=2$ supersymmetry and is invariant under the ${\rm SU(3)}\times {\rm U(1)}_R^{W}$ subgroup of ${\rm SO(8)}$.

The scale of  AdS$_4$ is set by $\cals P_*$. Hence we have
\begin{equation}\label{L4gCPW}
L_{\text{AdS}_4}^2 = - \frac{3}{g^2\mathcal P_*} = \begin{cases} 
\displaystyle {1\over 2g^2} & \quad \text{for SO(8)}\,,\\[6 pt]
\displaystyle \frac{2}{3\sqrt{3}g^2} & \quad \text{for\, W}\,.
\end{cases}
\end{equation}
This four-dimensional background uplifts to the CPW solution \cite{Corrado:2001nv} of the eleven-dimensional supergravity. A more detailed discussion of these (and other) AdS$_4$ vacua in this truncation as well as the spectrum of scalar excitations around them can be found in \cite{Bobev:2010ib}. 

A crucial fact that motivates much of the discussion in this paper   is that there exists a supersymmetric gravitational domain wall solution which connects the two AdS$_4$ vacua described above \cite{Ahn:2000aq,Bobev:2009ms}. This domain wall is the holographic dual realization of the RG flow  described in Section \ref{subsec:ABJMmass}, which connects the ABJM SCFT   to the mABJM SCFT.

There are two further consistent truncations of the supergravity model described above that are of interest for our discussion. The first one is the STU-model obtained by setting the hyperscalar, $z$, to zero and retaining  the three complex scalars, $z_i$, and the four Abelian gauge fields, $A^\alpha$. For a discussion of this model in the present context, see for example  \cite{Benini:2015eyy}. 

The second truncation is to the  $\rm SU(3)\times U(1)_R$-invariant sector originally studied in \cite{Warner:1983vz} and recently discussed in \cite{Bobev:2010ib}. It is obtained by setting
\begin{equation}\label{}
z_1\eql z_2\eql z_3\eql -\bar{z}^\text{BHPW}\,,\qquad z\eql \zeta_2^\text{BHPW}\,.
\end{equation}
The superscript BHPW refers to the scalars in \cite{Bobev:2010ib}, where one also has to set  $\zeta_1^\text{BHPW}=0$. In addition, one must impose  $A^1=A^2=A^3$, which leaves only two Abelian fields in the truncation. 

\subsection{The BH Ansatz} 
\label{sec:ansatz}

Our goal is to study supersymmetric black hole solutions in  the supergravity model  presented above that are dual descriptions of the partial topological twists of the mABJM SCFT discussed in Section \ref{sec:CFT}.
These solutions should  interpolate between one of the two supersymmetric AdS$_4$ vacua, the $\SO(8)$-invariant vacuum in \eqref{AdS4SO8} or the $\SU(3)\times {\rm U}(1)$-invariant vacuum \eqref{cpwz} and a near horizon region   with the metric   of the form AdS$_2\times \Sigma_{\fg}$, where $\Sigma_{\fg}$ is a Riemann surface. As in other known examples of black holes solutions in $\rm AdS_4$ (see, e.g., \cite{Cacciatori:2009iz,DallAgata:2010ejj,Hristov:2010ri}),  we need to turn on   both scalar fields with nontrivial profiles,  as well as non-vanishing gauge fields  carrying the dyonic charges of the black hole at asymptotic infinity. In the presence of both the magnetic and electric charges, this turns out to be a difficult problem in general.
 
Fortunately,  the entropy of these black holes can be determined by a much simpler set-up, namely by studying the solutions  in the near horizon region only.  This is what we will do in the remainder of this section. We will return to the  more difficult problem of constructing full black hole solutions in Section~\ref{sec:magneticBH}, where we present both analytic and numerical solutions for magnetically charged black holes, but with vanishing electric charges.

To construct the near horizon AdS$_2\times \Sigma_{\fg}$ solutions of interest, we take the scalar fields, $z_i$ and $z$,  to be constant and the metric  of the form,
\begin{equation}\label{AnsatzAdS2}
ds^2 = e^{2f_0} ds_{{\rm AdS}_2}^2 + e^{2h_0} ds^2_{\Sigma_{\fg}}\;,
\end{equation}
where the unit radius metric on ${\rm AdS}_2$ is
\begin{equation}\label{}
ds_{{\rm AdS}_2}^2\eql {1\over r^2}(-dt^2+dr^2)\,,
\end{equation}
and $f_0$ and $h_0$ are real constants. 
Given the results of the analysis in \cite{Anderson:2011cz}, we expect that without a loss of generality we can use a constant curvature metric on $\Sigma_\fg$   given in \eqref{hdef}. 

The gauge field fluxes, $F^\alpha=dA^\alpha$, and their (local) potentials, $A^\alpha$, are 
\begin{equation}\label{FdAnsatz}
\qquad F^{\alpha} = e_{\alpha} {\rm vol}_{{\rm AdS}_2} + m_{\alpha} {\rm vol}_{\Sigma_{\fg}}\;,
\qquad 
A^\alpha\eql e_{\alpha} \,\omega_{\rm AdS_2}+m_\alpha \,\omega_{\Sigma_\fg}\,,\qquad \alpha=0,\ldots,3\,.
\end{equation}

One can now plug this Ansatz into the supersymmetry variations and the equations of motion of   maximal $\cals N=8$  gauged supergravity and derive a system of algebraic equations between the metric constants, the scalar fields and the magnetic and electric charges. This is a straightforward but tedious calculation summarized in Appendix~\ref{appendixB}. There we also show that the black holes we construct preserve 2 real supercharges which are enhanced to 4 in the near horizon AdS$_2$ region.

\subsection{Dyonic BH near horizon BPS equations}
\label{subsec:dyonicBPS}

The truncation of the equations of motion and the supersymmetry variations of $\cals N=8$ $d=4$ gauged supergravity discussed in Appendix~\ref{appendixB} yields four types of algebraic equations for the supersymmetric  near horizon dyonic black holes:\footnote{See, the following equations in Appendix~\ref{appendixB}: (i) \eqref{mconstr}, \eqref{a1eqs} and \eqref{e0eqss}; (ii)  \eqref{BPSfi} and \eqref{Phi0eqs}; (iii) \eqref{apsolf0}; (iv) \eqref{Gequation}. We set $\xi=-1$. }
\begin{itemize}
\item [(i)] Four real equations  for the electric and magnetic parameters, $e_\alpha$ and $m_\alpha$:
\begin{align}
e_0 & \eql 0\,,\label{e0eqs}\\
e_0-e_1-e_2-e_3 & \eql 0\,,\label{e123eqs}\\
m_0 & \eql -{\kappa\over 2g}\,,\label{m0eqs}\\
m_0-m_1-m_2-m_3 & \eql 0\,,\label{m123eqs}
\end{align}
where $\kappa=1$,  0 or $-1$ is the normalized curvature of the Riemann surface.
\item [(ii)] Four complex equations for the scalar dressed components, $\Phi_\alpha$,  of the fluxes:
\begin{align}
\cfb_0 & \eql  -2g\,\overline\fW\,,\label{Ph0eqs}\\[6 pt]
\cfb_i & \eql  -2g(1-|z_i|^2)\,D_{z_i}\fW\,,\qquad i=1,\ldots,3\,,\label{Phieqs}
\end{align}
where, cf.\ \eqref{holV}, 
\begin{equation}\label{defofW}
\fW\eql  e^{K_V/2}\,\cals V\,,
\end{equation}

The fluxes $\Phi_\alpha$ are defined by 
\begin{equation}\label{Smatactmn}
S_{\alpha\beta}\cfb_\beta \eql e^{-2h_0}m_\alpha+\i\,e^{-2f_0}e_\alpha\,,
\end{equation}
where 
\begin{equation}\label{Smatrixmn}
\begin{split}
S_{\alpha0} & \eql {1\over \sqrt 2}\, L^\alpha\,,\\ 
S_{\alpha i} & \eql -{1\over \sqrt 2} (1-|z_i|^2) \overline{D_{z_i}L^\alpha}\,.
\end{split}
\end{equation}
Here  $L^\alpha=e^{K_V/2}X^\alpha$ are the symplectic sections, cf.\ \eqref{defofLa}. The K\"ahler covariant derivative in 
\eqref{Phieqs} and  \eqref{Smatrixmn} is defined as%
\footnote{Note that $D_{z_i}= e^{-K_V/2}\partial_{z_i}e^{K_V/2}= e^{K_V/2}\nabla_{z_i}e^{-K_V/2}$, see \eqref{defnabla}.}
\begin{equation}\label{}
D_{z_i}\eql \partial_{z_i}+{1\over 2}\partial_{z_i}K_V\,.
\end{equation}
Note that $D_{z_i}L^\alpha=f_{z_i}{}^\alpha$, see \eqref{thefs}. 
\item [(iii)] One complex equation for the metric constant, $f_0$, and the phase $\Lambda$, 
\begin{equation}\label{eqf0La}
e^{-f_0-\i\,\Lambda}\eql \sqrt 2\,\i\,g\,\fW\,.
\end{equation}
\item [(iv)] A complex cubic constraint for the scalars, $z_i$,
\begin{equation}\label{thezconstr}
\cals C\equiv z_1z_2 z_3 +z_1z_2 +z_2 z_3+z_3 z_1-z_1-z_2-z_3-1\eql 0\,.
\end{equation}
Note that 
\begin{equation}\label{cubeconstr}
\cals C\eql \sqrt2\,\left( X^0-X^1-X^2-X^3 \right)\,,
\end{equation}
where $X^\alpha$ are the holomorphic sections \eqref{theXsmain}.
\end{itemize}

An indirect check of the consistency of these equations with the ones obtained for general dyonic black holes using the formalism of $\cals N=2$ gauged supergravity \cite{DallAgata:2010ejj,Hristov:2010ri,Halmagyi:2013sla} is to rewrite them  as ``attractor equations.'' This is briefly summarized in Appendix~\ref{appattractor}. 

Our task here  is to solve the equations (i)-(iv) so that we can compare directly the black hole entropy
\begin{equation}\label{}
S_{\text{BH}} = \frac{\text{Area}}{4G_N^{(4)}} =  {\pi |\fg-1|\over G_N^{(4)}}\, e^{2h_0}\,,
\end{equation}
with the twisted topological index  \eqref{mABJMttind}. The strategy is to solve   for the metric parameters, the magnetic and electric parameters, and the hyperscalar in terms of the three scalars, $z_i$, which  then will be mapped onto the fugacities, $u_i$, of the mABJM theory.

We start by acting with the matrix $S_{\alpha\beta}$ on \eqref{Ph0eqs} and \eqref{Phieqs}. Using \eqref{Smatactmn},  this yields 
\begin{equation}\label{abseqs}
e^{-2h_0}m_\alpha+\i\,e^{-2f_0}e_\alpha \eql -\sqrt 2g\Big[L^\alpha\,\overline\fW-\sum_{i=1}^3 g^{z_i\bar z_i}\,\overline{D_{z_i}L}{}^\alpha D_{z_i}\fW\Big]\,.
\end{equation}
The next step is to project these equations onto the real and imaginary part and then use \eqref{e0eqs}-\eqref{m123eqs}. 

To this end,  first note that $\fW$ given in \eqref{defofW}
can be rewritten entirely in terms of the symplectic sections, $L^\alpha$,
\begin{equation}\label{calWL}
\begin{split}
\fW & \eql -2\sqrt 2\,L^0 +{\sqrt 2\over 1-|z|^2}\,(L^0-L^1-L^2-L^3)\,,
\end{split}
\end{equation}
where the second term is proportional to the cubic constraint \eqref{cubeconstr},
\begin{equation}\label{Wwconstr}
\fW \eql -2\sqrt 2\,L^0 +{1\over 1-|z|^2}\,\fG\,,\qquad \fG\equiv e^{K_V/2}\,\cals C\,.
\end{equation}
We can use this to simplify the first term in the square bracket in \eqref{abseqs}.

Next, we have the ``useful relation'' \cite{Ceresole:1995ca} 
\begin{equation}\label{usefulrel}
\sum_{i=1}^3 g^{z_i\bar z_i}  D_{z_i}L^\alpha  \overline{D_{z_i}L}{}^\beta \eql -{1\over 2}\,(\cals I^{-1})^{\alpha\beta}-{\overline L}{}^\alpha L^\beta\,,
\end{equation}
where $\cals I_{\alpha\beta}$ is real, the imaginary part  of $\cals N_{\alpha\beta}$ in \eqref{Nalphabetadef}. 
Using those identities  in \eqref{abseqs}, we find
\begin{equation}\label{ealsol}
e^{-2f_0}e_\alpha\eql -4\,\i\,g\left (\,\overline{L}{}^0 L{}^\alpha-L^0\overline{L}{}^\alpha \,\right)\,,
\end{equation}
and 
\begin{equation}\label{malsol}
\begin{split}
e^{-2h_0}m_\alpha & \eql 4\,g\,\left (\,\overline{L}{}^0 L{}^\alpha+L^0\overline{L}{}^\alpha \,\right)\\[6 pt]
&\qquad +2g (\cals I^{-1})^{0\alpha}-{g\over 1-|z|^2}\left[(\cals I^{-1})^{0\alpha}-(\cals I^{-1})^{1\alpha}-(\cals I^{-1})^{2\alpha}-(\cals I^{-1})^{3\alpha}\right]\,.
\end{split}
\end{equation}

The solution \eqref{ealsol} is consistent with the equations \eqref{e0eqs} and \eqref{e123eqs} for the electric parameters. Indeed, the first one is satisfied manifestly, while the left hand side in the second one is proportional to the cubic constraint \eqref{cubeconstr}. 
Substituting  \eqref{malsol} for $\alpha=0$ in \eqref{m0eqs}, we solve for $e^{-2h_0}$ and then similarly \eqref{m123eqs} for $|z|$. Finally, from \eqref{ealsol} and \eqref{malsol} we obtain an explicit solution for all electric parameters and magnetic fluxes. 

This shows that the equations \eqref{e0eqs}-\eqref{eqf0La} have a unique solution for all the dyonic black hole parameters in terms of the scalars, $z_i$, that are constrained by \eqref{thezconstr}. The problem is that the expressions such as \eqref{malsol} are quite difficult to use because of the complicated form of the inverse matrix, $\cals I^{-1}$.

To obtain  simpler explicit expressions for $e^{-2h_0}$ and $|z|$, we go back to  \eqref{abseqs}. Taking a linear combination of these equations such that the left hand side vanishes using \eqref{e123eqs} and \eqref{m123eqs} and observing that the first terms on the right hand side sum up to the constraint, we are left with
\begin{equation}\label{}
2\sqrt 2\,\sum_{i=1}^3  g^{z_i\bar z_i}\, \overline{D_{z_i}\fG} D_{z_i}L^0-{1\over 1-|z|^2}\,\sum_{i=1}^3   g^{z_i\bar z_i} \,\overline{D_{z_i}\fG} D_{z_i}\fG\eql 0\,.
\end{equation}
Note that by the ``useful relation,'' the first term above is real. Let us introduce the shorthand notation  $\langle\,\cdot\,,\,\cdot\,\rangle$ for the scalar product defined by the sums. Then we have 
\begin{equation}\label{zsols}
{1\over 1-|z|^2}\eql 2\sqrt 2\,{\langle D\fG, DL^0\rangle \over  \langle{D\fG}, D\fG\rangle}\,.
\end{equation}

One can verify explicitly that 
\begin{equation}\label{}
\langle DL^0,DL^0\rangle \eql 3|L^0|^2\,.
\end{equation}
This allows to further simplify the solution for $e^{-2h_0}$ that follow from  \eqref{malsol} with $\alpha=0$ after using \eqref{zsols}. The result is
\begin{equation}\label{}
e^{-2h_0}\eql 8 g^2\kappa\left(2|L^0|^2-{|\langle D\fG, DL^0\rangle|^2\over \langle{D\fG}, D\fG\rangle}\right)\,.
\end{equation}

Finally, one can evaluate the scalar products above explicitly and use
\begin{equation}\label{}
D_{z_i}\fG \big|_{\cals C=0}\eql e^{K_V/2}\,\partial_{z_i}\cals C\,.
\end{equation}
Let us define
\begin{equation}\label{}
\Xi \eql \sum_{i=1}^3 (1-|z_i|^2){1-z_i\over 1-\bar z_i}{\partial_{z_i} \cals C}\,,\qquad 
\Gamma\eql \sum_{i=1}^3 (1-|z_i|^2)^2\left|{\partial_{z_i} \cals C}\right|^2\,,\qquad \Pi\eql \prod_{i=1}^3(1-z_i)\,.
\end{equation}
Then we find
\begin{equation}\label{solem2h0}
e^{-2h_0}\eql 8 g^2\,\kappa\,|L^0|^2 \left(2-{|\Xi|^2\over \Gamma}\right)\,,
\end{equation}
and\footnote{One can verify explicitly that
\begin{equation*}
{\overline \Pi}\,\Xi-\Pi\,\overline\Xi \eql 48\,\mathcal{C}+ \sum_{i=1}^3(1-|z_i|^2)\,\mathcal{C}\partial_{\bar{z}_i}\overline{\Pi}+4\sum_{i=1}^3\mathcal{C}\,\partial_{\bar{z}_i}\bar{\mathcal{C}}-6\sum_{i,j=1}^3\mathcal{C}\,\partial_{\bar{z}_i}\partial_{\bar{z}_j}\bar{\mathcal{C}}-{\rm c.c.}\;.
\end{equation*}
Hence ${\overline \Pi}\,\Xi$ is real for $\cals C=\bar{\cals C}=0$ as expected from the discussion above.}
\begin{equation}\label{oneoverzzb}
{1\over 1-|z|^2}\eql -{\overline \Pi\,\Xi\over \Gamma}\,,
\end{equation}
and, from \eqref{eqf0La} and \eqref{Wwconstr}, 
\begin{equation}\label{solef0}
e^{-f_0}\eql -4 \,\i\,g\, e^{\i\,\Lambda}\,L^0\,.
\end{equation}
This completes the solution for the AdS$_2\times\Sigma_\fg$ near horizon black holes in our model.

\subsubsection{Comments} 

The result of our analysis above is an explicit  solution  for the metric parameters and the hypermultiplet given in \eqref{solem2h0}-\eqref{solef0} as functions of the constrained vector multiplets' scalars. The solutions for the magnetic fluxes and the electric parameters can then be read-off from 
\eqref{abseqs} or more directly from \eqref{ealsol} and \eqref{malsol}. It should be noted that in this near horizon solution the hypermultiplet scalar appears only through its absolute value $|z|$. 

It is well known that the STU black holes with electric charges can exist only for nontrivial axions, that is complex scalar fields, $z_i$. The same is true in our model. Indeed, if we set $z_i=\bar z_i$ to be real, the electric parameters given by \eqref{ealsol} automatically vanish. 

 The regularity of a  solution requires that the left hand sides in \eqref{solem2h0} and \eqref{solef0} be strictly positive and, since $|z|<1$, the left hand side in \eqref{oneoverzzb} be greater than~1. This fixes the phase $\Lambda$ and the supersymmetric projectors \eqref{projLa}, and excludes the possibility of black hole solutions with toroidal ($\kappa=0$) horizons. Then we are left with two conditions
\begin{align}\label{h0cond}
\kappa\left(2-{|\Xi|^2\over \Gamma}\right) & >0\,,\qquad \kappa\eql \pm 1\,,\\
\label{zcond}
-{\overline \Pi\,\Xi\over \Gamma}-1 & >0\,.
\end{align}
It is easy to check numerically that both for spherical ($\kappa=1$) and hyperbolic ($\kappa=-1$) horizons, there exist constrained scalars, $z_i$, for which both inequalities are satisfied. 
Once more something interesting happens in the purely magnetic limit. One can check that for real $z_i$'s the ratio of the left hand sides in \eqref{h0cond} and \eqref{zcond} is constant and equal to $-2$. Hence
\begin{equation}\label{highgenus}
-2\kappa>0\qquad \text{for}\qquad z_i\eql \bar z_i\,,
\end{equation}
which excludes spherical horizons for purely magnetic black holes. We show in Section~\ref{sec:magneticBH} that the hyperbolic 
near horizon solutions with only magnetic fluxes indeed give rise to bona fide black holes with AdS$_4$ asymptotics.

Ideally one would like to know for which values of the electric and magnetic parameters, $e_\alpha$ and $m_\alpha$, there are regular near-horizon black hole solutions. This entails finding an explicit solution for the scalars $z_i$, $z$, and the metric parameters, $f_0$ and $h_0$, in terms of $e_\alpha$ and $m_\alpha$. We were not able to analyze explicitly this complicated algebraic problem for the general dyonic solutions above. However, in Section~\ref{sec:solspace} we show how to answer this question for the purely magnetic black holes.

\subsubsection{STU black holes}
\label{sssec:STUBH}

It is straightforward to extract from the supersymmetry variation in Appendix~\ref{appendixB} the STU-limit of our model and reproduce the black holes studied in \cite{Cacciatori:2009iz,Benini:2015eyy,Benini:2016rke}. The resulting BPS equations  can be summarized as follows:

\begin{itemize}
\item [(i)] The four equations \eqref{e0eqs}-\eqref{m123eqs} are replaced by two equations defining the topological twist
\begin{align}\label{eRstu}
e_0+e_1+e_2+e_3 & \eql 0\\[6 pt]
m_0+m_1+m_2+m_3 & \eql -{\kappa\over g}\,.\label{mRstu}
\end{align}

\item[(ii)] Equations \eqref{Ph0eqs}-\eqref{Phieqs} remain the same, except that 
\begin{equation}\label{}
\fW^\text{STU}\equiv \fW\big|_{z=\bar z=0}\eql -\sqrt 2(L^0+L^1+L^2+L^3)\,.
\end{equation}

\item[(iii)] Equation \eqref{eqf0La} is the same, but with $\fW^\text{STU}$.

\item[(iv)] There is no constraint, which was due to the hypermultiplet scalar.

\end{itemize}

The equations in  (ii) give rise to the analogue of \eqref{abseqs}. By taking the sum of those four equations one finds that \eqref{eRstu} is identically satisfied. Then from  \eqref{eRstu} one finds 
\begin{equation}\label{solh0stu}
e^{-2h_0}\eql -g^2\kappa (|\fW^\text{STU}|^2-\langle D\fW^\text{STU},D\fW^\text{STU}\rangle)\,.
\end{equation}
Substituting \eqref{eqf0La} and \eqref{solh0stu} in \eqref{abseqs} one finds $e_0,\ldots,m_3$.

\section{The duality}
\label{sec:DyonicComp}

Our goal now is to test directly the conjecture that the topologically twisted  dyonic/magnetic indices in Section~\ref{ssec:twistedindex} match the entropy of the near horizon black hole solutions above. We also compare directly the magnetic fluxes and electric charges on both sides of the duality.
  
The translation between the gravitational and field theory quantities of interest is provided by  the free energy,  $F_{S^3}$, on both sides of the duality. On the gravity side, the free energy of pure  AdS$_4$ with $S^3$ as an asymptotic boundary can be computed from an on-shell action which diverges unless properly regulated. With the correct counterterms described in \cite{Emparan:1999pm}, one finds
\begin{equation}\label{Fdef}
F_{S^3} = \frac{ \pi L_{\text{AdS}_4}^2}{2G_N^{(4)}}  \;,
\end{equation}
where $G_N^{(4)}$ is the four-dimensional Newton constant and $L_{\text{AdS}_4}$ is the radius of AdS$_4$.  This supergravity result agrees with  the free energy of the ABJM SCFT and the mABJM SCFT to leading order in $N$  given in \eqref{FS3ABJM} and \eqref{FS3CPW}, respectively. Indeed, it was shown in \cite{Jafferis:2011zi}  that the ratios  of these free energies and the radii of the corresponding AdS$_4$ vacua given in \eqref{L4gCPW} are  universal,
\begin{equation}\label{}
{F_{S^3}\over L_{\text{AdS}_4}^2}\eql {2\sqrt 2 \pi\over 3} g^2 N^{3/2}\,.
\end{equation}
Using these results and \eqref{normtg}, we arrive at the following string of equalities:
\begin{equation}\label{Sdeff}
S_{\text{BH}} \equiv \frac{\text{Area}}{4G_N^{(4)}} \eql \frac{\pi |\fg-1| }{G_N^{(4)}}\,e^{2h_0}   \eql 2 \, |\fg-1| {F_{S^3}\over  L_{\text{AdS}_4}^2}\,e^{2h_0}\eql
{4\sqrt 2 \pi\over 3}\,{|\fg-1|}\,g^2\, N^{3/2}\,e^{2h_0}\,.
\end{equation}
where ``Area'' is the area of the black hole horizon.

We have parametrized the field strengths \eqref{FdAnsatz} in terms of ``bare'' magnetic fluxes, $m_\alpha$, and the electric parameters, $e_\alpha$. Those are related to the actual magnetic and electric charges of the AdS$_4$ black holes by (see, e.g.,  \cite{Benini:2016rke}) 
\begin{equation}
\begin{split}\label{nqcharges}
n_\alpha \eql {1\over 4\pi |\fg-1|} \int_\Sigma F^\alpha\,,\qquad 
q_\alpha &= \frac{1}{4\pi |\fg-1|} \int_\Sigma \frac{\delta \mathcal{L}_\text{Max}}{\delta F^\alpha} \,,
\end{split}
\end{equation}
where $ \mathcal{L}_\text{Max}$ is  the Maxwell action  \eqref{Lmaxdef}. Starting with the Ansatz \eqref{FdAnsatz} and evaluating   the integrals  using \eqref{normtg}, we find (cf.~\cite{Halmagyi:2013sla}),
\begin{equation}\label{msandqssugra}
n_\alpha\eql m_\alpha\,,\qquad q_\alpha\eql - e^{2(h_0-f_0)} \mathcal{I}_{\alpha \beta} e_\beta + \mathcal{R}_{\alpha \beta} m_\beta \,.
\end{equation}
These are those magnetic fluxes and electric charges that should be matched with their field theory counterparts in Section~\ref{sec:CFT}.

\subsection{The ABJM SCFT  and STU supergravity}
\label{sec:abjmstu}

The equality between the twisted dyonic index in ABJM SCFT  and the entropy of the corresponding AdS$_4$ black holes in STU supergravity was first shown   in \cite{Benini:2016rke} by  mapping the extremization problem for the index \eqref{IdABJM} onto the BPS equations rewritten in the form of the ``attractor equations'' in \cite{DallAgata:2010ejj,Hristov:2010ri}.  Given the explicit form for the extremized index  derived in Section~\ref{sec:dabjm} and the entropy given by \eqref{Sdeff} and \eqref {solh0stu}, we can  now verify  that equality  directly. 

Following \cite{Benini:2016rke}, let us consider the map between the complex fugacities, $u_\alpha$,  in Section~\ref{sec:dabjm}  and the scalar fields, $z_i$,  at the black hole horizon given by  
\begin{equation}\label{bhzus}
u_{\alpha}(z) \eql {2 X^{\alpha-1}\over X^0+X^1+X^2+X^3}\,,\qquad \alpha\eql 1,\ldots,4\,,
\end{equation}
which automatically solves the constraint  \eqref{extdinabjm}.
Setting 
\begin{equation}\label{}
\Delta_\alpha\eql |u_\alpha |\,,\qquad \theta_\alpha\eql \mathop{\rm Arg}(u_\alpha)\,,
\end{equation}
we find that indeed 
\begin{equation}\label{eqISabjm}
\cals I_D^\text{ABJM}(\fn_\alpha(u),\fq_\alpha(u),u_\alpha)\eql S_{\rm BH}^\text{STU}(z_i)\,.
\end{equation}
The twisted index on the left hand side is evaluated using \eqref{extdinabjm}, which corresponds to the analytic continuation defined by \eqref{analcont}. 
 
By comparing the magnetic fluxes \eqref{thenabjm} and  
the electric charges \eqref{theqabjm} with the  ones obtained from \eqref{abseqs} using \eqref{eRstu} and \eqref{mRstu} in STU supergravity, we find that 
\begin{equation}\label{eqnnabjm}
\fn_\alpha\eql 2g\,|\fg-1|\,m_{\alpha-1}\,,
\end{equation}
and 
\begin{equation}\label{eqqqabjm}
\fq_{\alpha}\eql {2\sqrt 2\over 3} \,g\,|\fg-1|\,N^{3/2}\,q_{\alpha-1}\,,
\end{equation}
where $\alpha=1,\ldots,4$. The relations \eqref{eqnnabjm} and \eqref{eqqqabjm}   agree with the ones  proposed  in \cite{Benini:2016rke}. This confirms the BHZ conjecture for the ABJM SCFT and the STU black holes.

We have verified \eqref{eqISabjm}-\eqref{eqqqabjm} by evaluating both sides numerically for a large number of randomly chosen values of the near horizon scalars, $z_i$. The result is that the extremized index,  the magnetic fluxes and electric charges for the  particular branch for the square-root in \eqref{IdABJM} defined by \eqref{analcont} agree with the entropy, the fluxes and the charges on the supergravity side in the entire domain of the scalars,  $|z_i|<1$. This includes values of $z_i$ for which the supergravity solution may not be regular, such as when the entropy is negative.

\subsection{The mABJM SCFT and W supergravity}
\label{sselecchar}

We have seen in Sections~\ref{sec:mabjmdind} and \ref{sec:manjmditc} that the extremized dyonic twisted index in the  mABJM SCFT theory could be obtained in two ways. On the one hand, one can work entirely within the mABJM theory, which yields the extremized index \eqref{mABJMttind}, the   magnetic fluxes \eqref{nncpw} and the electric charges \eqref{qqcpw} corresponding to {\it three}  gauge fields for the unbroken ${\rm U}(1)^3$ global symmetry. On the other hand, one can start with the ABJM SCFT and extremize the dyonic twisted index while imposing two constraints on the fugacities, where the second constraint comes from the mass deformation from ABJM to mABJM. This leads to the same result for the extremized index as in  \eqref{mABJMttind}. However,  now  there are  four magnetic fluxes \eqref{thevns} and four electric charges \eqref{mdefqabjm} corresponding to the ${\rm U}(1)^4$ global symmetry in ABJM. The mass deformation fixes one of the magnetic fluxes, which together with the topological twist condition,  can be used to determine  the real parts of the two Lagrange multipliers  to find the full agreement between the remaining three magnetic fluxes in both calculations. However, there remains  an ambiguity in the solution for the electric charges due to the shift symmetry \eqref{shiftsymm}. In this section we will compare the field theory results with the supergravity calculations in Section~\ref{subsec:dyonicBPS} and, in particular, clarify the ambiguity of the electric charges found in Section~\ref{sec:manjmditc}. 

\subsubsection{The dyonic twisted index and the entropy}
\label{sec:mabjmentropy}

To obtain the mapping between the fugacities $u_2,u_3,u_4$ in mABJM and the near horizon scalars, $z_1,z_2,z_3$, in our supergravity model, we note that the cubic constraint \eqref{thezconstr} is equivalent to $u_1=1$ in \eqref{bhzus}. Then   $u_2,u_3,u_4$ automatically satisfy \eqref {uconsmabjm}. One can also arrive at this mapping by the following change of variables. Observe that the M\"obius transformation,
\begin{equation}\label{ztildedef}
z_i\qquad \longrightarrow\qquad \tilde z_i\eql {1-z_i\over 1+z_i}\,,\qquad \Re \,\tilde z_i>1\,,
\end{equation}
turns the cubic constraint \eqref{thezconstr} into 
\begin{equation}\label{ncubconstr}
\tilde{\cals C}\equiv \tilde z_i\tilde z_2\tilde z_3 -\tilde z_1-\tilde z_2-\tilde z_3\eql 0\,, 
\end{equation}
which can be rewritten in the following suggestive form,
\begin{equation}\label{}
{1\over \tilde z_2\tilde z_3}+{1\over \tilde z_1\tilde z_3}+{1\over \tilde z_1\tilde z_2}\eql 1\,.
\end{equation}
Hence, it is natural to define
\begin{equation}\label{construsub}
u_{i}\eql {\tilde z_{i-1}\over\tilde z_1\tilde z_2\tilde z_3}\,,\qquad i=2,3,4\,,
\end{equation}
which reproduces \eqref{bhzus} modulo the constraint \eqref{thezconstr}.

Using the substitution \eqref{construsub}, we find that the extremized twisted index in mABJM \eqref{mABJMttind} and the entropy given by  \eqref{Sdeff} and \eqref{solem2h0} are the same,
\begin{equation}\label{}
\cals I^\text{mABJM}_{D}(\fn_i(u),\fq_i(u);u_i)\eql S_\text{BH}(z_i)\qquad \text{for}\qquad \cals C(z_i)\eql 0\,,
\end{equation}
in the entire domain of the scalars, $|z_i|<1$, where the cubic contraint \eqref{thezconstr} is satisfied. 

\subsubsection{Duality for the  magnetic fluxes and  electric charges}
\label{sec422}

An initial puzzle when comparing the magnetic fluxes and the electric charges  in mABJM with the ones in the dual (W-) supergravity is that the latter appears to have four vector fields, while there are only three fluxes and three charges in  \eqref {nncpw} and  \eqref{qqcpw}, respectively. The resolution is that at the W-critical point, which is the gravity dual for mABJM, one of the vector fields,
\begin{equation}\label{}\label{massA}
A^{(m)}\eql A^0-A^1-A^2-A^3\,,
\end{equation}
becomes massive and must be set to zero. This leaves us with three vector fields, $A^1$, $A^2$ and $A^3$ and the corresponding  three magnetic fluxes, $n_i$, and three electric charges, $q_i$, in \eqref{nqcharges} to compare. 

Using the map \eqref{construsub} between the constrained scalars, $z_i$, and the fugacities, $u_i$,   we find the same relation \eqref{eqnnabjm} between the magnetic fluxes \eqref {nncpw} in mABJM and their gravity duals, 
\begin{equation}\label{nabjmcom}
\fn_i\eql 2g\,|\fg-1|\,n_{i-1}\,,\qquad i=2,3,4\,.
\end{equation}

The comparison of the electric charges is more subtle. We must first impose the massive condition \eqref{massA} in the Maxwell Lagrangian to find the scalar matrices $\widetilde{\cals I}_{ij}$ and $\widetilde{\cals R}_{ij}$ for the vector fields, $A^i$, 
\begin{equation}\label{}
\begin{split}
\widetilde{\cals I}_{ij} & \eql \cals I_{ij}+\cals I_{00}+\cals I_{i0}+\cals I_{0j}\,,\qquad 
\widetilde{\cals R}_{ij} \eql \cals R_{ij}+\cals R_{00}+\cals R_{i0}+\cals R_{0j}\,,\\
\end{split}
\end{equation}
and then calculate the electric charges, $\tilde q_i$, using \eqref{msandqssugra}.  Comparing with the field theory charges in  \eqref{qqcpw}, which we denote by $\tilde \fq_i$, we   find
\begin{equation}\label{qabjmcomp}
\tilde \fq_i\eql {2\sqrt 2\over 3}g\,|\fg-1|\,N^{3/2}\,\tilde q_{i-1}\,,\qquad i=2,3,4\,,
\end{equation}
which is the same as \eqref{eqqqabjm}. As before, the comparison is carried out by a numerical substitution and both \eqref{nabjmcom} and \eqref{qabjmcomp}  hold for all scalars, $z_i$, satisfying the cubic constraint.

\subsubsection{More on electric charges }
\label{sec:eleccharg}

We have argued in  Section~\ref{sec:manjmditc} that the dyonic twisted index and the magnetic fluxes in mABJM could be obtained unambiguously by performing simultaneously the mass deformation and the topological twist in ABJM. However, the resulting electric charges were determined only up to the 1-parameter shift symmetry \eqref {shiftsymm}. One way to fix that symmetry was to set one of the charges to zero, see \eqref{fq1cond}, to obtain a complete agreement with the purely mABJM charges. 

Another possibility, which we will discuss now, is to compare the four electric charges \eqref{mdefqabjm} with the four electric charges in our supergravity model that are present before imposing the massive constraint \eqref{massA} on the vector fields. With all four vector fields present, the corresponding charges, $q_\alpha$, are given by \eqref{msandqssugra}. By a direct substitution, we find  that 
\begin{equation}\label{qabjmtwoL}
\begin{split}
\fq_1-\nu_1 & \eql {2\sqrt 2\over 3}g\,|\fg-1|\,N^{3/2}\,q_0\,,\\
\fq_i+\nu_1 & \eql {2\sqrt 2\over 3}g\,|\fg-1|\,N^{3/2}\,q_{i-1}\,,\qquad i=2,3,4\,.
\end{split}
\end{equation}
Hence for $\nu_1=0$ we find the same relation between the charges as in \eqref {eqqqabjm}. Given \eqref{nu12s}, which followed from the reality of the extremized twisted index \eqref{IdABJM2c}, we see that a complete match between mass-deformed, topologically twisted ABJM theory and our supergravity model requires that both Lagrange multipliers, $\lambda_1$ and $\lambda_2$ , 
in \eqref{IdABJM2c} be real. Once again, the agreement between the field theory and the dual supergravity charges holds for all allowed values of the near horizon scalars.

\section{Magnetic black holes}
\label{sec:magneticBH}

So far we have constructed a large family of supersymmetric AdS$_2\times \Sigma_{\fg}$ solutions which can be interpreted  as near horizon limits of supersymmetric dyonic black holes in our gauged supergravity model. In addition, we have shown that the entropy associated with these near-horizon backgrounds is the same as the large $N$ limit of the dyonic topologically twisted index of the mABJM SCFT. Now we try to be more explicit and focus on a class of supersymmetric solutions of our model that do not have electric charges. This allows for a much more explicit analysis of the BPS equations. In particular, in addition to the near-horizon solutions discussed above, we are able to find fully-fledged black hole backgrounds.

To this end we modify the Ansatz employed in Section \ref{sec:ansatz} by setting the electric charge parameters, $e_{\alpha}$, in \eqref{FdAnsatz} to zero. It is consistent then to take the four complex scalars $z_{1,2,3}$ and $z$ to have constant phases. As discussed around \eqref{highgenus}, the BPS equations impose then that the Riemann surface is hyperbolic, so we set $\kappa=-1$ in this section.

We find it convenient to use the following reparametrization of the complex scalar fields, cf.\ \eqref{thezs} and \eqref{thezetas},
\begin{equation}\label{thezss}
z\eql \tanh\chi\,e^{\,\rm{i}\,\psi}\,,\qquad z_i\eql \tanh\lambda_i\,e^{\,{\rm i}\,\varphi_i}\,,\quad i=1,2,3\,,
\end{equation}
where $\chi,\lambda_i\geq 0$,  $2\pi > \psi,\varphi_i\geq 0$ are real valued fields.

Since we are interested in  solutions that can be asymptotic to the ${\rm SU(3)}\times {\rm U(1)}$ invariant vacuum, we choose the following phases in \eqref{thezs}, cf.\ \eqref{cpwz},
\begin{equation}\label{psivarphi}
\psi = \frac{\pi}{2}\;, \qquad  \varphi_1=\varphi_2=\varphi_3=\pi\;.
\end{equation}
The metric for the black hole solutions of interest takes the following form
\begin{equation}\label{AnsatzAdS2r}
ds^2 = e^{2f(r)}(-dt^2+dr^2) + e^{2h(r)} ds^2_{\Sigma_{\fg}}\;,
\end{equation}
with the same metric on the Riemann surface as in \eqref{AnsatzAdS2}.  The four real scalars are in general functions of 
the radial variable, $\chi(r)$ and $\lambda_i(r)$.

With this Ansatz at hand, one can  analyze the supersymmetric variations \eqref{spin12} and \eqref{spin32} of the $\cals N=8$ gauged  supergravity and find BPS equations for the metric functions and the scalars. To write these equations in a compact form,  we find it convenient to introduce the following positive variables:\footnote{Notice that for real $z_i$ we have $x_i=\tilde{z}_i$ where $\tilde{z}_i$ are defined in \eqref{ztildedef}.}
\begin{equation}\label{xlambda}
x_i\equiv e^{2\lambda_i}\,,\qquad i=1,2,3\,,
\end{equation}
as well as the ``real superpotential'' (see \eqref{holV}, \eqref{defofW}, and \eqref{psivarphi}) 
\begin{equation}\label{eq:W}
\mathcal{W} \equiv \mathfrak{W}|_{2\psi=\varphi_i=\pi}= \frac{2 x_{1} x_{2} x_{3} \sinh ^2(\chi )-\cosh ^2(\chi ) (x_{1} x_{2} x_{3}+x_{1}+x_{2}+x_{3})}{ 2\sqrt{x_{1}x_2x_3}} \, .
\end{equation}
Then the potential in \eqref{potential} can be written as
\begin{equation}
\mathcal{P} = \frac{1}{2}\left(\frac{\partial \mathcal{W}}{\partial\chi}\right)^2+\frac{1}{2}\sum_{i=1}^3\left(\frac{\partial \mathcal{W}}{\partial\lambda_i}\right)^2-\frac{3}{2}\,\mathcal{W}^2\;,
\end{equation}
and the  BPS equations are given by
\begin{equation}\label{magnBPS}
\begin{split}
\frac{df}{dr} &=  \frac{g}{\sqrt{2}} e^f\, \mathcal{W} - e^{f-2h} \,\mathcal{H} \, ,  \qquad\qquad  \frac{dh}{dr} = \frac{g}{\sqrt{2}} e^f \,\mathcal{W} + e^{f-2h} \,\mathcal{H} \, ,\\[6 pt]
\frac{d\chi}{dr} &= -\frac{g}{\sqrt{2}}\, e^f\, \frac{\partial \mathcal{W}}{\partial \chi} \, ,  \hspace{77 pt} \frac{d\lambda_i}{dr} = -\frac{g}{\sqrt{2}} \,e^f\, \frac{\partial \mathcal{W}}{\partial \lambda_i} - e^{f-2h}\, \frac{\partial \mathcal{H}}{\partial \lambda_i} \,,
\end{split}
\end{equation}
where 
\begin{equation}\label{eq:H}
\mathcal{H} = \frac{1}{2 \sqrt{2\,x_{1} x_{2} x_{3}}} \left(m_0+m_{1} x_{2} x_{3}+m_2 x_{1} x_{3}+m_3 x_{1} x_{2}\right)\;.
\end{equation}
In addition to the equations in \eqref{magnBPS},  one has to impose the constraints \eqref{m0eqs} and \eqref{m123eqs} on the magnetic fluxes.

Our goal is to find supersymmetric black hole solutions with regular horizons  to the BPS equations \eqref{magnBPS}, \eqref{m0eqs}, \eqref{m123eqs} and to analyze their entropy.

\subsection{${\rm AdS}_2\times \Sigma_\fg$ solutions} 
\label{AdS2xSigmag}

We begin with a classification of the possible ${\rm AdS}_2\times \Sigma_\fg$ solutions, which should correspond to the near-horizon limits of the supersymmetric black holes of interest. To this end,  we take the familiar Ansatz \eqref{AnsatzAdS2} for the metric and real, constant scalar fields, $x_i$ and $\chi$. For that  radial dependence of the metric functions and the scalars, the BPS equations reduce to a set of algebraic equations, which can be solved following the procedure outlined in Section~\ref{subsec:dyonicBPS}. 

Just as before, for solutions with nonzero\footnote{As discussed above, setting $\chi=0$ reduces our supergravity model to the STU model and thus the supersymmetric black hole solutions with $\chi=0$ reduce to the ones discussed in \cite{Benini:2015eyy}.} $\chi$, the scalars must obey the cubic constraint  \eqref{thezconstr}, which now takes the form 
\begin{equation}\label{sumxmxc}
x_1 +x_2 +x_3 \eql x_1 x_2 x_3\,.
\end{equation}
This is the same as the constraint   \eqref{ncubconstr} for real $z_i$, cf. \eqref{xlambda}.

Using \eqref{abseqs},  the magnetic fluxes can be expressed in terms of the scalars as follows:
\begin{equation}\label{solal}
\begin{split}
m_1 & \eql {m_0\over x_1x_2+x_1x_3+x_2x_3}\Big(1+{x_2\over x_3}+{x_3\over x_2}\Big)\,,\\[6 pt]
m_2 & \eql {m_0\over x_1x_2+x_1x_3+x_2x_3}\Big(1+{x_3\over x_1}+{x_1\over x_3}\Big)\,,\\[6 pt]
m_3 & \eql {m_0\over x_1x_2+x_1x_3+x_2x_3}\Big(1+{x_1\over x_2}+{x_2\over x_1}\Big)\,,
\end{split}
\end{equation}
where $m_0$ satisfies \eqref{m0eqs}. The other constraint \eqref{m123eqs} on the magnetic fluxes is then automatically satisfied modulo the cubic constraint \eqref{sumxmxc}. 

Ideally, one would like to invert \eqref{solal} to find the scalar fields in terms of the magnetic fluxes, that should be thought of as the physical parameters specifying a solution.
However, the inversion is tedious and not very insightful, so we choose to write our solutions as above in terms of the scalar fields.

Using  \eqref{solal}, we can solve for $f_0$, $h_0$ and the scalar, $\chi$, in terms of the scalars, $x_i$:
\begin{equation}\label{fhchi}
\begin{split}
e^{2h_0} &= \frac{1}{2g^2}\frac{1}{x_1 x_2+x_2 x_3+x_3 x_1} \left(x_1 x_2 x_3-\frac{1}{x_1}-\frac{1}{x_2}-\frac{1}{x_3}\right) \, , \\[6 pt]
{\rm csch}^2\chi &= 1+{x_1^2+x_2^2+x_3^2\over x_1x_2+x_1x_3+x_2x_3}\,,\\[6 pt]
e^{f_0} &= \frac{1}{g \sqrt{2 x_1 x_2 x_3}} \,, 
\end{split}
\end{equation}
where  $x_i$ are constrained by \eqref{sumxmxc}. These equations can also be obtained directly from  \eqref{solem2h0}, \eqref{oneoverzzb}, and \eqref{solef0} by restricting the complex scalars, $z_i$, as in \eqref{thezss} and \eqref{psivarphi}.

\subsubsection{AdS$_2$ solution space} \label{sec:solspace}

\begin{figure}[t]
\begin{center}
\includegraphics[width=3.5in]{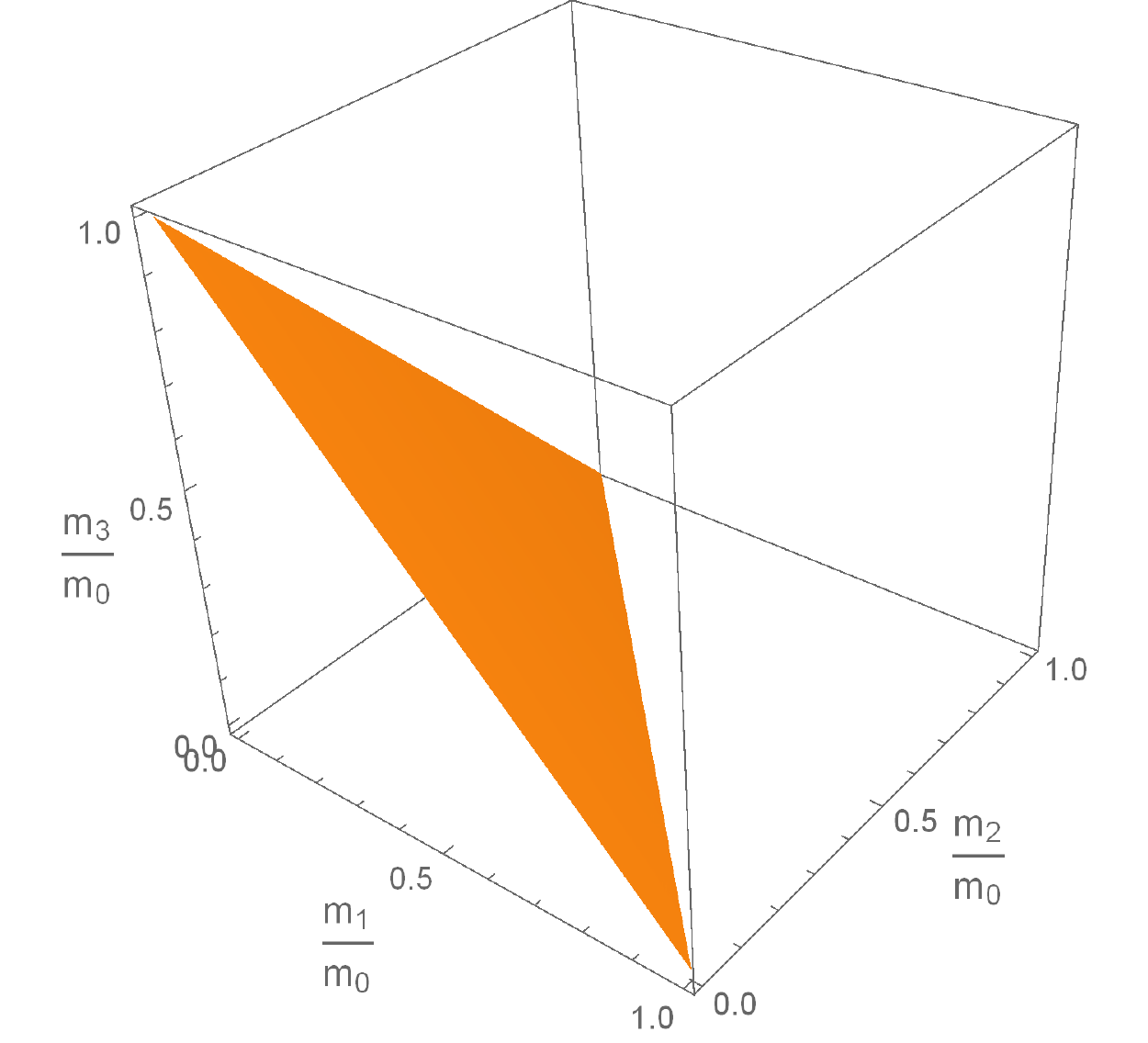}
\caption{The range of magnetic fluxes giving rise to regular AdS$_2\times\Sigma_{\fg}$ solutions.}
\label{fig:positivity}
\end{center}
\end{figure}

Although we have solved the algebraic BPS equations, we still need to analyze for what range of the magnetic fluxes, $m_{\alpha}$, we have a regular well-defined horizon. By that we mean a solution for which the scalars $\chi$ and $\lambda_i$, as well as $f_0$ and $h_0$, are real.

The magnetic fluxes, $m_\alpha$,  and the scalar fields, $x_i$, are related by \eqref{solal}. The scalar fields must be positive, $x_i>0$, and are constrained by \eqref{sumxmxc}. For $\kappa=-1$, we have $m_0>0$ (see \eqref{m0eqs}) and thus the constraint \eqref{m123eqs}, combined with the positivity of $x_i$, leads to the following region in the $m_{1,2,3}$ magnetic flux space:
\begin{equation}\label{CPWconstr}
m_1+m_2+m_3\eql m_0 \,,\qquad \frac{m_i}{m_0} > 0\,.
\end{equation}
This region is the triangle shown in Figure~\ref{fig:positivity}. As noted above, the relation between the magnetic fluxes in \eqref{CPWconstr} combined with \eqref{solal} ensures that the constraint \eqref{sumxmxc} is satisfied. We have checked numerically that for every value of the magnetic fluxes  inside the region specified by \eqref{CPWconstr}, there is a corresponding regular AdS$_2\times \Sigma_{\fg}$  solution, that is    the scalars $x_i$ and  $\chi$,  and the metric functions $e^{f_0}$ and $e^{h_0}$ are real and positive. For the magnetic fluxes at the boundary of the triangle (both the edges and the vertices)   in Figure \ref{fig:positivity},  one has to analyze the regularity of the solution with more care since the relations \eqref{solal} become singular. By solving the algebraic BPS equations directly, we find that there are no regular AdS$_2\times \Sigma_{\fg}$ solutions for these  ``boundary'' values of $m_{\alpha}.$

It is instructive to compare the region of the allowed magnetic fluxes in Figure~\ref{fig:positivity} with the region in which the  STU model regular magnetic black holes discussed in \cite{Benini:2015eyy} exist. Using the results in Section \ref{sssec:STUBH} and the positivity of the scalars, $x_i$, one can show that for the purely magnetic STU model black holes with hyperbolic horizons, i.e.\ $\kappa=-1$, the magnetic fluxes must obey the inequality 
\begin{equation}\label{STUmagineq}
(m_0+m_1-m_2-m_3)(m_0-m_1+m_2-m_3)(m_0-m_1-m_2+m_3) < 0\;.
\end{equation}
When evaluated on the surface $m_0=m_1+m_2+m_3$ relevant for our discussion,  \eqref{STUmagineq} reduces to  the constraint $m_1m_2m_3<0$. Given \eqref{CPWconstr} and \eqref{m0eqs}, we thus conclude that the magnetic STU model black holes have magnetic fluxes that lie outside the orange region  in Figure~\ref{fig:positivity}.

\subsubsection{Black hole entropy}
\label{subsec:genBHmag}

For general magnetic fluxes, it is a non-trivial exercise to write the entropy as a function of the magnetic fluxes since one has to invert the algebraic equations  \eqref{solal}. Therefore we adopt a different strategy. The key observation is that one can use the relation \eqref{nabjmcom} between the field theory and supergravity magnetic fluxes   to show that the extremized values of the R-charges, $\bDelta _{i}$, given in \eqref{n234CPW}, are related to the scalars, $x_i$, obtained by solving \eqref{solal}, by  
\begin{equation} \label{deltax}
\bDelta _{i+1} = \frac{x_i}{x_1 x_2 x_3} \, .
\end{equation}
This relation is the same as \eqref{construsub} evaluated for real scalars and fugacities. Thus we can use \eqref{Sdeff} and \eqref{fhchi} to express the entropy of the black hole in terms of the scalars, $x_i$. Then the relation \eqref{deltax} allows us to compare the entropy to the topologically twisted index for the mABJM SCFT as written in \eqref{twistedCPW}. Implementing this procedure leads to the following expression for the entropy of a general magnetic black hole in our model:
\begin{equation}
\label{SBHgenmag}
S_\text{BH} = \frac{3\sqrt{3}}{2}\frac{1}{x_1 x_2+x_2 x_3+x_3 x_1} \left(x_1 x_2 x_3-\frac{1}{x_1}-\frac{1}{x_2}-\frac{1}{x_3}\right)(\fg-1) F_{S^3} \,,
\end{equation}
where we used  \eqref{FS3CPW}, \eqref{L4gCPW}, and \eqref{Sdeff}. Now, we can compare \eqref{SBHgenmag} to the twisted index of the mABJM SCFT in \eqref{twistedCPW}. Indeed, using the relation \eqref{deltax}, we find that the black hole entropy is equal to the topologically twisted index for all magnetic black hole solutions.

In Section \ref{subsec:magnexpl},  we computed explicitly the twisted index of the mABJM SCFT for a particular choice of magnetic fluxes. Let us now describe the supergravity dual to this setup. To this end we set
\begin{equation}\label{SU2a}
m_1 = m_2 = m_0\, \fn \, .
\end{equation}
The remaining charges $m_0$ and $m_3$ are then fixed in terms of the constant $\fn$ by \eqref{m0eqs} and \eqref{m123eqs}. Thus, for a given choice of $\Sigma_{\fg}$, we are left with a one-parameter family of AdS$_2$ solutions. The positivity constraints \eqref{CPWconstr} imply that in order to have a regular horizon we should have $0<\fn<1/2$. Combining \eqref{SU2a} with \eqref{solal} implies that the scalar fields should obey
\begin{equation}
\frac{x_2}{x_3}+\frac{x_3}{x_2} = \frac{x_1}{x_3} + \frac{x_3}{x_1} \,.
\end{equation}
This equation has two solutions: 
\begin{equation}
\text{Branch 1: } x_1 x_2 = x_3^2 \, , \qquad \text{Branch 2: } x_1=x_2 \, .
\end{equation}
Solving \eqref{solal} for $x_i$ in terms of $\fn$ breaks each of the branches in two more branches\footnote{The $x_i$ must be positive so we discard solutions where the $x_i$ take negative values.} which we denote by the subscript $\pm$.
For Branch 1$_\pm$ we find
\begin{align}\label{x1}
x_1 = \frac{1-\mathfrak{n}\pm\sqrt{(1+\mathfrak{n}) (1-3\mathfrak{n})}}{2 \mathfrak{n}^{3/2}} \, , \quad x_2 = \frac{1-\mathfrak{n}\mp\sqrt{(1+\mathfrak{n}) (1-3\mathfrak{n})}}{2 \mathfrak{n}^{3/2}} \, , \quad  x_3 = \frac{1}{\sqrt{\mathfrak{n}}}\, ,
\end{align}
and for Branch 2$_\pm$ we find
\begin{equation}\label{x2}
x_1 = x_2 = \frac{1}{\sqrt{\fn \mp \sqrt{(1+\fn)(\fn-1/3)}}} \, , \quad x_3 =  \frac{2}{x_1-x_1^{-1}} \, .
\end{equation}
The scalar, $\chi$,  as well as the metric constants, $f_0$ and $h_0$, for each of the four branches can be determined by plugging the expressions for $x_i$ above in \eqref{fhchi}. Finally we have to impose that for each of the branches $\chi$ is real and $x_i$, $e^{2f_0}$ and $e^{2h_0}$ are positive. This restricts the range of the flux parameter, $\fn$, as follows 
\begin{equation}
\text{Branch 1$_\pm$:} \quad 0<\fn<\frac{1}{3} \, , \qquad \text{Branch 2$_\pm$:} \quad \frac{1}{3}<\fn<\frac{1}{2} \, .
\end{equation}
Note that for each value of $\fn$, there are two corresponding near-horizon solutions.
Thus fixing $\fn$ does not specify a unique black hole solution -- one should additionally provide the scalar and metric functions profiles.
In the IR this amounts to selecting a $\pm$ branch, while in the UV one should specify the falloff conditions on the scalar fields. In Section \ref{numerical} we construct numerically the full black hole solution for Branch 1$_+$ and one can do the same for the other branches.

Having inverted \eqref{solal}, we are ready to compute the entropy for this class of near-horizon backgrounds. To do this we combine \eqref{fhchi}, \eqref{x1}, \eqref{x2}, \eqref{Fdef}, and \eqref{Sdeff} to find
\begin{align}\label{entropySU2}
\text{Branch 1$_\pm$: } S_\text{BH} &= \frac{3\sqrt{3\fn}(1-\fn)}{2} (\fg-1) F_{S^3} \, , \\
\text{Branch 2$_\pm$: } S_\text{BH} &=\frac{3}{2\sqrt{2}} \frac{(1 - 2\fn)\left(1+3 \fn^2 \pm(1-3 \fn) \sqrt{(1+\fn) \left(\fn-\frac{1}{3}\right)}\right)}{\sqrt{(1 - 2\fn)\left(1-3 \fn^2 \mp(1-3 \fn) \sqrt{(1+\fn) \left(\fn-\frac{1}{3}\right)}\right)}}  (\fg-1) F_{S^3} \, ,
\end{align}
Comparing these supergravity results to the field theory computation \eqref{twistSU2} we find that the black hole entropy and the twisted index agree perfectly.

One might wonder why did we venture into such an explicit analysis of this particular class of magnetic black hole horizons when we have already shown in Section \ref{sselecchar} that the topologically twisted index matches the black hole entropy for a more general class of dyonic black holes. The key point we want to stress here is that the supersymmetric black holes  are parametrized by their electric and magnetic charges and one has to carefully study the allowed values of these charges for which a regular black hole horizon exists. Unfortunately, the algebraic equations that determine the supersymmetric AdS$_2$ solutions are complicated and do not allow for an  analytic solution of this problem. The  example studied in this section reveals explicitly the somewhat involved branch structure of the space of regular black holes parametrized by the electric and magnetic charges. The successful comparison between the topologically twisted index and the black hole entropy   in Section \ref{sselecchar} was somewhat implicit and did not allow for such an insight.

\subsection{The universal solution} \label{universal}

There is a special type of solution to the BPS equations \eqref{magnBPS} for which the scalars do not flow as a function of the radial coordinate. This is the supergravity dual to the \emph{universal solution} described in Section \ref{ssec:twistedindex}, which arises from a topological twist purely along the superconformal R-symmetry. This black hole solution was discussed in \cite{Romans:1991nq,Caldarelli:1998hg} in the context of minimal four-dimensional gauged supergravity and its universality was emphasized recently in \cite{ABCMZ,BC}.\footnote{See also \cite{Guarino:2017jly} for a recent discussion of this universal solution.} The near-horizon limit of this solution is part of the class of solutions described in Section \ref{AdS2xSigmag} -- specifically it is obtained by setting $x_i=\sqrt{3}$ and $\fn=1/3$ in \eqref{SU2a}. Since the scalar fields do not flow, we set them at their Warner AdS$_4$ vacuum values \eqref{cpwz}. Note that this is consistent with the BPS equations in \eqref{magnBPS}. Using \eqref{m0eqs}, \eqref{m123eqs}, and \eqref{SU2a} with $\fn=1/3$ leads to the following magnetic fluxes
\begin{equation}\label{}
m_0=3m_1=3m_2=3m_3 = -\frac{\kappa }{2g} \,.
\end{equation}
For these values of the scalar fields and charges, we find that $\cals W$ and $\cals H$ take on the following constant values:
\begin{equation}
\cals W_{*} = -3^{3/4} \,, \qquad \cals H_{*} = -\frac{\kappa}{\sqrt{2}\,3^{3/4}g} \,.
\end{equation}
To find the metric functions it is useful to trade the radial coordinate $r$ in \eqref{AnsatzAdS2r} for a new coordinate $r'$ implicitly defined by
\begin{equation}
e^{2f}dr^2=e^{-2f}dr'^2 \,.
\end{equation}
In the new radial variable, the BPS equations \eqref{magnBPS} for the metric functions read
\begin{align} \label{BPSrho}
\frac{d f}{d r'} = \frac{g}{\sqrt{2}} \,e^{-f} \,\mathcal{W} - e^{-f-2h} \,\mathcal{H} \, , \qquad \frac{d h}{d r'}  = \frac{g}{\sqrt{2}} \,e^{-f}\, \mathcal{W} + e^{-f-2h} \,\mathcal{H} \, .
\end{align}
One can check that 
\begin{equation}\label{constmot}
\mathcal{J}\equiv \frac{g}{\sqrt{2}} \,  e^{-f+h}\,\mathcal{W}+ e^{-f-h} \,\mathcal{H} \, ,
\end{equation}
is a constant of motion for the system \eqref{BPSrho}, 
\begin{equation}\label{Ih1}
{d \cals J\over d r'}\eql 0\,.
\end{equation}
Moreover, using \eqref{BPSrho}, $\mathcal{J}$ can be written as
\begin{equation}\label{Ih2}
\mathcal{J} = \frac{d e^{h}}{d r'} \, .
\end{equation}
Combining \eqref{Ih1} and \eqref{Ih2}, we find that $e^{h}=c_1 r'+c_0$, where $c_1$ and $c_0$ are integration constants.
Plugging this into the BPS equation for $h$, \eqref{BPSrho}, and solving for $f$ gives
\begin{equation}
e^f=\frac{1}{c_1}\Big(\frac{g}{\sqrt{2}}\,\mathcal{W}\, e^h+\mathcal{H}\, e^{-h}\Big) \, .
\end{equation}
Finally, using those results in  the metric, we find
\begin{equation}\label{universalBH}
ds^2 = -\left(\frac{\rho}{L_{\text{AdS}_4}}+\frac{\kappa L_{\text{AdS}_4}}{2\rho}\right)^2 dt'^2 + \left(\frac{\rho}{L_{\text{AdS}_4}}+\frac{\kappa L_{\text{AdS}_4}}{2\rho}\right)^{-2} d\rho^2 + \rho^2 ds_{\Sigma_\fg}^2 \,,
\end{equation}
where $L_{\text{AdS}_4}$ is the scale of the Warner AdS$_4$ vacuum defined in \eqref{L4gCPW}, $\rho \equiv c_1 r' +c_0$ and $t'\equiv t/c_1$. 

For $\kappa=0$, i.e.\ $\fg=1$, the gauge field vanishes and the solution is simply AdS$_4$ in  Poincar\'e coordinates. For $\kappa=+1$, i.e.\ $\fg=0$, the metric has a naked singularity at $\rho=0$. For $\kappa=-1$, we find a hyperbolic black hole with a regular horizon at $\rho_0=L_{{\rm AdS}_4}/\sqrt{2}$.
The metric is normalized such that in the UV, i.e. $\rho \rightarrow \infty$, it approaches AdS$_4$ with a hyperbolic boundary and radius $L_{\text{AdS}_4}$.
In the IR, i.e. $\rho \rightarrow \rho_0$, the metric approaches
\begin{equation}
ds^2 = \frac{L_{\text{AdS}_4}^2}{4} \left( ds^2_{\text{AdS}_2} + 2 ds^2_{\Sigma_\fg} \right) \, .
\end{equation}
This near-horizon solution is part of the larger class of AdS$_2\times\Sigma_\fg$ solutions described in Section~\ref{AdS2xSigmag}.
The entropy for the hyperbolic black hole \eqref{universalBH} can thus be obtained from \eqref{entropySU2} by setting $\fn=1/3$.
Then one finds
\begin{equation}\label{SugraUS}
S_\text{BH} = (\fg-1) F_{S^3}^{{\rm mABJM}} \, ,
\end{equation}
where $F_{S^3}^{{\rm mABJM}}$, the free energy for the AdS$_4$ Warner vacuum with $S^3$ as asymptotic boundary, is given in \eqref{Fdef}.
Thus we find an exact match between the universal results: \eqref{SugraUS} from   supergravity and \eqref{IunivCPW} from field theory.

\subsection{Numerical black hole solutions} 
\label{numerical}

In this section we present a numerical analysis of the BPS equations \eqref{magnBPS}. We will do so for the choice of magnetic fluxes $m_1=m_2=m_0 \fn$ discussed in Section \ref{subsec:genBHmag} and furthermore restrict to a scalar field profile that corresponds to  Branch 1$_+$ in \eqref{x1}.
There is no obstruction for repeating the same analysis for general choices of magnetic fluxes and branches.

We find it useful to define $p\equiv f+h$ and use it as the new radial variable. Taking the sum of the BPS equations for $f$ and $h$ in \eqref{magnBPS} allows us to write
\begin{equation}
\frac{d}{d r} = \frac{d p}{d r} \frac{d }{d p} = \sqrt{2}\, e^f g\, \mathcal{W} \frac{d}{d p} \, .
\end{equation}
The BPS equations in terms of $p$ reduce to the following system of five first order non-linear ODEs for the functions $h(p)$, $\chi(p)$ and $\lambda_i(p)$:
\begin{equation}
\begin{split}
\frac{d h}{d p} &= \frac{1}{2}\left(1 +\sqrt{2} \frac{ \cals H}{g \cals W} e^{-2h}\right) \, ,\\[6 pt]
\frac{\partial \chi}{\partial p} &=  -\frac{1}{2 \mathcal{W}} \frac{\partial \mathcal{W}}{\partial \chi} \, , \\[6 pt]
\frac{\partial \lambda_i}{\partial p} &=  -\frac{1}{2g\mathcal{W}} \left(g\frac{\partial \mathcal{W}}{\partial \lambda_i} + \sqrt{2}e^{-2h} \frac{\partial \mathcal{H}}{\partial \lambda_i} \right) \, .
\end{split}
\end{equation}
Once these equations are solved, the function $f(p)$ can be found  using the identity $f=p-h(p)$. 

In order to perform the numerical analysis, it is most convenient to specify the boundary conditions in the IR at the AdS$_2\times \Sigma_{\fg}$ horizon.
This is where we choose to restrict ourselves to Branch 1$_+$ by taking the following IR boundary conditions:
\begin{align}\label{IRcond}
\lambda_1^\text{IR} &= \frac{1}{2}\log\left(\frac{1-\mathfrak{n}+\sqrt{(1+\mathfrak{n}) (1-3\mathfrak{n})}}{2 \mathfrak{n}^{3/2}}\right) \, , & \lambda_3^\text{IR} &= \frac{1}{2}\log\left(\frac{1}{\sqrt{\mathfrak{n}}}\right) \, , \\[6 pt] \label{IRcondd}
\lambda_2^\text{IR} &= \frac{1}{2}\log\left(\frac{1-\mathfrak{n}-\sqrt{(1+\mathfrak{n}) (1-3\mathfrak{n})}}{2 \mathfrak{n}^{3/2}}\right) \, , & \chi^\text{IR} &= \textrm{arctanh}{\sqrt{\fn}} \, , \\[6 pt]
h^\text{IR} &= \frac{1}{2}\log\left(\frac{\sqrt{\fn}(1-\fn)}{2g^2}\right) \,,
\end{align}
where we used \eqref{xlambda}, \eqref{x1} (with $\pm \rightarrow+$) and \eqref{fhchi}. We note that in the radial coordinate $p$ the IR AdS$_2$ region is at $p\to -\infty$ and the UV AdS$_4$ is at $p\to \infty$. We use a numerical implementation in Mathematica by starting with these initial conditions in the IR and numerically integrating towards the UV.
To move away from the near-horizon solution we perturb the scalar fields slightly from their IR values.
However, arbitrary perturbations will generally result in singular solutions.

\begin{figure}[t]
\centering
\includegraphics[width=.5\textwidth]{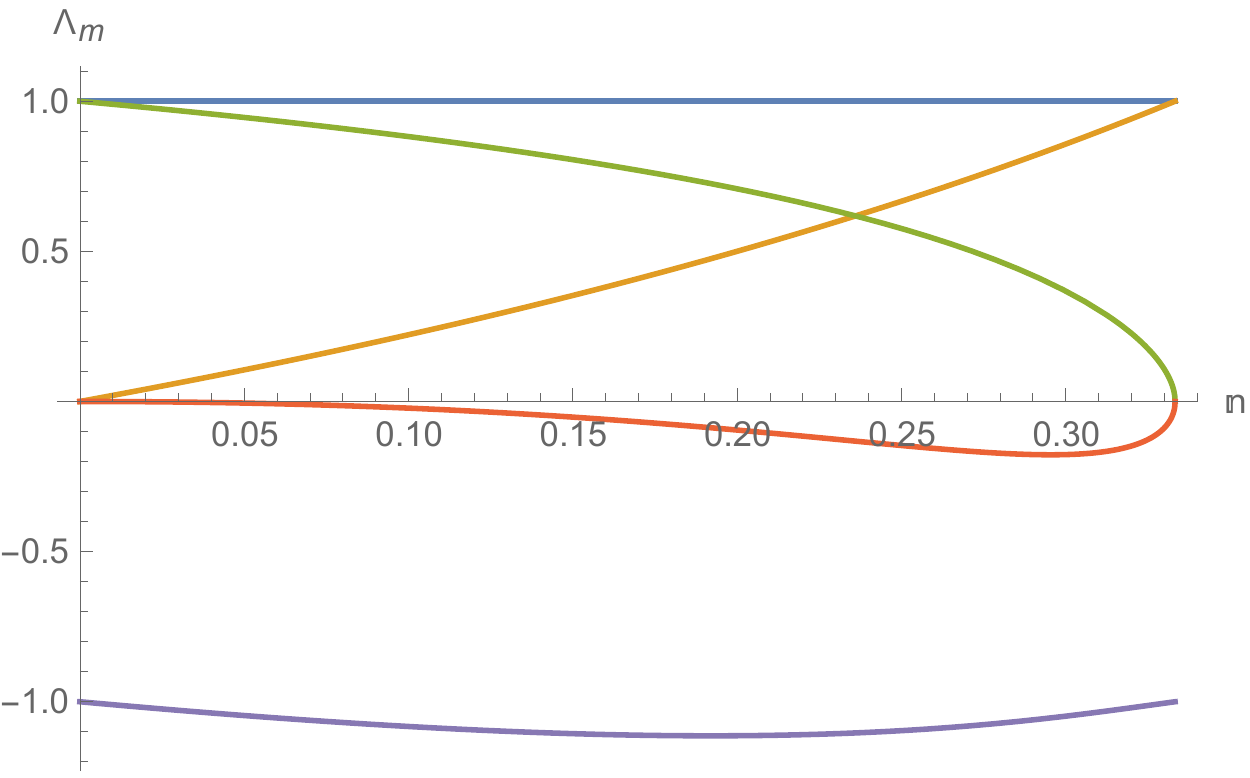}
\caption{Eigenvalues, $\Lambda_m$, of $M_{nm}$ as a function of $\mathfrak{n}$.}
\label{fig:Meig}
\end{figure}

To find the allowed perturbations that produce regular asymptotically AdS$_4$ solutions let us define the fields $\phi_n$ as
\begin{equation}\label{initial}
\lambda_i = \lambda_i^{\text{IR}}+\phi_{\lambda_i}\, , \quad \chi = \chi^{\text{IR}} + \phi_\chi \, , \quad h = h^\text{IR} + \phi_h \, ,
\end{equation}
and expand the BPS equation to first order in $\phi_n$.
This produces the following set of linear equations
\begin{equation}
\frac{\partial \phi_n}{\partial p} = M_{nm} \phi_m \, , \quad m,n \in \{\lambda_1,\lambda_2,\lambda_3,\chi,h\} \, ,
\end{equation}
where $M_{nm}$ is a matrix that depends on the IR values of the scalar and metric fields. Negative eigenvalues of the matrix $M_{nm}$ correspond to directions in field space which lead to singular solutions. Therefore we have to choose the deformation in \eqref{initial} in the direction along the  positive eigenvalues.
The matrix $M_{mn}$ always has two negative and three positive eigenvalues in the allowed range of $\fn$, i.e.\ $0<\fn<1/3$, as shown in Figure \ref{fig:Meig}.
Note that two eigenvalues degenerate at $\mathfrak{n}=\sqrt{5}-2$. Since this is an irrational number it cannot correspond to a properly quantized magnetic flux and we do not discuss it further. This leads us to set up the initial conditions as \eqref{initial} with
\begin{equation}
\phi_n = \epsilon_{(1)} \varphi_{n}^{(1)} + \epsilon_{(2)} \varphi_{n}^{(2)} +\epsilon_{(3)} \varphi_{n}^{(3)} \, ,
\end{equation}
where $\epsilon_{(1)}$, $\epsilon_{(2)}$, $\epsilon_{(3)}$, are small parameters, i.e. $\epsilon_{(a)}\ll1$ for $a=1,2,3$, and $\varphi^{(1)}_n$, $\varphi^{(2)}_n$, $\varphi^{(3)}_n$, are orthonormal  eigenvectors corresponding to the three positive eigenvalues.

\begin{figure}[t]
\centering
        \includegraphics[width=.7\textwidth]{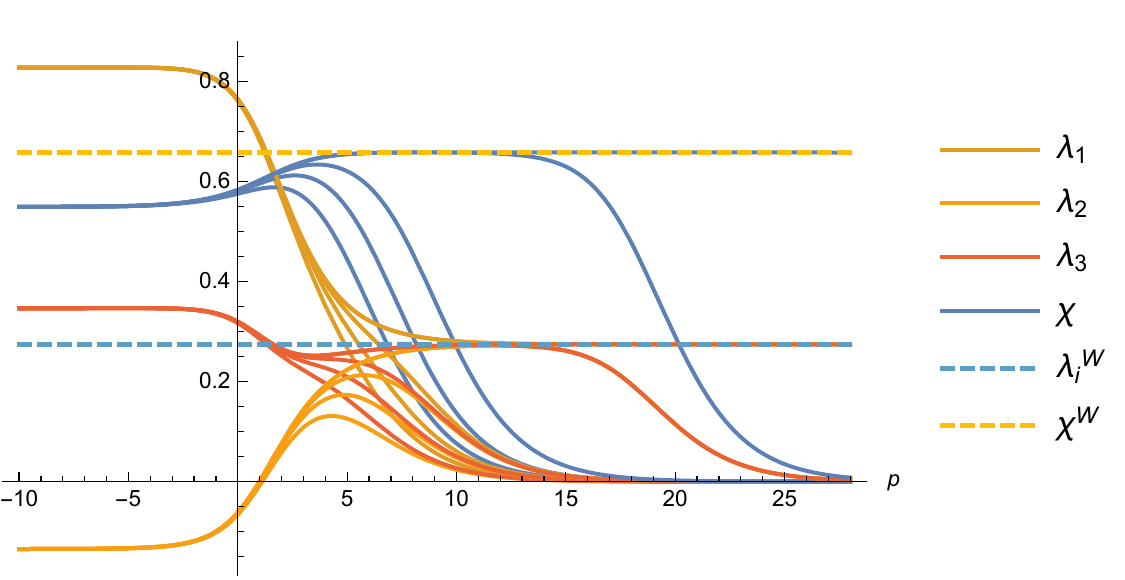}
        \caption{Examples of scalar profiles for $\fn=1/4$ fine tuned to approach the Warner AdS$_4$ fixed point. The dashed lines correspond to the fixed point values for the scalars given in \eqref{cpwz}.}
        \label{fig:scalar12}
\end{figure}

\begin{figure}[H]
    \centering
        \includegraphics[width=.6\textwidth]{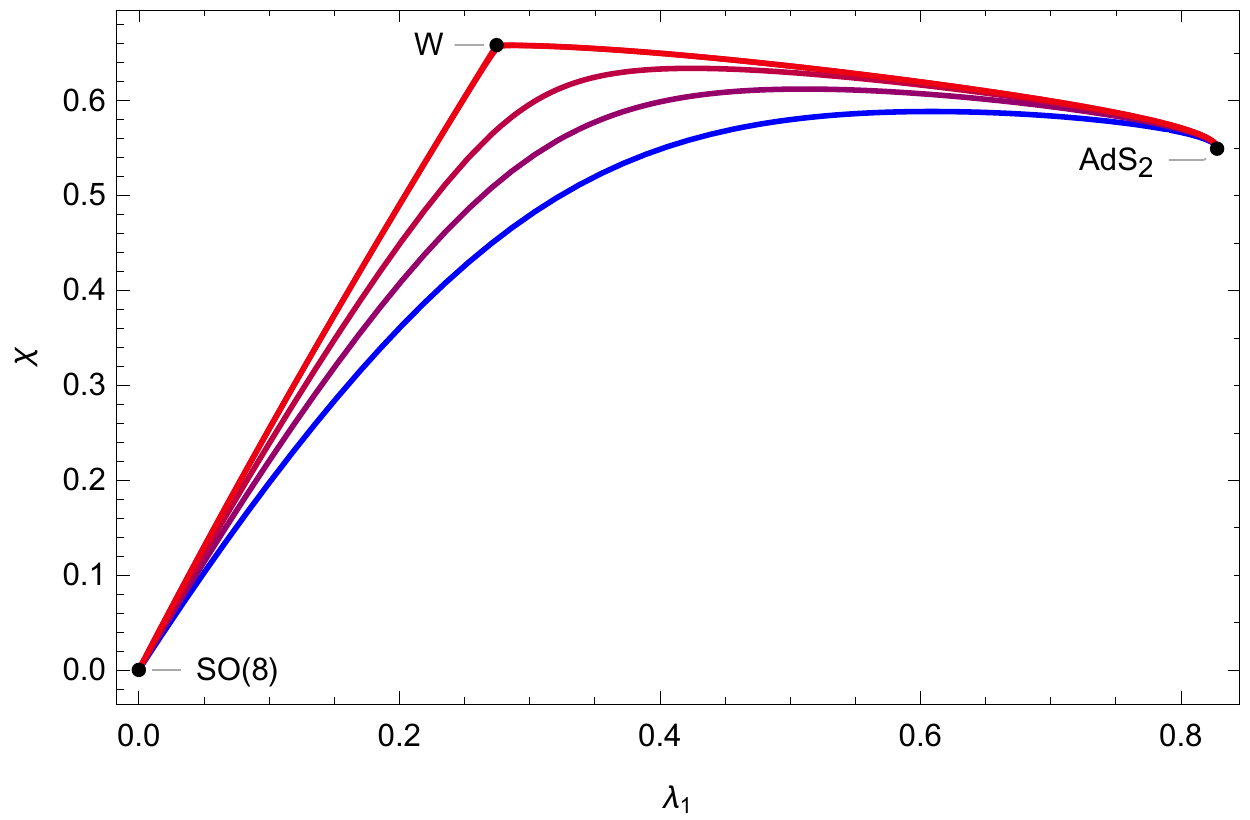}
        \caption{Examples of scalar profiles for $\fn=1/4$ projected onto the $\chi,\lambda_1$ plane. The parameters correspond to those of Figure \ref{fig:scalar12}.}
        \label{fig:chilambda12}
\end{figure}

\begin{figure}[t]
    \centering
        \includegraphics[width=.8\textwidth]{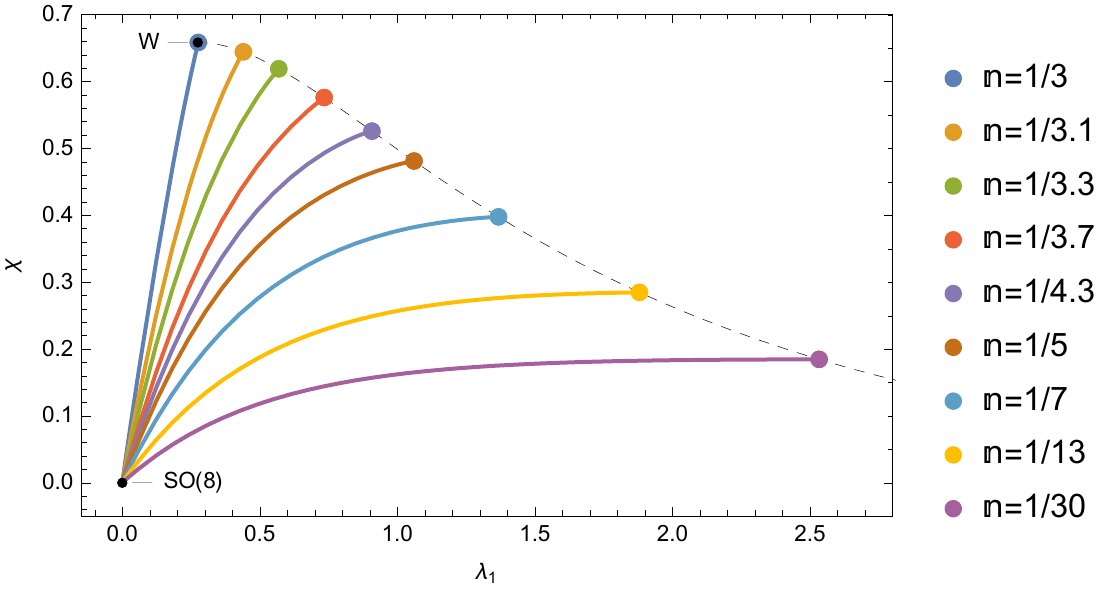}
        \caption{Examples of scalar profiles for various values of $\fn$ ranging from $\fn=\frac{1}{3}$ to $\fn=\frac{1}{30}$, projected onto the $\chi,\lambda_1$ plane. The dashed line corresponds to $\chi^\text{IR}(\lambda_1^\text{IR})$ using \eqref{IRcond} and \eqref{IRcondd}.}
        \label{fig:chilambdan}
\end{figure}

Figures \ref{fig:scalar12} and \ref{fig:chilambda12} showcase the numerical analysis for $\mathfrak{n}=1/4$, $|\epsilon_{(a)}| \approx 10^{-4}$, the IR value of the radial coordinate $p_{\rm IR}=-10$ and the UV value at around $p_{\rm UV}=28$. The Warner point is numerically unstable, but by finely tuning $\epsilon_{(a)}$, one can get very close to the Warner fixed point as indicated in Figure \ref{fig:chilambda12} and Figure \ref{fig:scalar12}.
Figure \ref{fig:scalar12} shows that the scalar fields take on the Warner values \eqref{cpwz} and stay there longer as we tune more finely towards the Warner AdS$_4$ vacuum.
Eventually however, the flow will always move back to the ${\rm SO(8)}$ AdS$_4$ vacuum which is numerically stable. As visible from Figure \ref{fig:chilambda12}, the solutions shown in Figure \ref{fig:scalar12} lie very close to the Warner fixed point. Rather than keeping $\mathfrak{n}$ fixed and taking different initial conditions, we can also vary $\fn$ to produce Figure \ref{fig:chilambdan}.

\noindent 

For the sake of brevity of the presentation here we focused on a detailed analysis of the numerical solutions that asymptote to the $1_+$ branch. However,  we have also found similar numerical solutions for the $1_{-}$ and $2_{\pm}$ branches discussed in Section \ref{subsec:genBHmag}. In fact, the solutions for the $1_{-}$ branch are identical to the ones for the $1_{+}$ branch upon an interchange of the scalar fields $\lambda_1$ and $\lambda_2$.

It is clear from Figure \ref{fig:chilambda12} that for a fixed AdS$_2\times \Sigma_\fg$ near horizon geometry there is a one real parameter family of black hole solutions that asymptote to the ${\rm SO(8)}$ AdS$_4$ vacuum of the four-dimensional supergravity. Only a single member of that family of solutions, illustrated by the red curve in Figure \ref{fig:chilambda12}, asymptotes to the Warner AdS$_4$ vacuum. This may be viewed as a violation of black hole uniqueness and is a feature absent in the known asymptotically black holes solutions of four-dimensional $\mathcal{N}=2$ gauged supergravity without hypermultiplets. We believe that the violation of black hole uniqueness is due to the presence of the hypermultiplet scalar field that has a non-trivial profile in the black hole solution and is charged under the ${\rm U(1)}_{m}$ gauge field. 

A different perspective on this continuous family of solutions is offered by the dual holographic QFT. The ${\rm SO(8)}$ AdS$_4$ vacuum is dual to the ABJM SCFT. In the field theory setup discussed in Section \ref{sec:CFT} we are deforming this SCFT in two ways. We turn on the superpotential mass term  \eqref{CPWsuperp} and in addition we perform the partial topological twist described in Section~\ref{ssec:twistedindex}. These two deformations are relevant and are thus associated with dimensionful parameters, namely the mass, $m$, and the length scale, $\ell_{\Sigma}$, of the Riemann surface, $\Sigma_\fg$.\footnote{We have fixed  $\ell_{\Sigma}$ throughout this paper by normalizing the curvature of $\Sigma_\fg$ to $\kappa$.} Therefore we should expect a one-parameter family of RG flows emerging from the ABJM SCFT in the UV labelled by the dimensionless parameter $m\ell_{\Sigma}$. The one-parameter family of supersymmetric black holes we have constructed is the holographic dual realization of this family of RG flows. A very similar picture was presented in \cite{Bobev:2014jva} for a deformation of the four-dimensional $\mathcal{N}=4$ SYM, which is a combination of a superpotential mass term and a partial topological twist.

\section{Conclusions}
\label{sec:Conclusions}

In this paper we studied the topologically twisted index of the mABJM SCFT which can be thought of as an interacting IR fixed point arising from the ABJM theory deformed by an $\mathcal{N}=2$ preserving mass term. We exploited the gravitational dual of the mABJM SCFT to construct static supersymmetric black hole solutions of the maximal four-dimensional $\rm SO(8)$ $\mathcal{N}=8$ gauged supergravity. In addition, we showed explicitly that the planar limit of the topologically twisted index is equal to the entropy of these black holes to leading order in $N$. Our results can be viewed as an extension of the results and conjectures in \cite{Benini:2015eyy,Benini:2016rke} which employed the topologically twisted index of the ABJM theory to account for the entropy of supersymmetric asymptotically AdS$_4$ black holes in the STU model of four-dimensional $\mathcal{N}=2$ gauged supergravity. A distinct feature of our new supergravity solutions is that they can be viewed as solutions to a particular four-dimensional $\mathcal{N}=2$ gauged supergravity coupled to three vector multiplets and one hyper multiplet. Thus our solutions constitute rare examples of supersymmetric asymptotically AdS$_4$ black holes with non-trivial profiles for hyper multiplet scalars. Our work opens several interesting avenues to explore.

We have provided an exhaustive classification of the supersymmetric dyonic AdS$_2$ solutions in our supergravity model.  However we have discussed fully-fledged black holes only when the electric charges vanish. Given that we have numerous dyonic AdS$_2$ backgrounds with entropy that matches exactly the topologically twisted index of the dual SCFT, it is natural to expect that one can construct full black hole solutions that interpolate between the AdS$_4$ Warner vacuum and these near-horizon AdS$_2$ geometries. Due to the complexity of the BPS equations in our supergravity model, it is likely that such black hole solutions can only be constructed numerically. 

In our analysis we focused on static solutions of the BPS equations. It is known that there are also rotating supersymmetric black holes asymptotic to AdS$_4$ (see, e.g., \cite{Cacciatori:2009iz}). It is natural to conjecture that our supergravity truncation also contains similar solutions and it would be very interesting to construct them explicitly and to understand their entropy microscopically. 

Finally, it should  be possible to construct non-supersymmetric black hole solutions in our supergravity theory. A large class of such solutions was found in \cite{Klemm:2012yg,Chow:2013gba} in  the STU model. One might hope that similar methods can be applied to our truncation, although the presence of hypermultiplet scalars may complicate the construction.

It would be interesting to uplift our black hole solutions to M-theory by combining the uplifts of the STU-model \cite{Cvetic:1999xp,Azizi:2016noi} with the uplift of the Warner critical point \cite{Corrado:2001nv}.
Such M-theory backgrounds may further elucidate the structure of supersymmetric wrapped M2-branes and potentially allow for constructing generalizations of our solutions. A particular generalization may proceed as follows. It was pointed out in \cite{Corrado:2001nv} that the CPW AdS$_4$ vacuum of eleven-dimensional supergravity can be generalized by changing the topology of the internal squashed~$S^7$. The SU(3) invariance of the Warner AdS$_4$ vacuum is realized in eleven dimensions by an explicit $\mathbb{CP}^2$ submanifold of $S^7$. It was observed in \cite{Corrado:2001nv} that one can substitute this four-manifold with any other K\"ahler-Einstein four-manifold $\mathcal{M}_4$ while still preserving $\mathcal{N}=2$ supersymmetry. If this manifold has at least one ${\rm U(1)}$ isometry some of our black hole solutions can probably be generalized. We expect that  $\mathcal{M}_4 = \mathbb{CP}^1 \times \mathbb{CP}^1$ should be the simplest generalization to study.\footnote{See Section 4.3 of \cite{Jafferis:2011zi} for a field theory discussion of this type of generalization.} The universal flow solution discussed in Section~\ref{universal}  should be easy to find for any $\mathcal{M}_4$. 

Finally, we would like to point out that our consistent truncation of the four-dimensional $\mathcal{N}=8$  gauged supergravity may find other applications in the context of holography. One potentially fruitful avenue for further study is to look for Euclidean solutions that are asymptotic to the Warner AdS$_4$ vacuum with an $S^3$ boundary and have non-trivial scalar profiles. These backgrounds should be generalizations of the solutions discussed in \cite{Freedman:2013ryh} and can be viewed as a holographic analog of the $F$-maximization procedure applied to the mABJM SCFT. More precisely, these putative supergravity solutions should describe massive deformations  of the mABJM $\mathcal{N}=2$ SCFT on $S^3$ that break conformal invariance but preserve supersymmetry. It would be interesting to construct these backgrounds explicitly.

\bigskip
\bigskip
\leftline{\bf Acknowledgements}
\smallskip
\noindent We are grateful to P. Marcos Crichigno, Fri\dh rik Gautason, Alessandra Gnecchi, Adolfo Guarino, Kiril Hristov, and Alberto Zaffaroni for interesting discussions. The work of NB is supported in part by an Odysseus grant G0F9516N from the FWO. The work of VSM is supported by a doctoral fellowship from the Fund for Scientific Research - Flanders (FWO) and in part by the European Research Council grant no. ERC-2013-CoG 616732 HoloQosmos. NB and VSM are also supported by the KU Lueven C1 grant ZKD1118 C16/16/005, by the Belgian Federal Science Policy Office through the Inter-University Attraction Pole P7/37, and by the COST Action MP1210 The String Theory Universe. KP was supported in part by DOE grant DE-SC0011687. KP would like to thank the Instituut voor Theoretische Fysica, KU Leuven for hospitality and support during the initial and final stages of this project.



\appendix
\section{Conventions}
\label{appconv}

Throughout this paper we consider smooth Riemann surfaces $\Sigma_{\fg}$ of genus $\fg$. We put a constant curvature metric on $\Sigma_{\fg}$ of the form
\begin{equation}\label{hdef}
ds^{2}_{\Sigma_{\fg}}=H(x,y)^2\,(dx^{2}+dy^{2})\,,\qquad {H(x,y)} = \begin{cases}\displaystyle \frac{2}{1 + x^2 +y^2} & \text{for } \fg = 0 \,,\\    \sqrt{2\pi} & \text{for } \fg = 1\,, \\ \displaystyle{1\over y} & \text{for } \fg >1\;.\end{cases}  
\end{equation}
The volume form 
\begin{equation}\label{volSigma}
{\rm vol}_{\Sigma_{\fg}}\equiv H^2\,dx\wedge dy\,,
\end{equation}
integrates to:
\begin{equation}\label{normtg}
\int_{\Sigma_\fg}{\rm vol}_{\Sigma_{\fg}}=2\pi\eta_{\Sigma} \,, \qquad \eta_{\Sigma} =\begin{cases}
               2|\fg-1| &\qquad\text{for $\fg\neq 1$}\,,\\
              1 &\qquad\text{for $\fg= 1$}\,.
            \end{cases}
\end{equation}
The normalized  curvature of $\Sigma_{\fg}$ is denoted  by $\kappa=1,0$ and $-1$ for $\fg=0$, $\fg=1$, and $\fg>1$, respectively. 
We also use the following locally defined   potential,  $\omega_{\Sigma_\fg}$, for the volume form:
\begin{equation}\label{potSigma}
\omega_{\Sigma_\fg}\eql \begin{cases}\displaystyle
{2(xdy-ydx)\over 1+x^2+y^2} & \text{for~$\fg=0$}\,,\\
\pi(xdy-ydx) & \text{for~$\fg=1$}\,,\\\displaystyle
{dx\over y} & \text{for~$\fg>1$}\,,
\end{cases}\qquad \quad d\omega_{\Sigma_\fg}={\rm vol}_{\Sigma_{\fg}} \,.
\end{equation}

The AdS$_2$ metric is
\begin{equation}\label{}
ds^2_\text{AdS$_2$}\eql {1\over r^2}({-dt^2+dr^2})\,,
\end{equation}
with the volume form and the potential given by
\begin{equation}\label{volAdS}
\text{vol}_\text{AdS$_2$}\eql{dt\wedge dr\over r^2}\,,\qquad \omega_\text{AdS$_2$}\eql {dt\over r}\,.
\end{equation}

To conform with the prevailing custom, we use different index conventions in the field theory and the supergravity sections of the paper. In Section~\ref{sec:CFT}, the indices labelling the R-charges and fugacities are $\alpha,\beta =1,\ldots,4$ and $i,j=2,3,4$, while in Sections~\ref{sec:Sugra} and \ref {sec:magneticBH} the range of the same indices now labelling the dual scalar fields and the vector potentials is $\alpha,\beta=0,\ldots,3$ and $i,j=1,2,3$. In other words,
\begin{equation}\label{}
\alpha^\text{QFT}\eql \alpha^\text{SUGRA}+1\,,\qquad i^\text{QFT}\eql i^\text{SUGRA}+1\,,
\end{equation}
and so on.

\section{The U(1)$^2$-invariant truncation}
\label{appendixA}

In this appendix we present  a $\rm U(1)^2$-invariant truncation of the de Wit-Nicolai $\cals N=8$ gauged supergravity in four dimensions \cite{deWit:1982bul}, where $\rm U(1)^2$ is the Cartan subgroup of the standard $\rm SU(3)\subset SO(6)\subset SO(8)$. On the one side, this truncation can be viewed as a generalization of the $\rm SU(3)$-invariant truncation originally studied in \cite{Warner:1983vz} and recently, in more detail, in \cite{Bobev:2010ib}. On the other side, it    generalizes the  $\rm U(1)^4$-invariant truncation, where $\rm U(1)^4$ is the Cartan subgroup of $\rm SO(8)$, to the STU-model \cite{Behrndt:1996hu,Duff:1999gh,Cvetic:1999xp}. The resulting theory is a  matter coupled  $\cals N=2$ gauged supergravity specified by the geometric data that naturally arise from  the two simpler truncations.  To determine those data, we use the same method as in Appendix B of \cite{Bobev:2010ib}, that is we compare judiciously chosen terms in supersymmetry variations and  in the action of the truncated $\cals N=8$ theory  with those in $\cals N=2$ supergravity. We work here with the original formulation of  gauged $\cals N=2$ supergravity as given in   \cite{Andrianopoli:1996cm,Andrianopoli:1996vr}.
 
Let $T_{12}$, $T_{34}$, $T_{56}$ and $T_{78}$  denote the four standard Cartan generators of $\rm SO(8)$, where   $T_{ij}$  is the generator of rotation in the $(ij)$-plane with charge one. Then the  two symmetry generators  are
\begin{equation}\label{Csu3gen}
 {1\over 2}(T_{12}-T_{34})\,,\qquad   {1\over \sqrt 3}(T_{12}+T_{34}-2\,T_{56})\,,
\end{equation}
under which the 8 gravitini, $\psi_\mu{}^i$, and the corresponding supersymmetries, $\epsilon^i$, of the full theory transform with the charges
\begin{equation}\label{decom8}
\bfs 8_v\quad \longrightarrow\quad  (1,\coeff 1 {\sqrt{3}})+ (1,-\coeff  1 {\sqrt{3}})+ (-1,\coeff 1 {\sqrt{3}})+ (-1,-\coeff 1 {\sqrt{3}})+(0,\coeff 2 {\sqrt 3})+(0,-\coeff 2 {\sqrt 3})+(0,0)+(0,0)\,.
\end{equation}
The two invariant gravitini and supersymmetries are, respectively, the chiral $\psi^{7,8}$ and $\epsilon^{7,8}$ and their complex conjugates  $\psi_{\,7,8}$ and $\epsilon_{7,8}$ of opposite chirality.

The unbroken gauge symmetry  is given by the commutant of the generators \eqref{Csu3gen} in $\rm SO(8)$. Clearly, it is the Cartan subgroup, $\rm U(1)^4$, of $\rm SO(8)$. The corresponding gauge fields, $A^{ij}$, are the same as in the STU-model and consist of the graviphoton and   three gauge fields in vector multiplets. We find it  convenient to work in the same  symplectic frame as in \cite{Duff:1999gh}, which is specified by the following  canonical gauge fields, $A^\alpha$, $\alpha=0,\ldots,3$,
\begin{equation}\label{Avfields}
\begin{split}
A^{12} & \eql {1\over 2}(A^0+A^1-A^2-A^3)\,,\qquad 
A^{34}  \eql {1\over 2}(A^0-A^1+A^2-A^3)\,,\\
A^{56} & \eql {1\over 2}(A^0-A^1-A^2+A^3)\,,\qquad 
A^{78}  \eql {1\over 2}(A^0+A^1+A^2+A^3)\,.
\end{split}
\end{equation}

In the symmetric gauge,  the scalar  56-bein of the $\cals N=8$ supergravity is  given by 
\begin{equation}\label{56bein}
\cals V\eeql \left(\begin{matrix}
u_{ij}{}^{IJ} & v_{ijKL} \\ v^{klIJ} & u^{kl}{}_{KL}
\end{matrix}\right)\eql  \exp \left(\begin{matrix}
0 & \phi_{ijkl}\\
 \phi^{ijkl} & 0
\end{matrix}\right)\in {\rm E_{7(7)}/SU(8)}\,,
\end{equation}
where 
\begin{equation}\label{selfphi}
\phi_{ijkl}\eql {1\over 24}\epsilon_{ijklmnpq}\phi^{mnpq}\,,\qquad \phi^{ijkl}\eql(\phi_{ijkl})^*\,,
\end{equation}
are completely antisymmetric complex self-dual scalar fields.   We find that the $\rm U(1)^2$-invariant nonvanishing $\phi_{ijkl}$'s are given by
\begin{equation}\label{}
\begin{split}
\phi_{1278}& \eql -{1\over 2}\, \lambda_1\, e^{\i\,\varphi_1}\,,\qquad \phi_{3478}\eql -{1\over 2} \,\lambda_2\, e^{\i\,\varphi_2}\,,\qquad \phi_{5678}\eql -{1\over 2}\, \lambda_3 \,e^{\i\,\varphi_3}\,,\\[3 pt]
\phi_{1234} & \eql -{1\over 2}\, \lambda_3 \,e^{-\i\,\varphi_3}\,,\qquad \phi_{1256}\eql -{1\over 2}\,\lambda_2\, e^{-\i\,\varphi_2}\,,\qquad \phi_{3456}\eql -{1\over 2}\, \lambda_1\, e^{-\i\,\varphi_1}\,,\\[3 pt]
\phi_{1357} & \eql -\phi_{1467}\eql -\phi_{2367}\eql -\phi_{2457}\eql  {1\over 4}(\chi_1\cos\psi_1+\i\,\chi_2\sin\psi_2)\,,\\[3 pt]
\phi_{1367} & \eql \phi_{1457}\eql \phi_{2357}\eql -\phi_{2467}\eql -{1\over 4}(\chi_1\sin\psi_1-\i\,\chi_2\cos\psi_2)\,,\\[3 pt]
\phi_{1368}& \eql \phi_{1458}\eql \phi_{2358}\eql -\phi_{2468}\eql -{1\over 4}(\chi_1\cos\psi_1-\i\,\chi_2\sin\psi_2)\,,\\[3 pt]
\phi_{1358}& \eql -\phi_{1468}\eql -\phi_{2368}\eql -\phi_{2458}\eql -{1\over 4}(\chi_1\sin\psi_1+\i\,\chi_2\cos\psi_2)\,,
\end{split}
\end{equation}
where the  five complex fields,
\begin{equation}\label{thezs}
z_i\eql {\lambda_i\tanh|\lambda_i|\over |\lambda_i|}\,e^{\i\,\varphi_i}\,,\qquad i=1,2,3\,,
\end{equation}
and
\begin{equation}\label{thezetas}
\zeta_r\eql {\chi_r\tanh\sqrt{\chi_1^2+\chi_2^2}\over \sqrt{\chi_1^2+\chi_2^2}}\,e^{\i\,\psi_r}\,,\qquad r=1,2\,,
\end{equation}
parametrize the  special K\"ahler manifold, $\cals M_{V}$, of three vector multiplets and the quaternionic K\"ahler manifold, $ \cals M_{H}$, of the universal hypermultiplet, respectively. The two K\"ahler manifolds are
\begin{equation}\label{}
\cals M_{V}\times \cals M_{H} \eql \left[{\rm SU(1,1)\over U(1)}\right]^3\times {\rm SU(2,1)\over SU(2)\times U(1)}\,,
\end{equation}
with the  standard metrics
\begin{equation}\label{}
g_{z_i\bar z_j}dz_id\bar z_j\eql {dz_1d\bar z_1\over (1-|z_1|^2)^2}+{dz_2d\bar z_2\over (1-|z_2|^2)^2}+{dz_3d\bar z_3\over (1-|z_3|^2)^2}\,,
\end{equation}
and
\begin{equation}\label{}
g_{\zeta_i\bar\zeta_j}d\zeta_id\bar\zeta_j\eql {d\zeta_1d\bar\zeta_1+d\zeta_2d\bar\zeta_2\over 1-|\zeta_1|^2-|\zeta_2|^2}+{(\zeta_1 d\bar\zeta_1+\zeta_2 d\bar\zeta_2)(\bar\zeta_1d\zeta_1+\bar\zeta_2 d\zeta_2)\over (1-|\zeta_1|^2-|\zeta_2|^2)^2}\,,
\end{equation}
and the corresponding K\"ahler potentials
\begin{align}\label{KvP}
K_{V} & \eql -\log\big[ (1-|z_1|^2)(1-|z_2|^2)(1-|z_3|^2)\big]\,,\\[3 pt]
K_{H} & \eql -\log(1-|\zeta_1|^2-|\zeta_2|^2)\,.\label{KhP}
\end{align}
Note that $\cals M_V$ is invariant under, $\rm U(1)^4$, and is the scalar manifold of the STU-model. In turn, by construction, $\cals M_H$ is invariant under the symmetries \eqref{Csu3gen} and in fact under the full $\rm SU(3)$. It is thus the same as the hypermultiplet in the $\rm SU(3)$-invariant truncation \cite{Bobev:2010ib}.

As expected, there are eight invariant chiral spin-1/2 fields:
\begin{equation}\label{spin12v}
\chi^{127}\,,\quad \chi^{128}\,,\quad \chi^{347}\,,\quad \chi^{348}\,,\quad \chi^{567}\,,\quad \chi^{568}\,,
\end{equation}
\begin{equation}\label{spin12h}
\chi^{135}\eql -\chi^{146}\eql -\chi^{236}\eql -\chi^{245}\,,\qquad  
\chi^{246}\eql -\chi^{136}\eql -\chi^{145}\eql -\chi^{235}\,,
\end{equation}
and their complex conjugates. By examining the supersymmetry variations of the scalar fields,\footnote{For supersymmetry variations in the $\cals N=8$ theory, see (3.1)-(3.5)  and Section 5 in  \cite{deWit:1982bul}. For the supersymmetry variations in $\cals N=2$ supergravity, see, e.g.,\ (4.18)-(4.25) in \cite{Andrianopoli:1996vr}.}  one can check that \eqref{spin12v} belong to the three vector multiplets, while \eqref{spin12h} to the hypermultiplet.

Similarly, from the supersymmetry variations of the vector fields \eqref{Avfields} into spin-3/2 fields, we read off the symplectic sections 
\begin{equation}\label{defofLa}
L^\alpha \eql e^{K_V/2} X^\alpha\,,\qquad \alpha=0,\ldots,3\,,
\end{equation}
where the holomorphic sections, $X^\alpha$, are explicitly given by 
\begin{equation}\label{theXs}
\begin{split}
X^0 & \eql {1\over 2\sqrt 2}\,(1-z_1)(1-z_2)(1-z_3)\,,\qquad 
X^1   \eql {1\over 2\sqrt 2}\,(1-z_1)(1+z_2)(1+z_3)\,,\\
X^2 & \eql {1\over 2\sqrt 2}\,(1+z_1)(1-z_2)(1+z_3)\,,\qquad  
X^3   \eql {1\over 2\sqrt 2}\,(1+z_1)(1+z_2)(1-z_3)\,,\\
\end{split}
\end{equation}
and $K_V$ is the  K\"ahler potential  \eqref{KvP}. The specific normalization in \eqref{theXs} is fixed by imposing identities of the special K\"ahler geometry and by matching terms in the  $\cals N=8$ and $\cals N=2$ actions. 

As a consistency check we verify that the vectors $f_{z_j}{}^\alpha$, that follow from the  supersymmetry variations of the vector fields into spin-1/2 fields, are indeed given by
\begin{equation}\label{thefs}
f_{z_j}{}^\alpha\equiv D_{z_j}L^\alpha\eql \big(\partial_{z_j}+{1\over 2}\partial_{z_j}K_V\big)\,L^\alpha,
\end{equation}
where the derivative on the right hand side is the usual  K\"ahler covariant derivative.

The prepotential in the STU-model as a function of the holomorphic sections \eqref{theXs} is 
\begin{equation}\label{prep}
F\eql -2\,\i \sqrt{X^0X^1X^2X^3}\,,
\end{equation}
It is determined  by solving 
\begin{equation}\label{derprep}
F\eql {1\over 2}X^\alpha F_\alpha\,,\qquad F_\alpha\eql {\partial F\over\partial X^\alpha}\,,
\end{equation}
and requiring that $F$ be homogenous of degree two as a function of $X^\alpha$'s. 
The overall normalization  can be verified  by the matching of Maxwell actions (see, \eqref{N2act} below), in which the coupling of the scalars, $z_i$, to the gauge fields is given by the second derivatives of the prepotential,
\begin{equation}\label{twoderprep}
F_{\alpha\beta}\equiv {\partial^2F\over \partial X^\alpha\partial X^\beta}\,.
\end{equation}
One should note that a priori the prepotential \eqref{prep} and its derivatives \eqref{derprep} and \eqref{twoderprep}  have a sign ambiguity, in particular when evaluated  as functions of the scalars, $z_i$. That ambiguity is removed by setting
\begin{equation}\label{}
\sqrt{X^0X^1X^2X^3}\eql {1\over 8}(1-z_1^2)(1-z_2^2)(1-z_3^2)\,,
\end{equation}
which follows from  the corresponding $\cals N=8$ expressions.

The gauging in the $\cals N=2$ supergravity is determined by the action of the gauge symmetries  on the scalar manifolds.  As we have already noted above, there are just two $\rm U(1)$'s that act nontrivially on $\cals M_H$, and hence the gauging is the same as in the $\rm SU(3)$-invariant truncation. By comparing  
\eqref{Avfields} with (2.38) and (2.40) in  \cite{Bobev:2010ib}, we find the Killing vectors, $K_\alpha$,  corresponding to the gauge fields $A^\alpha$ in \eqref{Avfields},
\begin{equation}\label{}
\begin{split}
K_0 & \eql 2\,\i\, \zeta_1\partial_{\zeta_1}+\i\,\zeta_2\partial_{\zeta_2}+\text{c.c.}\,,\\[6 pt]
K_1 & \eql K_2\eql K_3\eql -\i\,\zeta_2\partial_{\zeta_2}+\text{c.c.}\,.
\end{split}
\end{equation}
The corresponding moment maps, $P_\alpha=(P_\alpha ^1,P_\alpha ^2,P_\alpha ^3)$, can be  read-off from (B.39) and (B.40) in \cite{Bobev:2010ib} and become quite simple if one sets one of the hyperscalars to zero. In particular, for $\zeta_1=0$ and $\zeta_2=z$, we have
\begin{equation}\label{}
P^\alpha\eql P^2_\alpha\eql 0\,,\qquad P^3_0\eql -{1-2 \,|z|^2\over 1-|z|^2}\,,\qquad P^3_1\eql P^3_2\eql P^3_3\eql -{1\over 1-|z|^2}\,.
\end{equation}
This completes the list of geometric data for the $\cals N=2$ supergravity that arises from this truncation.

As a consistency check we verify explicitly that the $\cals N=8$ bosonic action reduces to the canonical $\cals N=2$ action for the invariant fields. The latter reads
\begin{equation}\label{N2act}
\begin{split}
e^{-1}\cals L_{\cals N=2}& \eql {1\over 2}\,R-g_{z_i\bar z_j}\partial_\mu z_i\partial^\mu\bar z_j-g_{\zeta_i\bar\zeta_j}\nabla_\mu \zeta_i\nabla^\mu\bar\zeta_j\\[6 pt]
& \quad  +{1\over 4}\Big(\ \cals I_{\alpha\beta}F_{\mu\nu}^\alpha F^{\beta\,\mu\nu}-\cals R_{\alpha\beta}F_{\mu\nu}^\alpha \widetilde F^{\beta\,\mu\nu}\Big)-g^2\,\cals P\,,
\end{split}
\end{equation}
where
\begin{equation}\label{}
\nabla_\mu \zeta_i\eql \partial_\mu \zeta_i+g A_\mu^\alpha K_\alpha{}^{\zeta_i}\,,
\end{equation}
is the covariant derivative of the scalar fields, $\cals R_{\alpha\beta}$ and $\cals I_{\alpha\beta}$ are, respectively, the real and imaginary part of the matrix
\begin{equation}\label{}
\cals N_{\alpha\beta}\eql \overline F_{\alpha\beta}+2 \,\i\,{(\Im F_{\alpha\gamma})(\Im F_{\beta\delta})X^\gamma X^\delta\over (\Im F_{\gamma\delta})X^\gamma X^\delta}\,,\qquad \cals N_{\alpha\beta}\eql \cals R_{\alpha\beta}+\i\,\cals I_{\alpha\beta}\,,
\end{equation}
and\footnote{The ``dot'' denotes the summation over the vector index of the moment maps.}  
\begin{equation}\label{}
\cals P \eql 4\, g_{ab} K_\alpha{}^{a}K_\beta{}^{b}\overline L^\alpha L^\beta+g^{z_i\bar z_j}f_{z_i}{}^\alpha f_{\bar z_j}{}^\beta P_\alpha\cdot P_\beta-3\, \overline L^\alpha L^\beta P_\alpha\cdot P_\beta\,.
\end{equation}
is the scalar potential.

For the vector fields with constant curvatures as in \eqref{FdAnsatz}, the Maxwell equations reduce to the following system of algebraic equations:
\begin{equation}\label{Maxabs}
\begin{split}
-{(1+|\zeta_1|^2)|\zeta_2|^2\over (1-|\zeta_1|^2-|\zeta_2|^2)^2}\,m_0+{(1-|\zeta_1|^2)|\zeta_2|^2\over(1-|\zeta_1|^2-|\zeta_2|^2)^2}\,(m_1+m_2+m_3) & \eql 0\,,\\[6 pt]
{|\zeta_1|^2 (4-|\zeta_2|^2)+|\zeta_2|^2\over  (1-|\zeta_1|^2-|\zeta_2|^2)^2}\,m_0-{(1+|\zeta_1|^2)|\zeta_2|^2\over (1-|\zeta_1|^2-|\zeta_2|^2)^2}\,(m_1+m_2+m_3) & \eql 0\,,
\end{split}
\end{equation} 
with the same equations for the electric parameters, $e_\alpha$. 
 
\section{Derivation of the near horizon BPS equations}
\label{appendixB}

\subsection*{C.1.\ The general set-up}
In this appendix  we outline the main steps of the truncation of the fermion supersymmetry variations in  $\mathcal{N}=8$ gauged supergravity \cite{deWit:1982bul} to the ${\rm U}(1)^2$-invariant sector,  and the derivation of the BPS equations for the dyonic black holes  used in Section~\ref{subsec:dyonicBPS}. The main difference with  the similar truncations discussed previously, such as   the STU-model in \cite{Duff:1999gh} or   the $\SU(3)\times {\rm U}(1)^\text{W}$-invariant truncation in \cite{Ahn:2000aq} or \cite{Corrado:2001nv}, is the presence of a nontrivial electric field, which precludes futher truncation to real scalar fields.  At the same time,  we  simplify our calculation by restricting to the  ${\rm U(1)}^3$-invariant bosonic fields of the  ${\rm AdS}_2\times\Sigma_\fg$  near horizon black hole Ansatz  in Section~\ref{sec:ansatz}. Hence we use the metric \eqref{AnsatzAdS2} with constants $f_0$ and $h_0$, constant scalars, $z_i$ and $z$, $i=1,2,3$, and constant electric and magnetic fluxes \eqref{FdAnsatz}.

The spin-1/2 and spin-3/2 supersymmetry variations of the $\cals N=8$ theory are given by:
\begin{align}\label{spin12}
\delta\chi^{ijk}& \eql -\cals A_\mu{}^{ijkl}\gamma^\mu\epsilon_l+{3\over 2}\gamma^{\mu\nu}\overline F{}_{\mu\nu}^{-[ij}\epsilon^{k]}
-2g A_{2l}{}^{ijk}\epsilon^l\,,\\[6 pt]
\delta\psi_\mu{}^i & \eql 2D_\mu\epsilon^i+{1\over 4}\sqrt 2 \,\overline{ F}{}^-_{\nu\rho}{}^{ij}\gamma^{\nu\rho}\gamma_\mu\epsilon_j+\sqrt 2 g A_1{}^{ij}\gamma_\mu\epsilon_j\,,\label{spin32}
\end{align}
and their complex conjugates. 
We refer the reader to  \cite{deWit:1982bul} for the  definitions and explicit formulae for the covariantized scalar kinetic tensor, $\cals A_\mu{}^{ijkl}$, and the scalar $A$-tensors, $A_1{}^{ij}$ and $A_{2l}{}^{ijk}$, which are constructed from the scalar 56-bein \eqref{56bein}. 

The gauge fields enter the variations \eqref{spin12} and \eqref{spin32} both through the ``bare'' potential, $A^{IJ}$, in the scalar kinetic tensor and the covariant derivative,  as well as through\footnote{Note that the ``bar'' here does not mean complex conjugation.}  $\overline F{}^-{}^{ij}$, which are the anti-self-dual field strengths ``dressed'' with  the scalar fields. They can be expressed in terms of the field strengths, $F^{IJ}$, by solving    the following system of equations:
\begin{equation}\label{defFbarm}
F^-{}^{IJ}\eql (u_{ij}{}^{IJ}+v_{ijIJ})\,\overline F{}^-{}^{ij}\,,\qquad   F^{-}{}^{IJ}\eql {1\over 2}(F^{IJ}-\i\,*F^{IJ})\,.
\end{equation}
The symmetry of the truncation guarantees that  $\overline F{}^-{}^{ij}$, as an $\rm SO(8)$ tensor, has the same structure as the vector potential, $A^{IJ}$, in \eqref{Avfields}, with  $A^\alpha$ replaced with $\overline F{}^{-\alpha}$. However, in the absence of a closed form general solution to \eqref{defFbarm}, one has to perform the calculation explicitly. 
The result simplifies if we use the following linear combinations: 
\begin{equation}\label{}
\begin{split}
\overline F{}^{-0}+\overline F{}^{-1}+\overline F{}^{-2}+\overline F{}^{-3} 
& \eql -{\i\over 2}\, \cfb_0 \,(e^{2f_0}\,\text{vol}_{\rm AdS_2}+\i\,e^{2h_0}\,\text{vol}_{\Sigma_\fg}) \,,\\
\overline F{}^{-0}+\overline F{}^{-1}-\overline F{}^{-2}-\overline F{}^{-3} & \eql  -{\i\over 2}\, \cfb_1 \,(e^{2f_0}\,\text{vol}_{\rm AdS_2}+\i\,e^{2h_0}\,\text{vol}_{\Sigma_\fg})\,,\\
 \overline F{}^{-0}-\overline F{}^{-1}+\overline F{}^{-2}-\overline F{}^{-3}& \eql  -{\i\over 2}\, \cfb_2 \,(e^{2f_0}\,\text{vol}_{\rm AdS_2}+\i\,e^{2h_0}\,\text{vol}_{\Sigma_\fg})\,,\\
\overline F{}^{-0}-\overline F{}^{-1}-\overline F{}^{-2}+\overline F{}^{-3}& \eql  -{\i\over 2}\, \cfb_3 \,(e^{2f_0}\,\text{vol}_{\rm AdS_2}+\i\,e^{2h_0}\,\text{vol}_{\Sigma_\fg})\,,
\end{split}
\end{equation}
that also arise in the supersymmetry variations below. Then the complex constants, $\cfb_\alpha$, are related to the electric and magnetic parameters, $e_\alpha$ and $m_\alpha$, in \eqref{FdAnsatz} by
\begin{equation}\label{Smatact}
S_{\alpha\beta}\cfb_\beta \eql e^{-2h_0}m_\alpha+\i\,e^{-2f_0}e_\alpha\,,
\end{equation}
where 
\begin{equation}\label{Smatrix}
\begin{split}
S_{\alpha0} & \eql {1\over 2}\sqrt 2\, L^\alpha\,,\\ 
S_{\alpha i} & \eql -{1\over 2}\sqrt 2 (1-|z_i|^2) \overline{D_{z_i}L^\alpha}\,,\qquad \alpha=0,\ldots,3\,,\quad i=1,\ldots,3\,.
\end{split}
\end{equation}
The symplectic sections, $L^\alpha$, have been defined in \eqref{defofLa} and their  K\"ahler covariant derivative in \eqref{thefs}.

Finally, it has been observed in \cite{Halmagyi:2013sla} that the BPS equations for near horizon black holes in $\cals N=2$ supergravity coupled to hypermultiplets must be supplemented by Maxwell equations, which    impose  massive constraints on the electric and magnetic parameters.  The same holds in our model. Indeed, setting $\zeta_1=0$ and $\zeta_2=z$ in \eqref{Maxabs} we find that, cf.\ \eqref{massA},
\begin{equation}\label{mconstr}
\begin{split}
e^{(m)}& \equiv e_0-e_1-e_2-e_3\eql 0\,,\\[6 pt]
m^{(m)} & \equiv m_0-m_1-m_2-m_4\eql 0\,.
\end{split}
\end{equation}
Implementing those constraints from the start simplifies the derivation of the BPS equations considerably. 

The spinor fields, $\chi^{ijk}$ and $\psi_\mu{}^i$, and the supersymmetry parameters, $\epsilon^i$, in \eqref{spin12} and \eqref{spin32} are $\gamma^5$-chiral. Since we are using a real representation of the $\gamma$-matrices, which makes $\gamma^5$ to be pure imaginary, the complex conjugation, which lowers/raises the $\SU(8)$ indices $i,j,k,\ldots$, changes the $\gamma^5$-chirality, for example
\begin{equation}\label{proj5}
\gamma^5 \epsilon^j\eql \epsilon^j\,,\qquad \gamma^5\epsilon_j\eql -\epsilon_j\,,\qquad \gamma^5\equiv \i\,\gamma^0\gamma^1\gamma^2\gamma^3\,.
\end{equation}
In particular, \eqref{proj5} implies
\begin{equation}\label{proj23}
\gamma^2\gamma^3\epsilon^j\eql -\i\,\gamma^0\gamma^1\epsilon^j\,,\qquad 
\gamma^2\gamma^3\epsilon_j\eql \i\,\gamma^0\gamma^1\epsilon_j\,.
\end{equation}
It follows from \eqref{decom8} that there are two U(1)$^2$-invariant supersymmetry parameters, $\epsilon^7$ and $\epsilon^8$. In the following we set 
\begin{equation}\label{noninveps}
\epsilon^1\eql\ldots\eql \epsilon^6\eql 0\,,
\end{equation}
and relabel  $\epsilon^7$ and $\epsilon^8$ as $\epsilon^1$ and $\epsilon^2$, respectively. 

The condition for a supersymmetric solution is that \eqref{spin12} and \eqref{spin32} vanish.  Here, we are interested in solutions for which the Killing spinors, $\epsilon^i$, of unbroken supersymmetries are constant along the Riemann surface and  the usual Killing spinors along AdS$_2$. In the coordinate system we are using, this  means that $\epsilon^i$ do not  depend on $t$, $x$ and $y$ and satisfy
\begin{equation}\label{raddep}
\partial_r\epsilon^i\eql -{1\over 2r}\epsilon^i\,,\qquad i=1,2\,,
\end{equation}
along the radial coordinate, $r$. In addition to the spinors obeying \eqref{raddep} there are also the ``conformal Killing spinors'' in AdS$_2$ dual to the S-type supercharges in the 1d superconformal quantum mechanics.

\subsection*{C.2\quad The spin-1/2 variations}

After imposing \eqref{noninveps} in  \eqref{spin12}, the only nonvanishing variations are for the U(1)$^2$-invariant  fields in \eqref{spin12v} and \eqref{spin12h}. Setting the variations of \eqref{spin12v} to zero and using \eqref{proj23}, yields  three pairs of equations of the form
\begin{equation}\label{spin12va}
\begin{split}
\cfb_i \epsilon^1+2\,\i\,g\,\fF_i\,\gamma^0\gamma^1\epsilon^2 & \eql 0\,, \\[6 pt]
\cfb_i \epsilon^2-2\,\i\,g\,\fF_i\,\gamma^0\gamma^1\epsilon^1 & \eql 0\,,\qquad i=1,\ldots,3\,,
\end{split}
\end{equation}
where
\begin{equation}\label{}
\begin{split}
\fF_1 & \eql  e^{K_S/2} \Big[{|z|^2\over 1-|z|^2}(1-\bar z_1)(1-z_2)(1-z_3)+{2\over 1-|z|^2}(\bar z_1-z_2z_3) \Big]\,, \end{split}
\end{equation}
with $\fF_2$ and $\fF_3$ obtained by the other two cyclic permutations of $z_1$, $z_2$ and $z_3$. Equations \eqref{spin12va} reduce to the projector
\begin{equation}\label{proj12}  
\gamma^0\gamma^1\epsilon^1\eql \i\,\xi \,\epsilon^2\,,\qquad \gamma^0\gamma^1\epsilon^2\eql -\i\,\xi \, \epsilon^2\,,\qquad \xi=\pm 1\,,
\end{equation}
and three BPS equations
\begin{equation}\label{BPSfi}
\cfb_i\eql -2\xi g\,\fF_i\,,\qquad i=1,\ldots,3.
\end{equation}

Using the massive constraints \eqref{mconstr},  the  variations of \eqref{spin12h} simplify to
\begin{equation}\label{Gvars12}
\begin{split}
\fGG\,\Big[{z-\bar z\over 1-|z|^2}\,\epsilon^1-\i\,{z+\bar z\over 1-|z|^2}\epsilon^2\Big]\eql 0\,,\\[6 pt]
\fGG\,\Big[{z+\bar z\over 1-|z|^2}\,\epsilon^1-\i\,{z-\bar z\over 1-|z|^2}\epsilon^2\Big]\eql 0\,,\\
\end{split}
\end{equation}
where
\begin{equation}\label{}
\fGG\eql e^{K_S/2}\big[(1-z_1)(1-z_2)(1-z_3)+2(z_1z_2z_3-1)\big]\,.
\end{equation}
Clearly, \eqref{Gvars12} vanish identically when we turn-off the hypermultiplet and thus are  absent in the STU-model. For a nontrivial hypermultiplet, $z\not=0$, they imply the BPS equation
\begin{equation}\label{Gequation}
\fGG\eql 0\,,
\end{equation}
which is a cubic constraint on the scalars, $z_i$. 

\subsection*{C.3\quad The spin-3/2 variations}

We now turn to the spin-3/2 variations  \eqref{spin32}. Using  \eqref{proj23}, \eqref{proj12} and \eqref{mconstr} in the variations $\gamma^2\delta\psi_x{}^{7,8}+\gamma^3\delta\psi_y{}^{7,8}$, we find 
\begin{equation}\label{a1eqs}
m_0\eql \kappa\,{\xi\over 2g}\,,
\end{equation}
where $\kappa=\pm 1$ or 0 is the normalized curvature of $\Sigma_\fg$, see Appendix~\ref{appconv}. The difference of the two variations yields 
\begin{equation}\label{Phi0eqs}
\cfb_0\eql 2g\xi\,\overline \fW\,,
\end{equation}
where
\begin{equation}\label{}
\fW\eql  e^{K_V/2}\Big[{|z|^2\over 1-|z|^2}(1-z_1)(1-z_2)(1-z_3)+{2\over 1-|z|^2}(z_1z_2z_3-1)\Big]\,.
\end{equation}

The variations, $\delta\psi^{7,8}_r$, along the radial directions, assuming \eqref{raddep}, give
\begin{equation}\label{}
\begin{split}
\gamma^0\epsilon^1 & \eql   \sqrt 2\,\i\,g\xi\,e^{f_0}\fW\,\epsilon_2\,,\\[6 pt]
\gamma^0\epsilon^2 & \eql - \sqrt 2\,\i\,g\xi\,e^{f_0}\fW\,\epsilon_1\,.\\
\end{split}
\end{equation}
Taken together with their complex conjugates, they yield the projector
\begin{equation}\label{projLa}
\gamma^0\epsilon^1\eql e^{\i\,\Lambda}\epsilon_2\,,\qquad \gamma^0\epsilon^2\eql -e^{\i\,\Lambda}\epsilon_1\,,
\end{equation}
and the BPS equation
\begin{equation}\label{apsolf0}
e^{-f_0}\eql -  \sqrt 2\,\i\,g\xi\,e^{\i\,\Lambda}\,\fW\,,
\end{equation}
where $\Lambda$ is a constant.

Finally, using all the projectors as well as \eqref{Phi0eqs}
 and \eqref{apsolf0}, the variations $\delta\psi_t{}^{7,8}$ set
\begin{equation}\label{e0eqss}
e_0\eql 0\,.
\end{equation}
This concludes our truncation of the supersymmetry variations \eqref{spin12} and \eqref{spin32}.

\subsection*{C.4\quad Summary and comments}

We have shown that the truncation of the $\cals N=8$ supersymmetry variations and the Maxwell  equations resulted in:
\begin{itemize}
\item [(i)] Four real equations \eqref{mconstr}, \eqref{a1eqs} and \eqref{e0eqss} for the electric and magnetic parameters $e_\alpha$ and $m_\alpha$.
\item [(ii)] Four complex equations \eqref{BPSfi} and \eqref{Phi0eqs} for the scalar dressed components, $\Phi_\alpha$,  of the fluxes.
\item [(iii)] One complex equation \eqref{apsolf0} for the metric constant, $f_0$, and the phase $\Lambda$.
\item [(iv)] A complex cubic constraint \eqref{Gequation} for the scalars, $z_i$.
\end{itemize}

Using the geometric data of the corresponding $\cals N=2$ supergravity derived in Appendix~\ref{appendixA}, we have verified that our BPS equations above agree, modulo differences in conventions, with those derived for the near horizon black holes in general $\cals N=2$ $d=4$ gauged supergravities coupled to hypermultiplets  in \cite{Halmagyi:2013sla}. 
In fact, a comparison with the $\cals N=2$ formulae suggests some simplifications.  In particular, we have
\begin{equation}\label{}
\fF_i\eql -(1-|z_i|^2)\,D_{z_i}\fW\,,\qquad i=1,2,3\,.
\end{equation}
Those identities turn out  useful for solving the BPS equations in Section~\ref{subsec:dyonicBPS} using some standard identities of the special K\"ahler geometry \cite{Ceresole:1995ca,Andrianopoli:1996vr} and to rewrite them as  attractor equations in Appendix~\ref{appattractor}.

Finally, note that the equations above are invariant under \cite{Halmagyi:2013sla}
\begin{equation}\label{}
(e_\alpha,m_\alpha,\Lambda,\xi)\qquad \longrightarrow\qquad (-e_\alpha,-m_\alpha,\Lambda+\pi,-\xi)\,,
\end{equation}
so that we may set $\xi=-1$ for convenience.

\section{The attractor equations}
\label{appattractor}

In Section~\ref{sec:DyonicComp}, we obtained a match between the black hole entropy and topologically twisted index by explicitly solving the BPS equations and extremization equations for the supergravity scalar fields and field theory fugacities, respectively.
In this section we show an alternative method to achieve a match between the twisted index and the black hole entropy.
In particular, we will solve a subset of the BPS equations for $e^{2h_0}$ as a function of the scalar fields and the electric and magnetic charges and show that the remaining BPS equations imply that $e^{2h_0}$ is extremized with respect to the scalar fields. This allows for a comparison with the topologically twisted index and its extremization with respect to the fugacities and Lagrange multipliers. This procedure is the same as the AdS$_4$ black hole attractor  mechanism discussed in \cite{Benini:2016rke,DallAgata:2010ejj,Hristov:2010ri,Gnecchi:2013mta,Halmagyi:2013sla,Klemm:2016wng}.

It is crucial for our analysis to work with the electric charges, $q_\alpha$, defined in \eqref{msandqssugra}. As explained in Sections \ref{sec422} and \ref{sec:eleccharg}, equation \eqref{nqcharges} leads to different electric charges depending on whether the massive condition \eqref{massA} is imposed in the Maxwell Lagrangian before or after varying with respect to $F^\alpha$.  In this appendix we choose the latter and as a consequence we will compare  the resulting entropy with the field theory in Section \ref{sec:manjmditc}. 

Using \eqref{usefulrel} and \eqref{Smatrixmn}, we can write \eqref{Smatactmn} as 
\begin{equation}\label{fs}
\Phi_\alpha = 4 \,\i\, e^{-2h_0} \overline{S}_{\beta \alpha} \left(q_\beta - \overline{\mathcal{N}}_{\beta\sigma} m_\sigma\right) \,.
\end{equation}
Combining \eqref{fs} with the identity 
\begin{equation}
L^\beta \mathcal{N}_{\alpha\beta} = M_\alpha \equiv e^{K_V/2} \frac{\partial F}{\partial X^\alpha} \,,
\end{equation}
we can rewrite the BPS equation \eqref{Ph0eqs} as
\begin{equation}\label{set1}
\widehat{\mathcal{Z}} \equiv \sqrt{2} e^{-2h_0} \left(L^\alpha q_\alpha - M_\alpha m_\alpha\right) +  {\rm i} \,  g \, \fW = 0 \,.
\end{equation}
With the use of the identity \cite{Andrianopoli:1996vr}
\begin{equation}
D_{z_i}\left(L^\alpha \mathcal{N}_{\alpha\beta}\right) =\left( D_{z_i} L^\alpha \right) \overline{\mathcal{N}}_{\alpha\beta} \,,
\end{equation}
we can furthermore write \eqref{Phieqs} as
\begin{equation}\label{set2}
D_{z_i} \widehat{\mathcal{Z}} = 0 \,.
\end{equation}
Since only $\fW$ depends explicitly on $z$, we can similarly write \eqref{thezconstr} as
\begin{equation}\label{set3}
\frac{\partial \widehat{\mathcal{Z}}}{\partial z} = 0 \,.
\end{equation}

Now, we note that the equations \eqref{set1}, \eqref{set2} and \eqref{set3} imply  the following suggestive set of equations
\begin{equation}\label{e2h0rewrite}
\begin{split}
e^{2h_0} = \sqrt{2}\, {\rm i}\, \frac{L^\alpha q_\alpha - M_\alpha m_\alpha}{ g \,\fW} \,, \qquad\qquad \frac{\partial e^{2h_0}}{\partial z_i} = \frac{\partial e^{2h_0}}{\partial z} = 0 \,,
\end{split}
\end{equation}
where $e^{2h_0}$ is a function of the electric charges, $q_\alpha$, magnetic fluxes, $m_\alpha$, and the scalars, $z_i$ and $z$.
The BPS equations thus imply an extremization procedure for the metric coefficient $e^{2h_0}$ as a function of the scalar fields.
Using \eqref{calWL} and implementing the relation \eqref{bhzus} between $L^\alpha$ and $u_\alpha$ to write 
\begin{equation}\label{e2h0expl}
e^{2h_0} = \frac{1}{g} \frac{\sum_{\alpha=1}^4 \sqrt{u_1u_2u_3u_4}\,{ \frac{m_{\alpha-1}}{u_{\alpha}}}-{\rm i} \,u_\alpha q_{\alpha-1}}{2u_1-\frac{1}{1-z \bar z}\left(u_1-u_2-u_3-u_4\right)} \,.
\end{equation}

We are now in a position to see how our equation manipulations above pay off. Note that it follows trivially from  \eqref{bhzus} that the $u_\alpha$'s satisfy $\sum_{\alpha=1}^{4} u_\alpha = 2$. In addition the last equation in \eqref{e2h0rewrite} combined with \eqref{e2h0expl} implements the massive constraint $u_1=u_2+u_3+u_4$. Implementing the massive constraint in \eqref{e2h0expl} we can write
\begin{equation}\label{e2h0nice}
e^{2h_0} = \frac{1}{2g} \sum_{\alpha=1}^4 \left(\sqrt{u_1u_2u_3u_4}\,\frac{m_{\alpha-1}}{u_{\alpha}}-{\rm i}\, u_\alpha q_{\alpha-1} \right) \,.
\end{equation}
After identifying the field theory and supergravity charges as in \eqref{nabjmcom} and \eqref{qabjmcomp} we observe that the entropy \eqref{Sdeff} with $e^{2h_0}$ replaced by \eqref{e2h0nice} takes the same functional form as the twisted index \eqref{IdABJM2c}.
Equation \eqref{e2h0rewrite} then implies that the entropy is extremized with respect to the $u_\alpha$, ensuring that the same extremization principle applies to both the topologically twisted index and the black hole entropy.

We have not yet discussed the BPS constraints \eqref{e0eqs}-\eqref{m123eqs} on the charges.
In supergravity we start with four magnetic fluxes and four electric parameters which satisfy the above four constraints.
While the constraints act linearly on the electric parameters $e_\alpha$, they act non-linearly on the electric charges $q_\alpha$. In field theory we start off with four magnetic charges and four electric charges and implement the two constraints \eqref{massconstr} and \eqref{toptwist} on the magnetic charges.
The constraints \eqref{m0eqs} and \eqref{m123eqs} are equivalent to the constraints \eqref{massconstr} and \eqref{toptwist}.
One more constraint is imposed on the electric charges by imposing the index to be real.
Indeed, the BPS constraints \eqref{e0eqs}-\eqref{m123eqs} are crucial to ensure that $e^{2h_0}$ is real.
However, there is no further constraint on the electric charges in the field theory and there is in fact a shift symmetry which allows us to shift the electric charges by a free parameter as in \eqref{shiftsymm}.
The supergravity computation has thus fixed the shift symmetry in a particular way.
Explicit comparison shows that the shift symmetry is fixed such that the Lagrange multipliers $\lambda_1,\lambda_2$ in Section~\ref{sec:manjmditc} are real, i.e. $\nu_1=\nu_2=0$.

In conclusion, both in supergravity and field theory we are evaluating the same expression subject to the same extremization equations.
Imposing the Lagrange multipliers to be real then ensures that also the constraints on the charges coincide.
We can thus conclude that the black hole entropy and the extremized topologically twisted index are equal.



\bibliography{AdS2-CPW}
\bibliographystyle{JHEP}

\end{document}